\documentclass[aps,pra,twocolumn,showpacs,superscriptaddress,longbibliography]{revtex4-1}
\usepackage{graphicx} 
\usepackage{amsmath}
\usepackage{graphicx,epstopdf}
\usepackage{gensymb}
\newcommand{\be}{\begin{equation}}
\newcommand{\ee}{\end{equation}}
\newcommand{\bea}{\begin{eqnarray}}
\newcommand{\eea}{\end{eqnarray}}
\newcommand{\bse}{\begin{subequations}}
	\newcommand{\ese}{\end{subequations}}

\usepackage{color}
\usepackage[colorlinks,bookmarks=false,citecolor=darkblue,linkcolor=red,urlcolor=blue]{hyperref}

\definecolor{darkred}{rgb}{0.7,0.0,0.0}

\definecolor{darkblue}{rgb}{0,0.02,0.45}

\definecolor{darkgreen}{rgb}{0.02,0.45,0.0}

\definecolor{violet}{rgb}{0.8,0.2,0.6}

\begin{document}

\title{Bose-Einstein condensation of triplons close to the quantum critical point in the quasi-one-dimensional spin-$1/2$ antiferromagnet NaVOPO$_4$}

\author{Prashanta K. Mukharjee}
\affiliation{School of Physics, Indian Institute of Science Education and Research, Thiruvananthapuram-695551, India}
\author{K. M. Ranjith}
\author{B. Koo}
\author{J. Sichelschmidt}
\author{M. Baenitz}
\affiliation{Max Planck Institute for Chemical Physics of Solids, N$\ddot{o}$thnitzer Str. 40, 01187 Dresden, Germany}
\author{Y. Skourski}
\affiliation{Dresden High Magnetic Field Laboratory, Helmholtz-Zentrum Dresden-Rossendorf, 01314 Dresden, Germany}
\author{Y. Inagaki}
\affiliation{Department of Applied Quantum Physics, Faculty of Engineering, Kyushu University, Fukuoka 819-0395, Japan}
\affiliation{Ames Laboratory, U.S. Department of Energy, Iowa State University, Ames, Iowa 50011, USA}
\author{Y. Furukawa}
\affiliation{Ames Laboratory, U.S. Department of Energy, Iowa State University, Ames, Iowa 50011, USA}
\author{A. A. Tsirlin}
\affiliation{Experimental Physics VI, Center for Electronic Correlations and Magnetism, Institute of Physics, University of Augsburg, 86135 Augsburg, Germany}
\author{R. Nath}
\email{rnath@iisertvm.ac.in}
\affiliation{School of Physics, Indian Institute of Science Education and Research, Thiruvananthapuram-695551, India}
\date{\today}

\begin{abstract}
Structural and magnetic properties of a quasi-one-dimensional spin-$1/2$ compound NaVOPO$_4$ are explored by x-ray diffraction, magnetic susceptibility, high-field magnetization, specific heat, electron spin resonance, and $^{31}$P nuclear magnetic resonance measurements, as well as complementary \textit{ab initio} calculations. Whereas magnetic susceptibility of NaVOPO$_4$ may be compatible with the gapless uniform spin chain model, detailed examination of the crystal structure reveals a weak alternation of the exchange couplings with the alternation ratio $\alpha\simeq 0.98$ and the ensuing zero-field spin gap $\Delta_{0}/k_{\rm B} \simeq 2.4$~K directly probed by field-dependent magnetization measurements. No long-range order is observed down to 50\,mK in zero field. However, applied fields above the critical field $H_{c1}\simeq 1.6$\,T give rise to a magnetic ordering transition with the phase boundary $T_{\rm N} \propto {(H - H_{\rm c1})^{\frac{1}{\phi}}}$, where $\phi \simeq 1.8$ is close to the value expected for Bose-Einstein condensation of triplons. With its weak alternation of the exchange couplings and small spin gap, NaVOPO$_4$ lies close to the quantum critical point.

\end{abstract}

\maketitle

\section{\textbf{Introduction}}
Quantum phase transitions in low-dimensional antiferromagnets (AFM) remain one of the most fascinating topics in condensed-matter physics.\cite{Sachdev2007,Sachdev475,Sachdev173}
One-dimensional (1D) Heisenberg spin-$1/2$ chains are prone to quantum fluctuations that give rise to rich ground-state properties. An individual spin-$1/2$ chain does not show any long-range magnetic order and features gapless excitations. However, in real materials due to the presence of inherent interchain couplings a three-dimensional (3D) long-range-order (LRO) will usually occur, albeit at temperatures much lower than the intrachain coupling.\cite{Kojima1787} Alternating spin-$1/2$ chains, where two nonequivalent couplings interchange along the chain, are different. Their energy spectrum is gapped,\cite{Barnes11384,Yamauchi3729} and this	gap protects the system from long-range ordering, unless interchain couplings are strong enough to close the gap. 

In spin-gap materials, the elementary excitations are spin-1 triplons. A magnetic field reduces the gap and causes an AFM LRO when the field exceeds the critical value of the gap closing. This magnetic order is different from the classical magnetic ordering in many aspects and can be described as a Bose-Einstein Condensation (BEC) of dilute triplons.\cite{Rice760} Typically, BEC is being realized in gapped materials based on spin dimers with two-dimensional (2D) and 3D interdimer correlations at low temperatures.\cite{Giamarchi198,Nikuni5868,Aczel207203,Hirata174406} Another field-induced phenomenon observed in gapped spin systems is the Tomanaga-Lutinger Liquid (TLL) phase which is mostly expected for 1D systems.\cite{Klanjsek137207,Hong137207,Willenberg060407} Additionally, gapped spin systems also use to exhibit several other field-induced features such as Wigner crystallization of magnons, \cite{Horsch076403} magnetization plataeus, \cite{Kageyama3168} etc. Therefore, 1D materials with 3D interchain couplings and spin ladders, which are essentially coupled spin chains, can provide the opportunity to study both BEC and TLL physics in the same system.\cite{Mukhopadhyay177206,Thielemann020408,Willenberg060407} Such phases occur between two critical fields $H_{\rm c1}$ and $H_{\rm c2}$, which depend on the hierarchy of coupling strengths. Experimental access to these fields requires appropriate design of the material with an appreciably small spin gap which will allow for a complete exploration of the field ($H$) vs temperature ($T$) phase diagram. In the past, an ample number of low-dimensional magnets were studied extensively, but most of them feature a large spin gap. Therefore, a very high field is applied to study the field-induced effects in all these materials.\cite{Aczel207203,Jaime087203,Aczel100409,Samulon047202}

Herein, we have carried out a systematic investigation of the structural and magnetic properties of the quasi-1D compound NaVOPO$_4$ by performing the temperature dependent x-ray diffraction, magnetization, specific heat, ESR, and $^{31}$P NMR experiments. Our experiments are also accompanied by the density-functional band structure calculations. We compare NaVOPO$_4$ with the isostructural compounds NaVOAsO$_4$ and AgVOAsO$_4$ to understand the underlying structure-property relationship.\cite{Arjun014421,Ahmed224433,Tsirlin144412,Weickert2019} 

\begin{figure*}
	\includegraphics [scale = 0.8]{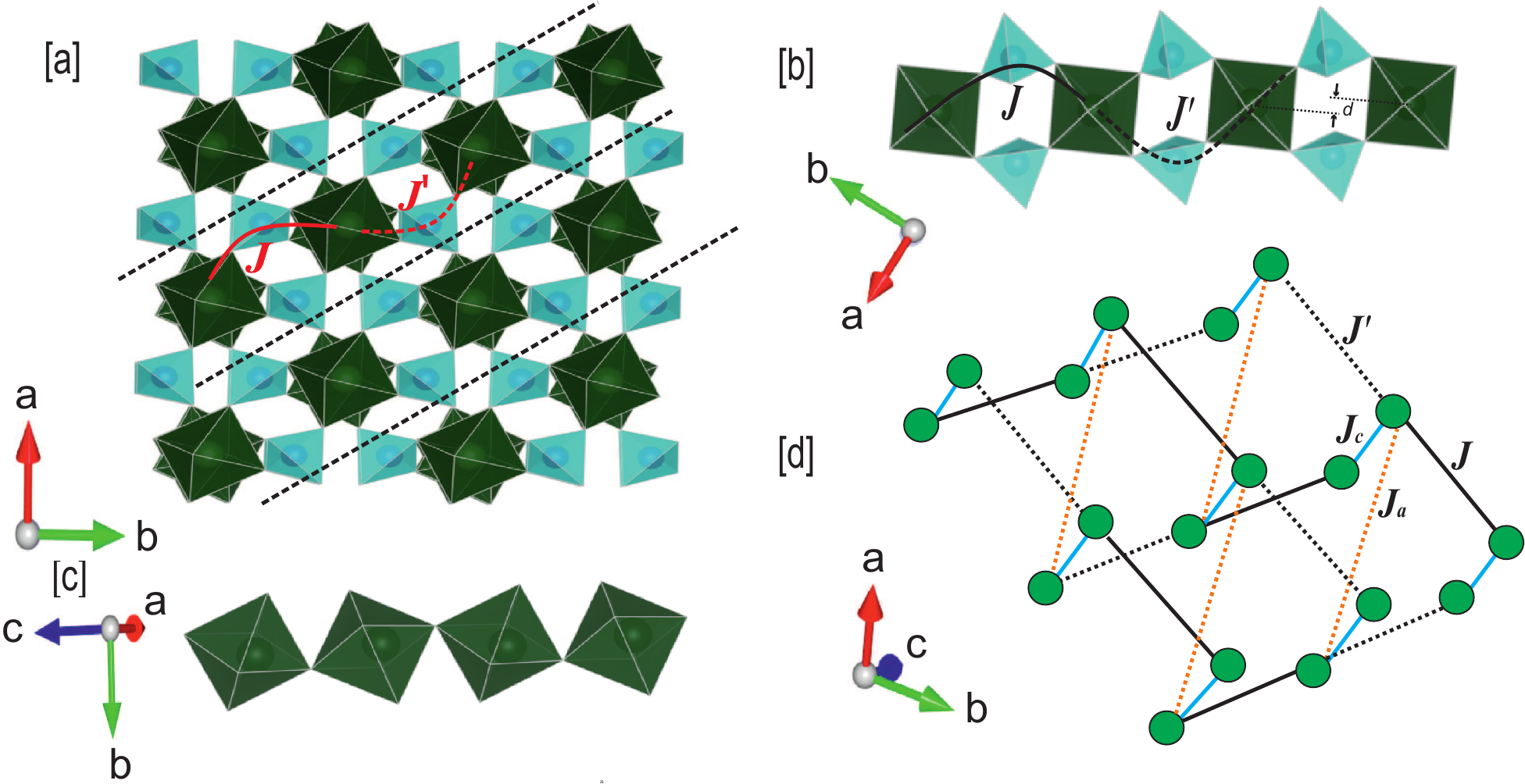}\\
	\caption{(a) Crystal structure of NaVOPO$_4$ projected in the $ab$-plane. The spin chains are separated by the black dashed lines. The VO$_{6}$ and PO$_{4}$ polyhedra are shown in green and blue colors, respectively. (b) A section of the magnetic (bond-alternating) chain with the intrachain couplings $J$ and $J^\prime$. (c) A representative structural chain extending along the $c$-axis (d) Sketch of the spin lattice showing all relevant exchange interactions.}
	\label{Fig1}
\end{figure*}

The aforementioned V$^{4+}$ phosphates and arsenates crystallize in the monoclinic crystal structure (space group $P2_{1}/c$). One peculiarity of this series is that the structural chains formed by the corner-sharing VO$_6$ octahedra do not represent the magnetic chains~\cite{Tsirlin144412,Weickert2019}. Band structure calculations reveal that the magnetic chains run along the extended V--O--As--O--V path, whereas the shorter V--O--V path along the structural chains gives rise to a weak and ferromagnetic (FM) interaction $J_{\rm c} \simeq -5$~K. There exists another weak AFM interaction $J_{\rm a} \simeq 8$~K, which along with $J_{\rm c}$ constitutes a frustrated 3D interaction network between the chains [same as Fig.~\ref{Fig1}(d)]. A quantitative comparison of the exchange couplings in NaVOAsO$_4$ with AgVOAsO$_4$ is made in Ref.~[\onlinecite{Arjun014421}]. The Na compound is found to be more close to the 1D regime due to the stronger intrachain and weaker interchain exchange couplings. Consequently, its spin gap is larger than in the Ag analogue.


The replacement of As$^{5+}$ (ionic radius $\simeq ~0.33$~\AA) with P$^{5+}$ (ionic radius $\simeq ~0.17$~\AA) leaves the crystal structure unchanged, but the lattice parameters are reduced, thus shrinking the unit cell volume from $378.72$~\AA$^3$ to $354.44$~\AA$^3$. Such a reduction of the cell volume may bring in a drastic change in the magnetic properties. Here we show that, compared to NaVOAsO$_{4}$ and AgVOAsO$_4$, NaVOPO$_{4}$ has a much lower value of $H_{\rm c1}$ and $\Delta_{0}/k_{\rm B}$, thus lying very close to the quantum critical point (QCP) between the 3D LRO and gapped ground state.

NaVOPO$_{4}$ contains one Na, V, P atom each and five nonequivalent oxygen atoms. Each V is coordinated by six oxygen atoms forming a distorted VO$_6$ octahedron. In each octahedron, four equatorial V--O bonds are within the range $1.98- 2.01$~\AA, while the alternating short and long bond distances of V with O(1) are $1.62~$ \AA and $2.13$~\AA, respectively. Similarly, each phosphorous atom forms a nearly regular tetrahedron bearing the P--O distances of $\sim 1.54$~\AA. The distorted VO$_6$ octahedra are corner-shared through O(1) forming parallel structural chains along the crystallographic $c$-axis [Fig.~\ref{Fig1}(c)]. Each PO$_4$ tetrahedron is connected to four nearest VO$_6$ octahedra through the O(5) and O(2) corners forming bond-alternating spin chains arranged perpendicular to each other and spread in the $ab$-plane. Figure~\ref{Fig1}(b) shows a section of the magnetic chain though the extended V--O--P--O--V path. The perfect PO$_4$ tetrahedra also bridge the neighbouring chains into a three-dimensional network [see Fig.~\ref{Fig1}(a)], similar to AgVOAsO$_{4}$ and NaVOAsO$_{4}$. The possible exchange couplings in this spin system are depicted in Fig.~\ref{Fig1}(d). The weak interchain couplings $J_{\rm c}$ and $J_{\rm a}$ along the V--O--V and V--O--P--O--V bonds, respectively, create frustrated interactions between the magnetic chains. Clearly, P atoms are strongly coupled to the V$^{4+}$ ions, while Na atoms positioned between the chains are weakly coupled to the magnetic ions.

\section{\textbf{Methods}}
Polycrystalline sample of NaVOPO$_4$ was prepared by a conventional solid-state reaction method using the stoichiometric mixture of NaPO$_3$ and VO$_2$ (Aldrich, 99.995\%). The NaPO$_3$ precursor was obtained by heating NaH$_2$PO$_4$.H$_{2}$O (Aldrich, 99.995\%) for 4~hrs at 400~$^{\circ}$C in air. The reactants were ground thoroughly, pelletized, and fired at 720~$^{\circ}$C for two days in flowing argon atmosphere with two intermediate grindings. Phase purity of the sample was confirmed by powder x-ray diffraction (XRD) recorded at room temperature using a PANalytical powder diffractometer (Cu\textit{K}$_{\alpha}$ radiation, $\lambda_{\rm avg}\simeq 1.5418$~{\AA}). In order to check if any structural distortion is present, temperature-dependent powder XRD was performed over a broad temperature range (15~K~$\leq T\leq 600$~K). For low-temperature measurements, a low-$T$ attachment (Oxford
Phenix) while for high-temperature measurements, a high-$T$ oven attachment (Anton-Paar HTK 1200N) to the x-ray diffractometer were used. Rietveld refinement of the acquired data was performed using the \verb"FULLPROF" software package.\cite{Carvajal55}

Magnetization ($M$) was measured as a function of temperature ($0.5$~K$\leq T \leq 380$~K) and magnetic field ($H$). The measurements above 2~K were performed using the vibrating sample magnetometer (VSM) attachment to the Physical Property Measurement System [PPMS, Quantum Design]. For $T \leq 2$~K, the measurements were done using an $^{3}$He attachment to the SQUID magnetometer [MPMS-7T, Quantum
Design]. Specific heat ($C_{\rm p}$) as a function of temperature was measured down to $0.38$~K using the thermal relaxation technique in PPMS under magnetic fields up to $14$~T. For $T \leq 2$~K, the measurements were performed using an additional $^{3}$He attachment to the PPMS. High-field magnetization was measured in pulsed magnetic field up to $60$~T at the Dresden High Magnetic Field Laboratory. The details describing the measurement procedure are given in Refs.~[\onlinecite{Tsirlin132407,Skourski214420}].

The ESR experiments were carried out on fine-powdered sample with a standard continuous-wave spectrometer from $3$~K to $290$~K. We measured the power $P$ absorbed by the sample from a transverse magnetic microwave field (X-band, $\nu\simeq 9.4$~GHz) as a function of the external magnetic field $H$. A lock-in technique was used to enhance the signal-to-noise ratio which results the derivative of the resonance signal $dP/dH$.

The NMR experiments on the $^{31}$P nuclei (nuclear spin $I = 1/2$ and gyromagnetic ratio $\gamma/2\pi=17.235$~MHz/T) were carried out using pulsed NMR technique over a wide temperature range $0.05$~K~$\leq T \leq 250$~K and at different magnetic fields. For $T\geq 2$~K, all the experiments were performed using a $^4$He cryostat and a Tecmag spectrometer while in the 0.05~K~$\leq T \leq 10$~K range, a $^3$He/$^4$He dilution refrigerator (Kelvinox, Oxford Instruments) with the resonant circuit inside the mixing chamber was used. The $^{31}$P NMR spectra as a function of temperature were obtained either by performing Fourier transform (FT) of the spin echo signal at a fixed field after a $\pi/2 - \pi$ pulse sequence or by sweeping the magnetic field keeping the transmitter frequency constant. The NMR shift, $K(T)=[\nu(T)-\nu_{\rm ref}]/\nu_{\rm ref}$, was determined by measuring the resonance frequency $\nu(T)$ of the sample with respect to the standard H$_3$PO$_4$ sample (resonance frequency $\nu_{\rm ref}$). The $^{31}$P nuclear spin-lattice relaxation rate ($1/T_1$) was measured using a standard saturation pulse sequence.

Magnetic exchange couplings were obtained from density-functional (DFT) band-structure calculations performed in the FPLO code~\cite{Koepernik1743} with the generalized gradient approximation (GGA) for the exchange-correlation potential.\cite{Perdew3865} Two complementary approaches to the evaluation of exchange couplings were used, as explained in Refs.~\onlinecite{Tsirlin144412,Arjun014421} and in Sec.~\ref{sec:model} below. Correlation effects in the V $3d$ shell were taken into account on the mean-field DFT+$U$ level with the on-site Coulomb repulsion $U_d=4$\,eV, Hund's exchange $J_d=1$\,eV, and around-mean-field double-counting correction.\cite{Arjun014421}

\section{\textbf{Results and Analysis}}
\subsection{X-ray Diffraction}
\begin{figure}[h]
	\includegraphics [width = 7cm]{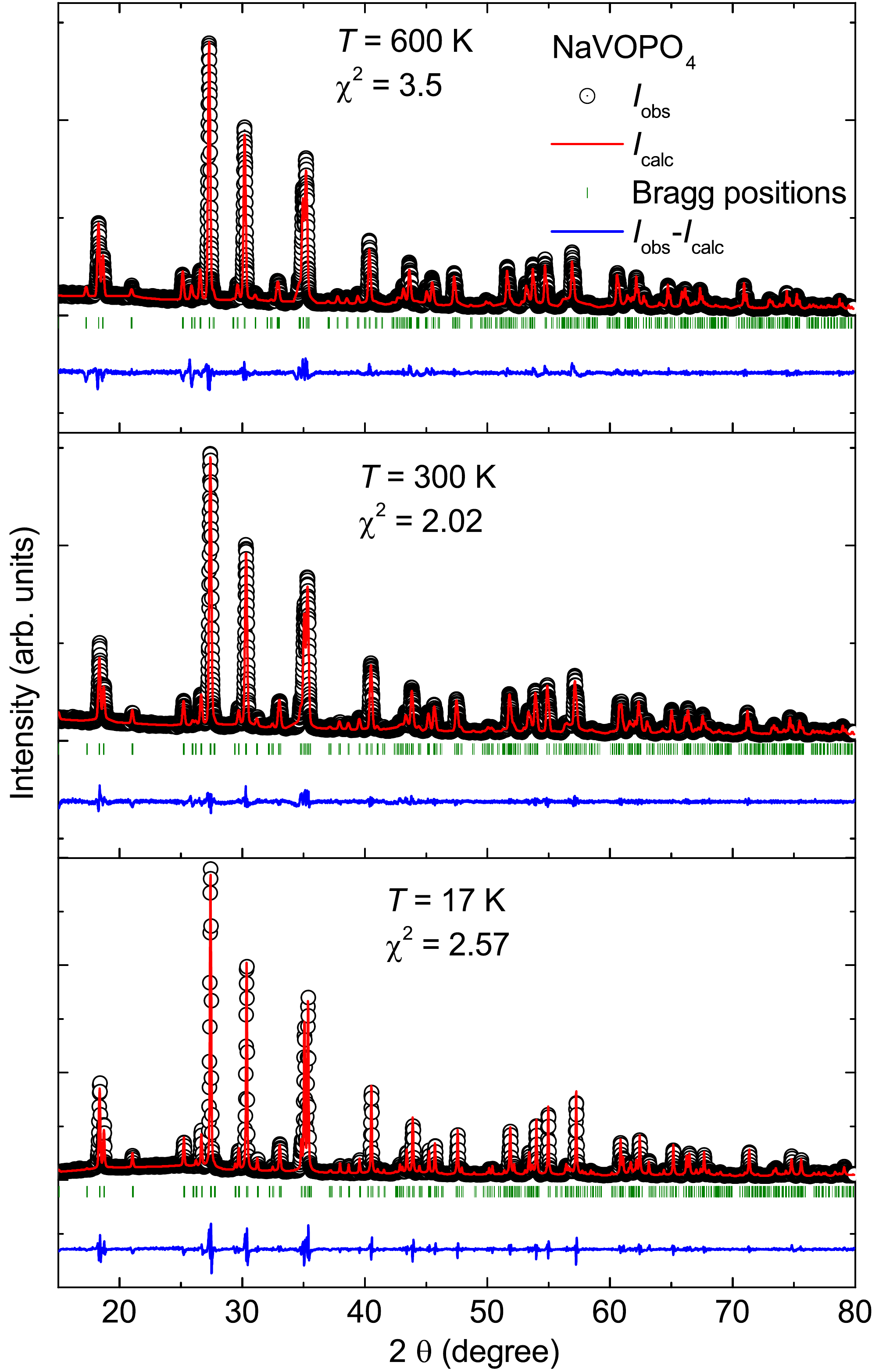}\\
	\caption{Powder x-ray diffraction data of NaVOPO$_4$ at $600$~K, $300$~K, and $17$~K. The solid lines denote the Rietveld refinement of the data. The Bragg peak positions are indicated by green vertical bars, and the bottom solid line indicates the difference between the experimental and calculated intensities.}
	\label{Fig2}
\end{figure}

Figure~\ref{Fig2} displays the powder XRD pattern of NaVOPO$_4$ at three different temperatures ($T = 600$~K, 300~K, and 17~K). The room-temperature diffraction pattern reveals that NaVOPO$_4$ crystallizes in the primitive monoclinic unit cell with the space group $P2_{1}/c$ and contains $Z = 4$ formula units per unit cell. The lattice parameters obtained from the Rietveld refinement at room temperature are $a = 6.520(1)$~\AA, $b = 8.446(1)$~\AA, $c = 7.115(1)$~\AA, $\beta = 115.260(8)^{\circ}$, and the unit cell volume $V_{\rm cell} \simeq 354.44$~\AA$^{3}$ which are consistent with the previous report.\cite{Lii67} 

\begin{figure}
	\includegraphics [width = \linewidth]{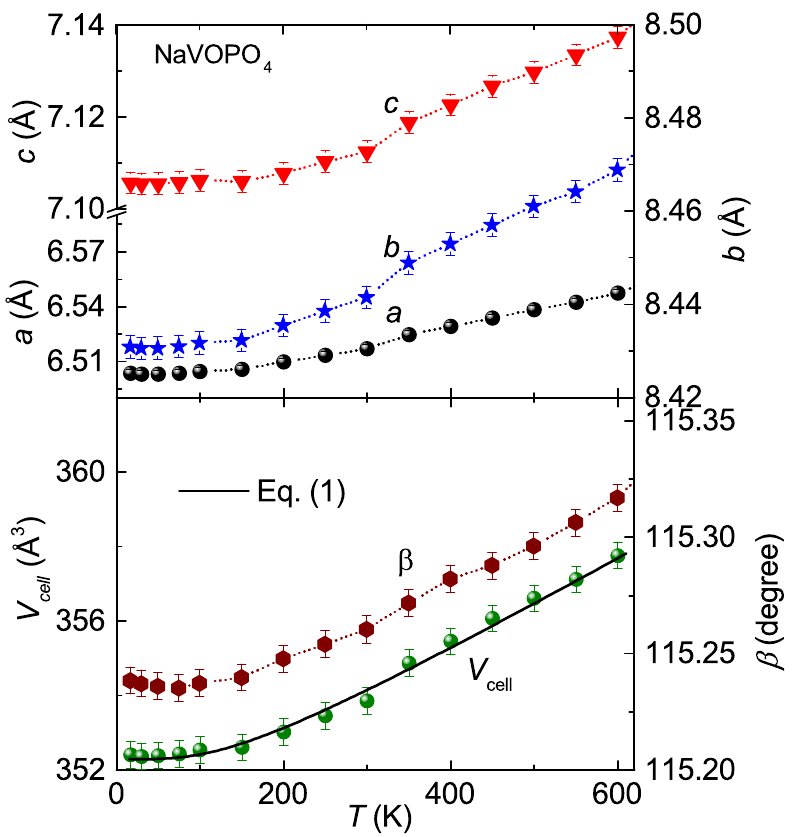}\\
	\caption{The lattice constants ($a$, $b$, $c$), monoclinic angle ($\beta$), and unit cell volume ($V_{\rm cell}$) are plotted as a function of temperature. The solid line is the fit of $V_{\rm cell}(T)$ by Eq.~\eqref{VcellvsT}.}
	\label{Fig3}
\end{figure}

Some of the spin-gap compounds show structural distortion upon cooling.\cite{Isobe1423,Hirota736,Isobe1178,Lepine3585} In our case, low-temperature XRD measurements did not show any noticeable changes, suggesting that there is no structural distortion down to at least 15~K. The temperature variation of the lattice parameters obtained from the Rietveld analysis is plotted in Fig.~\ref{Fig3}. All the lattice parameters decrease in a systematic manner with no sign of any structural distortion. To estimate the average Debye temperature ($\theta_{D}$), we fit the temperature variation of the unit cell volume [$V_{\rm cell}(T)$] using the equation\cite{Islam174432,Bag144436}
\begin{equation}
V(T)=\gamma U(T)/K_0+V_0,
\label{VcellvsT}
\end{equation}
where $V_{0}$ is the unit cell volume at $T = 0$~K, $K_{0}$ is the bulk modulus, and $\gamma$ is the Gr$\ddot{\rm u}$neisen parameter. The internal energy $U(T)$ can be expressed within the Debye approximation as
\begin{equation}
U(T)=9pk_{\rm B}T\left(\frac{T}{\theta_{\rm D}}\right)^3\int_{0}^{\theta_{\rm D}/T}\dfrac{x^3}{e^x-1}dx.
\end{equation}
In the above, $p$ stands for the total number of atoms inside the unit cell and $k_{\rm B}$ is the Boltzmann constant. The best fit of the data down to $T = 15$~K (lower panel of Fig.~\ref{Fig3}) was obtained with the parameters: $\theta_{D}\simeq 530$~K, $\frac{\gamma}{K_{0}}\simeq 6.73\times10^{-5}$ Pa$^{-1}$, and $V_{0} \simeq 352.3$~\AA$^{3}$.

\subsection{Magnetization}
\begin{figure}[h]
	\includegraphics [width = \linewidth]{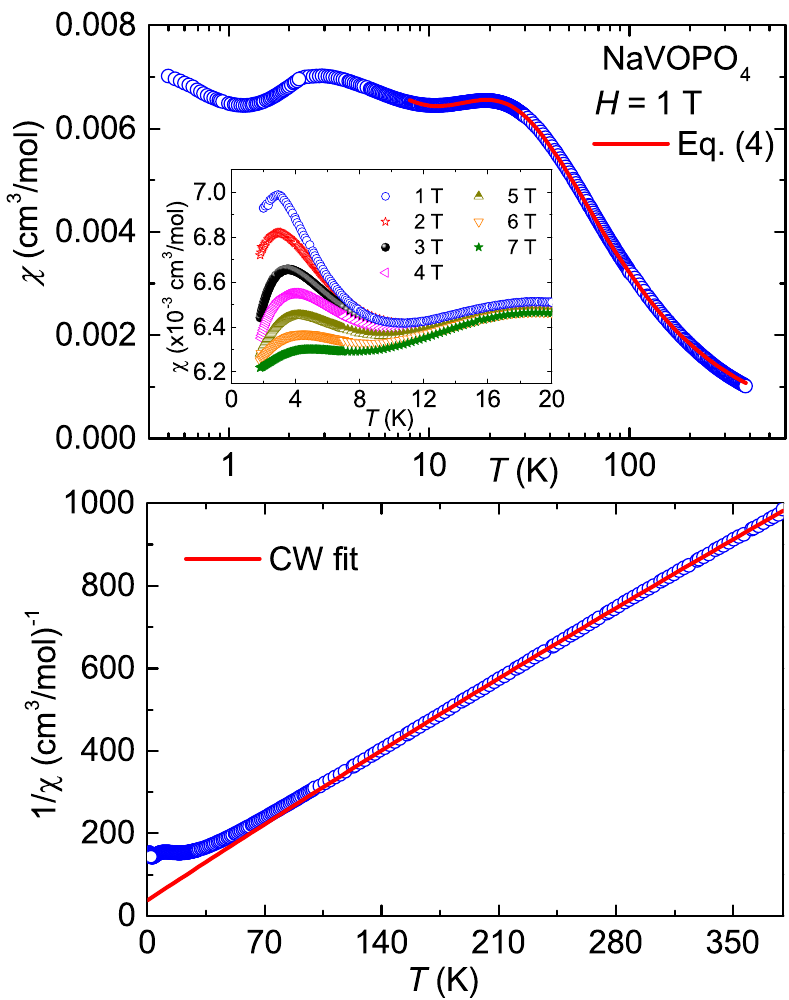}\\
	\caption{Upper panel: $\chi(T)$ measured at $H = 1 $~T. The solid line represents the fit using Eq.~\eqref{chi_alt}. Inset: $\chi(T)$ in the low temperature regime measured in different applied fields to focus on the low temperature broad maximum. Lower panel: $1/\chi$ vs $T$ and the solid line is the CW fit using Eq.~\eqref{cw}.}
	\label{Fig4}
\end{figure}
Magnetic susceptibility $\chi(T)~[\equiv M(T)/H$] measured in the applied field of $H = 1$~T is shown in the upper panel of Fig.~\ref{Fig4}. The inverse magnetic susceptibility $1/\chi(T)$ for the same field is shown in the lower panel of the figure. As the temperature is lowered, $\chi(T)$ increases in a Curie-Weiss manner as expected in the paramagnetic regime and then shows a broad maximum around $T_\chi^ {\rm {max}} \simeq 20$~K, indicative of short-range magnetic order and revealing magnetic one-dimensionality of NaVOPO$_4$. Although no clear indication of any magnetic LRO can be seen down to $0.5$~K, another broad feature is visible at around $4$~K. As shown in the inset of Fig.~\ref{Fig4}, this broad feature is field-dependent. Below 1.5~K, $\chi(T)$ shows an upturn likely caused by extrinsic paramagnetic impurities or defect spins present in the powder sample.\cite{Wolly137204} 

The $\chi(T)$ data in the paramagnetic region (above $150$~K) were fitted by the Curie-Weiss law (bottom panel of Fig.~\ref{Fig4})
\begin{equation}\label{cw}
\chi(T) = \chi_0 + \frac{C}{T - \theta_{\rm CW}},
\end{equation}
where $\chi_0$ accounts for the temperature-independent contribution consisting of the core diamagnetic susceptibility ($\chi_{\rm core}$) of the core electron shells and the Van-Vleck paramagnetic susceptibility ($\chi_{\rm VV}$) of the open shells of the V$^{4+}$ ions in the compound. The second term in Eq.~\eqref{cw} is the Curie-Weiss (CW) law with the CW temperature $\theta_{\rm CW}$ and Curie constant $C$. The fitted data in the temperature range $150$~K to $380$~K yields $\chi_0 \simeq 7 \times 10^{-5}$~cm$^3$/mol, $C \simeq 0.373$~cm$^3$K/mol, and $\theta_{\rm CW} \simeq -14.14$~K. From the value of $C$, the effective moment was calculated to be $\mu_{\rm eff} \simeq 1.72$~$\mu_{\rm B}$/V$^{4+}$ which is in close agreement with the expected spin-only value of 1.73~$\mu_{\rm B}$ for $S = 1/2$. This value of $\mu_{\rm eff}$ also corresponds to a $g$ value of $g \simeq 1.98$ which is nearly equal to the obtained $g$ value from the ESR experiment (discussed later). The negative value of $\theta_{\rm CW}$ indicates that the dominant exchange couplings between V$^{4+}$ ions are antiferromagnetic in nature. The $\chi_{\rm core}$ of NaVOPO$_4$ was estimated to be $-7.3 \times 10^{-5}$~cm$^3$/mol by adding the core diamagnetic susceptibility of the individual ions Na$^{1+}$, V$^{4+}$, P$^{5+}$, and O$^{2-}$.\cite{Selwood2013} The  $\chi_{\rm VV}$ was calculated to be $\sim 14.3\times 10^{-5}$~cm$^3$/mol by subtracting $\chi_{\rm core}$ from the experimental $\chi_{\rm 0}$ value.

To shed light on the nature of magnetic interactions in NaVOPO$_4$ and to estimate the corresponding exchange couplings, $\chi(T)$ was fitted by the following expression:
\begin{equation}\label{chi_alt}
\chi(T) = \chi_0 + \frac {C_{\rm imp}}{T}+\chi_{\rm spin},
\end{equation}
where the second term is the generic Curie law that accounts for the impurity contribution. $C_{\rm imp}$ stands for the Curie constant of the impurities and the third term $\chi_{\rm spin}$ is the expression for the spin susceptibility of a one-dimensional (1D) spin-$1/2$ Heisenberg AFM with uniform or alternating exchange couplings. Such an expression is available in the whole range of alternation ratios $\alpha$, $0\leq \alpha \leq 1$, and in a large temperature window ($k_{\rm B}T/J \geq 0.01$).\cite{Johnston9558} 

The fit of the $\chi(T)$ data down to $8$~K by Eq.~\eqref{chi_alt} taking $\chi_{\rm spin}$ for the alternating spin chain is shown in the upper panel of Fig.~\ref{Fig4} (solid line). The best fit of the data was obtained with the parameters: $\chi_0 \simeq 1.47 \times 10^{-4}$~cm$^3$/mol, $C_{\rm imp}\simeq 0.012$~cm$^3$K/mol, $g \simeq 1.95$ (fixed from ESR), alternation parameter $\alpha \simeq 0.98$, and the dominant exchange coupling $J/k_{B} \simeq 36.5$~K. The $C_{\rm imp}$ value obtained above is equivalent to $\sim 3.2$~\% impurity spins, assuming that they are of spin-$1/2$ nature. From the values of $J/k_{\rm B}$ and $\alpha$, the magnitude of the zero-field spin gap is estimated to be  $\Delta_{0}/k_{\rm B} = J/k_{\rm B} [(1- \alpha)^{3/4} (1+ \alpha)^{1/4}]$  $\simeq 2.4$~K.\cite{Barnes11384,Johnston9558} This spin gap corresponds to the critical field of $H_{\rm c1} = \Delta_{0}/(g \mu_{\rm B}) \simeq 1.74$~T.\cite{Tsirlin144412} It is to be noted that this formula overestimates the actual value of the spin gap since it neglects interchain couplings that tend to reduce the gap size. 

We have also fitted the bulk susceptibility data down to $8$~K taking $\chi_{\rm spin}$ for the uniform spin chain (not shown here). The fitting procedure was completed by fixing the $g$ value to 1.95 (from ESR) and the obtained parameters are $\chi_0 \simeq 1.59 \times 10^{-4}$~cm$^3$/mol, $C_{\rm imp}\simeq 0.013$~cm$^3$K/mol, and $J/k_{\rm B} \simeq 36.4$~K. Since both the alternating-chain and uniform-chain models fit the $\chi(T)$ data nicely  and yield similar values of the exchange coupling, it is apparent that the system lies close to the boundary between the uniform and alternating-chain models.
\begin{figure}[h]
	\includegraphics [width = \linewidth]{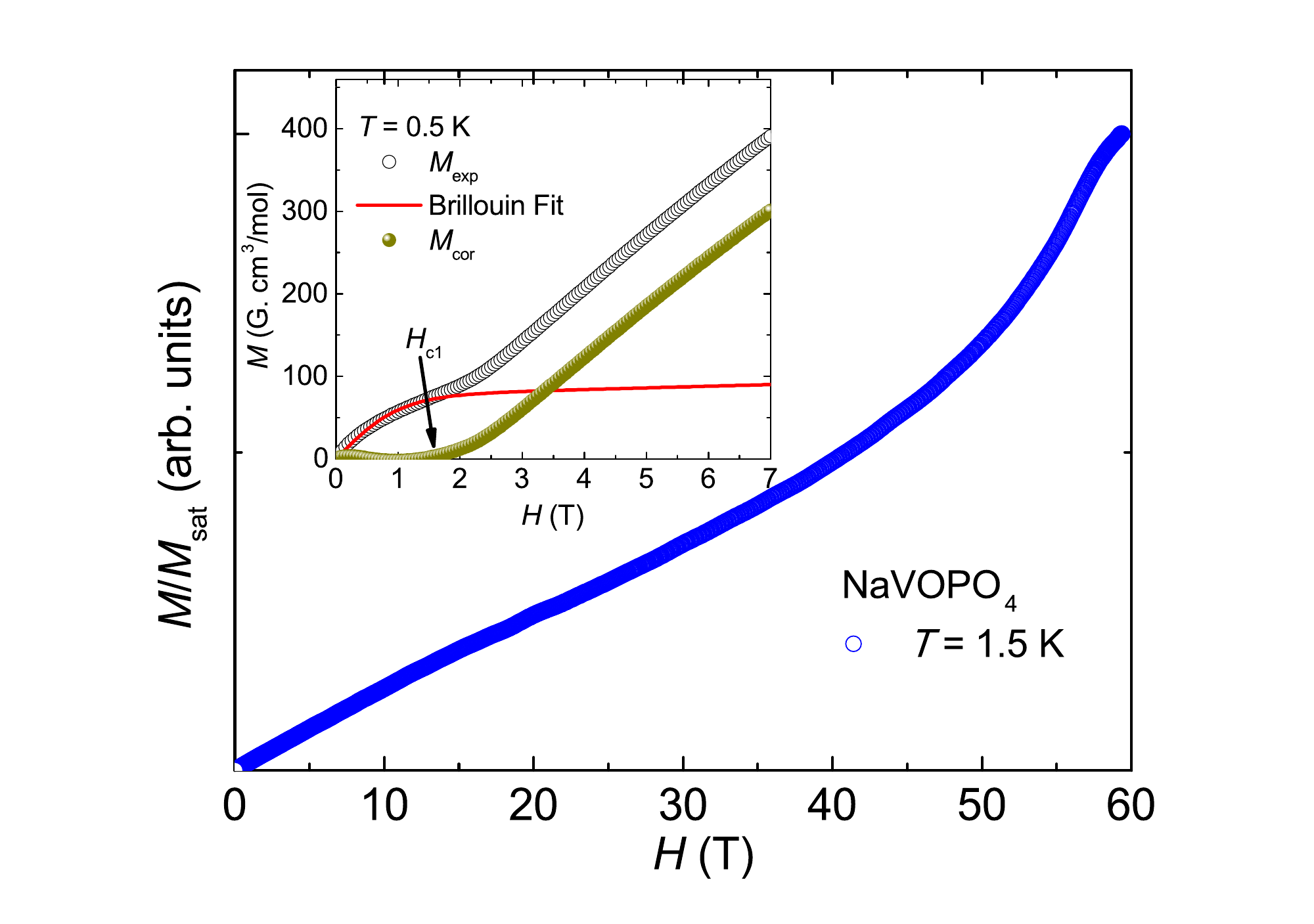}\\
	\caption{Magnetization $M$ vs field $H$ measured at $T = 1.5$~K, using pulsed magnetic field. Inset: $M$ vs $H$ measured at $T = 0.5$~K using MPMS. The solid line is the fit using Brillouin function [Eq.~\eqref{Brill}]. $M_{\rm cor}$ is the magnetization after subtraction of the impurity contribution.}
	\label{Fig5}
\end{figure} 
\begin{figure}[h]
	\includegraphics [width = \linewidth]{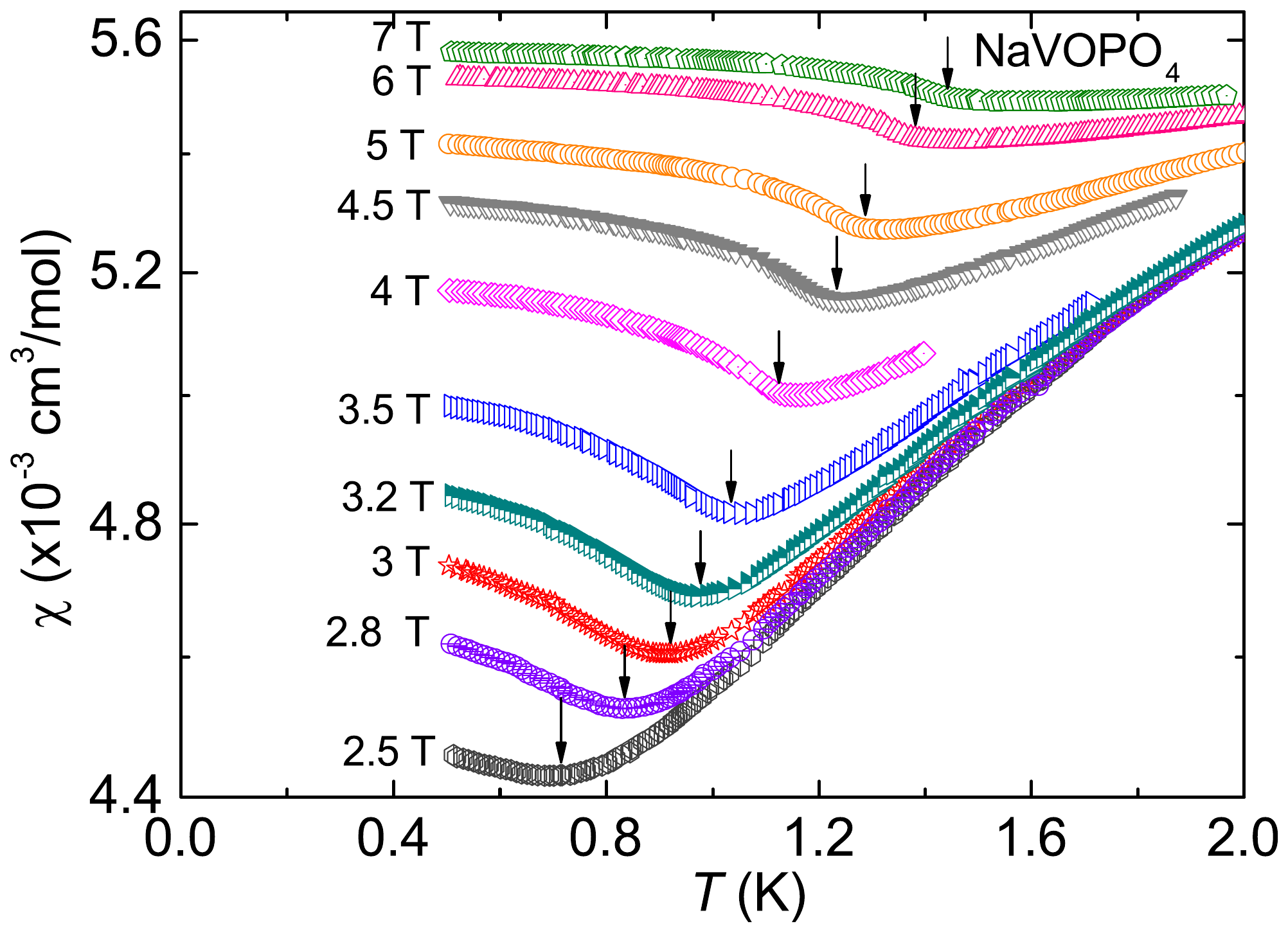}\\
	\caption{$\chi(T)$ measured in the low temperature region and in different applied fields. The downward arrows point to the field induced magnetic transition.}
	\label{Fig6}
\end{figure}

Figure~\ref{Fig5} presents the magnetization ($M$) vs field ($H$) curve at $T = 1.5$~K measured using pulsed magnetic field. $M$ increases almost in a linear fashion up to 40~T. Above 40~T, it exhibits a pronounced upward curvature and then approaches the saturation value at about $H_{\rm c2} \simeq 60$~T. Such a curvature in the magnetization data is a clear signature of strong quantum fluctuations, as anticipated for a spin-$1/2$ 1D system. The value of $H_{\rm c2} \simeq 60$~T corresponds to an exchange coupling of $J/k_{B} = g\mu_{\rm B}H_{\rm c2}/2k_{\rm B} \simeq 39.3$~K, assuming the 1D uniform spin chain model.\cite{Lebernegg174436} This value of $J/k_{\rm B}$ matches well with the one obtained from the $\chi(T)$ analysis.

Since the system lies at the phase boundary between the uniform-chain and alternating-chain models, it is expected that either the system should have a magnetic LRO or a spin gap at very low temperatures. In order to probe the ground state, we measured the magnetization isotherm at $T = 0.5$~K up to $7$~T (inset of Fig.~\ref{Fig5}). Below 2~T, it exhibits a dome-shaped feature above which the variation of $M$ is linear with $H$. This is a clear indication of the existence of a spin gap with the critical field of gap closing $H_{\rm c1} \leq 2$~T. Typically, in the gapped spin systems, the magnetization remains zero up to $H_{\rm c1}$. However, a non-zero value of $M$ below $H_{\rm c1}$ is likely due to the saturation of some extrinsic paramagnetic contributions and/or defects in our powder sample. Therefore, the data below $1.5$~T were fitted by \cite{Nath174513}
\begin{equation}
\label{Brill}
M = \chi H + N_{\rm A} f_{\rm imp} S_{\rm imp} g_{\rm imp} \mu_{\rm B} B_{\rm S_{\rm imp}}(x),
\end{equation}
where $\chi$ is the intrinsic susceptibility which was kept fixed, $f_{\rm imp}$ is the molar fraction of impurities, $S_{\rm imp}$ is the impurity spin, $g_{\rm imp}$ is the impurity $g$-factor, $N_{\rm A}$ is the Avogadro's number, $B_{\rm S_{\rm imp}}(x)$ is the Brillouin function with the modified argument $x = S_{\rm imp} g_{\rm imp} \mu_{\rm B} H/[k_{\rm B}(T-\theta_{\rm imp}$)].\cite{Kittel} The obtained fitting parameters are: $f_{\rm imp} \simeq 0.013(3)$~mol\%, $S_{\rm imp} \simeq 0.50(3)$, and $g_{\rm imp} \simeq~2.06(1)$.
The corrected magnetization ($M_{\rm cor}$) after the subtraction of this extrinsic contribution from the raw data is also presented in Fig.~\ref{Fig5}. Clearly, $M_{\rm cor}$ remains zero up to $H_{\rm c1} \simeq 1.6$~T that corresponds to the zero-field spin gap of $\Delta_{\rm 0}/k_{\rm B} \simeq 2$~K between the singlet ground state and the triplet excited states.

To check for the field-induced effects, if any, $\chi(T)$ was measured at low temperatures (0.5~K$\leq T \leq 2$~K) and in different applied fields. As shown in Fig.~\ref{Fig6}, it shows a change in slope at $\sim 0.7$~K in the $2.5$~T data. As the field increases, the transition shifts toward high temperatures marked by the downward arrows in Fig.~\ref{Fig6}. This cusp-like anomaly above $H\simeq 2.5$~T and its field variation are robust signatures of the field-induced magnetic LRO.\cite{Nikuni5868} The variation of $T_N$ with respect to $H$ is presented in Fig.~\ref{Fig15}. 

\subsection{ESR}
\begin{figure}[h]
	\includegraphics [width = \linewidth]{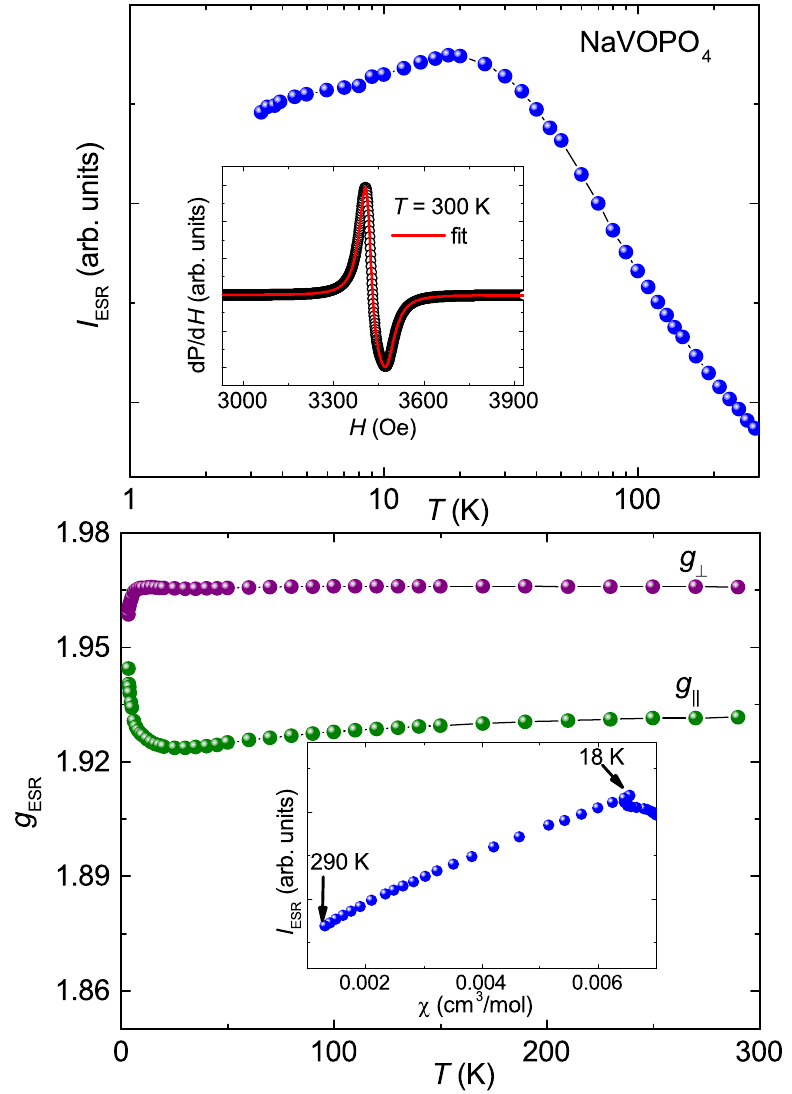}\\
	\caption{Upper Panel: Integrated ESR intensity vs temperature. Inset: ESR spectrum at room temperature measured at a microwave frequency of 9.4~GHz together with the powder-averaged Lorentzian fit (solid line). Lower Panel: Temperature variation of $g$ values (both perpendicular and parallel components) obtained from Lorentzian fit. Inset: $I_{\rm ESR}$ vs $\chi$ with temperature as an implicit parameter.}
	\label{Fig7}
\end{figure}

ESR experiment was performed on the powder sample and the results are displayed in Fig.~\ref{Fig7}. The inset in the top panel of Fig.~\ref{Fig7} shows a typical ESR powder spectrum at room temperature. The spectra at different temperatures could be fitted well using the powder-averaged Lorentzian line for the uniaxial $g$-factor anisotropy. The fit to the spectrum at room temperature reveals anisotropic $g$-tensor components: parallel component $g_\parallel \simeq 1.931$ and perpendicular component $g_\perp \simeq 1.965$. From these values the average $g$-factor is calculated to be $\bar{g} \simeq 1/3(g_\parallel + 2g_\perp) \simeq 1.951$. A slight deviation of the estimated $g$-factor ($\Delta g/g \simeq 0.02$) from the free-electron value ($g = 2$) is typical for V$^{4+}$ (spin-$1/2$) based compounds having the VO$_6$ octahedral coordination.\cite{Arjun174421,Tsirlin144412}

The temperature-dependent integrated ESR intensity ($I_{\rm ESR}$) is presented in the upper panel of Fig.~\ref{Fig7}. It resembles the bulk $\chi(T)$ behaviour with a pronounced broad maximum at around $T_{\rm ESR}^ {\rm {max}} \simeq 20$~K. When $I_{\rm ESR}$ is plotted as a function $\chi$, it indeed follows a straight-line behaviour (inset of the lower panel of Fig.~\ref{Fig7}) down to 18~K. Similar to the $\chi(T)$ data, $I_{\rm ESR}(T)$ also exhibits another weak but broad feature at around 4~K. As shown in the lower panel of Fig.~\ref{Fig7}, both $g$-components are constant over a wide temperature range down to $\sim 15$~K. The anomalous behavior below $15$~K might be related to the effect of defects/impurities.

\subsection{\textbf{Specific Heat}}
\begin{figure}[h]
	\includegraphics[width = \linewidth]{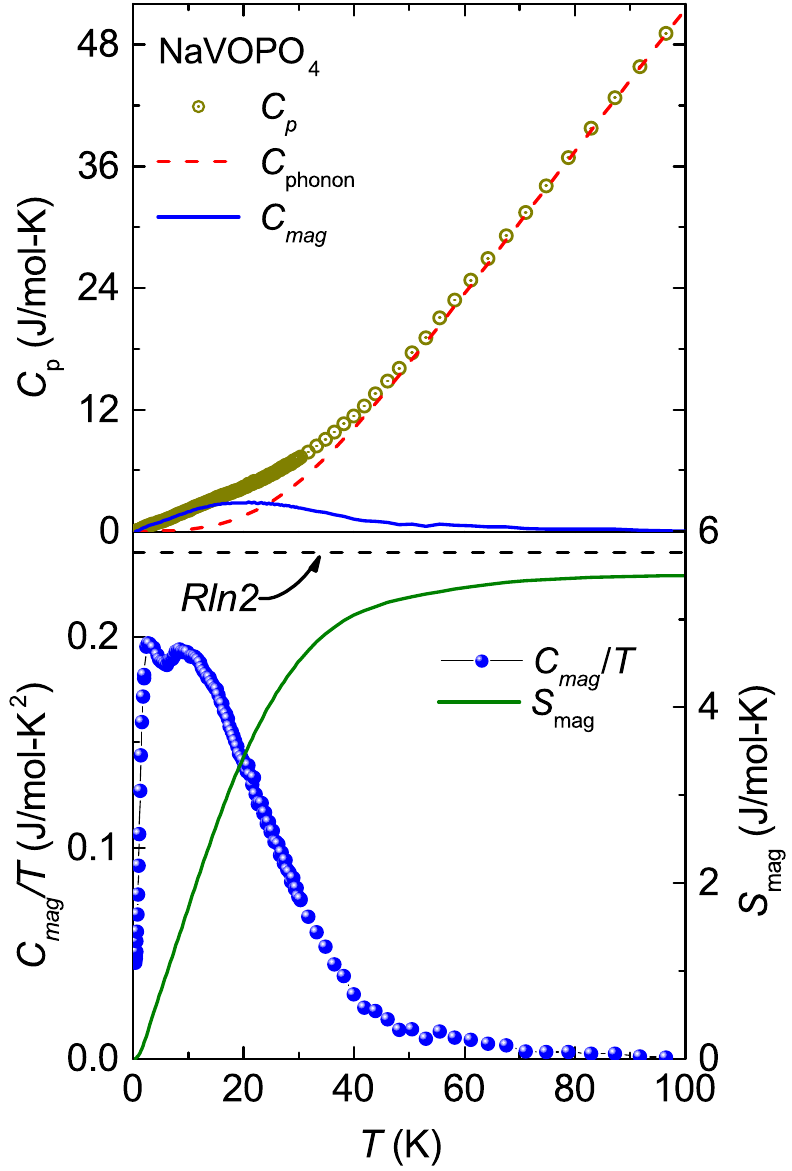}
	\caption{\label{Fig8} Upper Panel: Specific heat $C_{\rm p}$ vs $T$ of NaVOPO$_4$ in zero applied field. The dashed line is the phonon contribution to the specific heat ($C_{\rm ph}$) obtained using a Debye fit [Eq.~\eqref{Debye}]. The solid line indicates the magnetic contribution to the specific heat $C_{\rm mag}$. Lower Panel: Left \textit{y}-axis shows $C_{\rm mag}/T$ and right \textit{y}-axis shows the magnetic entropy $S_{\rm mag}$ vs $T$, respectively.}
\end{figure}
Specific heat [$C_{\rm p}(T)$] data measured under zero field are shown in the upper panel of Fig.~\ref{Fig8}. In the high-temperature region, $C_{\rm p}$ is entirely dominated by phonon excitations whereas the magnetic part dominates only at low temperatures. It shows a broad feature below about 20~K. There is no clear anomaly in the data down to $T = 0.38$~K, which excludes the possibility of any magnetic LRO in zero field. To quantify the magnetic contribution to the specific heat ($C_{\rm mag}$), we subtracted the phonon contribution ($C_{\rm ph}$) from the total measured specific heat $C_{\rm p}$. For this purpose, the experimental data at high temperatures ($T \geq 60$~K) were fitted by a linear combination of four Debye functions\cite{Niyaz214413}
\begin{equation}
\label{Debye}
C_{\rm ph}(T) = 9R\displaystyle\sum\limits_{\rm n=1}^{4} c_{\rm n} \left(\frac{T}{\theta_{\rm Dn}}\right)^3 \int_0^{\frac{\theta_{\rm Dn}}{T}} \frac{x^4e^x}{(e^x-1)^2} dx.
\end{equation}
Here, $R$ is the universal gas constant, the coefficients $c_{\rm n}$ account for the number of individual atoms, and $\theta_{\rm Dn}$ are the corresponding Debye temperatures. Since we have four different types of atoms with varying atomic masses and Debye temperature is inversely proportional to the atomic mass, we have used four Debye functions corresponding to the Na, V, P, and O atoms.

$C_{\rm mag}(T)$ extracted by subtracting $C_{\rm ph}(T)$ from $C_{\rm p}(T)$ is also shown in Fig.~\ref{Fig8}. The above procedure was verified by calculating the magnetic entropy $S_{\rm mag}$ through the integration of $C_{\rm mag}(T)/T$ which gives $S_{\rm mag} \simeq 5.6$~J/mol~K at 100~K (lower panel of Fig.~\ref{Fig8}). This value is very close to the expected entropy $S_{\rm mag} = R\ln 2 = 5.76$~J/mol~K for a spin-$\frac{1}{2}$ system. As shown in the upper panel of Fig.~\ref{Fig8}, $C_{\rm mag}$ has a broad maximum at $T_{\rm C}^{\rm max} \simeq 20$~K with an absolute value $C_{\rm mag}^{\rm max} \simeq 2.80$~J/mol-K. This value of $C_{\rm mag}^{\rm max}$ is indeed very close to the expected value $\sim 0.35R = 2.9$~J/mol~K for a uniform spin-$1/2$ chain.\cite{Bernu134409} Similarly, the value of $T_{\rm C}^{\rm max} \simeq 20$~K, which reflects the energy scale of the leading exchange interaction, corresponds to $J/k_{\rm B} = T_{\rm C}^{\rm max}/0.48 \simeq 41$~K in the framework of the uniform spin chain model.\cite{Johnston9558} It is indeed very close to the values obtained from the other experiments.
Furthermore, an anomaly in $C_{\rm mag}$ is visible at around $\sim 5$~K, similar to the $\chi(T)$ data suggesting that it is an intrinsic feature of the compound.
\begin{figure}
	\includegraphics[width = \linewidth]{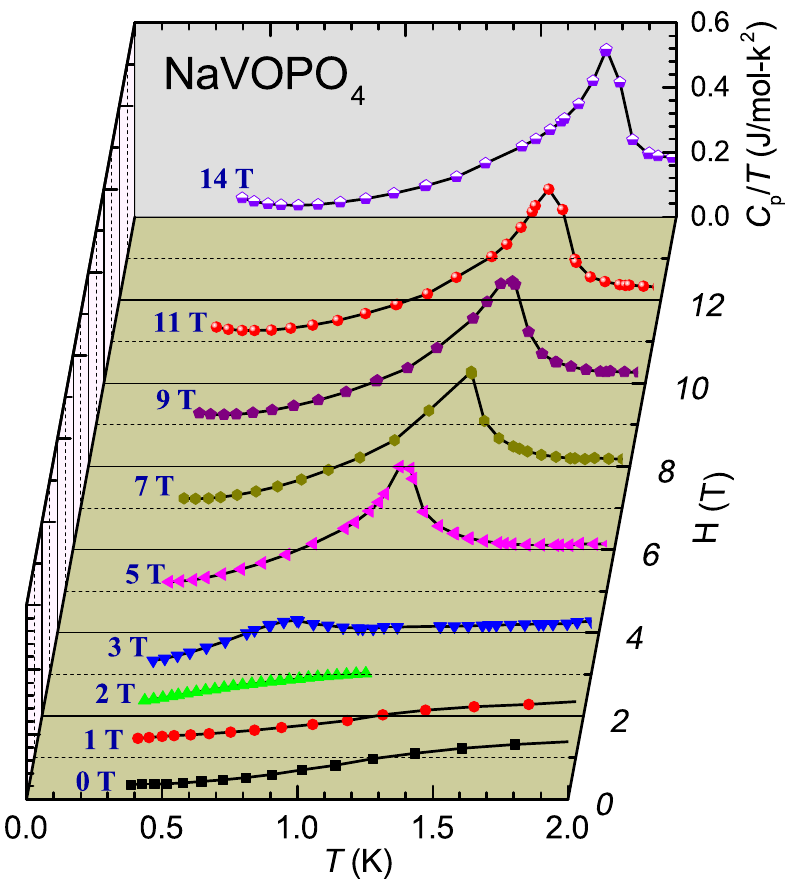}
	\caption{\label{Fig9} $C_{\rm p}/T$ vs $T$ of NaVOPO$_4$ in the low-temperature regime in different magnetic fields up to 14~T.}
\end{figure}

In addition, we have also measured $C_{\rm p}(T)$ in the low-temperature region (down to 0.38~K) in magnetic fields up to 14~T. Figure~\ref{Fig9} shows the data at few selected fields. In zero field, no anomaly associated with the magnetic LRO is found down to 0.38~K but it shows a rapid decrease below 1~K towards zero which could be due to the opening of a spin gap. An exponential fit below $1$~K yields $\Delta_{\rm 0}/k_{\rm B} \simeq 2.6$~K which is consistent with the value obtained from the $\chi(T)$ analysis. When the field is increased, no extra features are seen up to $2$~T. However, for $H \geq 2$~T, it displays a $\lambda$-type peak which moves toward high temperatures with increasing field. This can be ascribed to the crossover from a spin-gap state to a 3D LRO state under external magnetic field. $T_{\rm N}$ as a function of $H$ is plotted in Fig.~\ref{Fig15}.

\subsection{\textbf{$^{31}$P NMR}}
NMR experiments were performed over a wide temperature range down to $40$~mK to address several key questions, namely, (i) is the anomaly at $4$~K an intrinsic feature? (ii) what is the ground state in zero field? and (iii) does the system undergo field-induced magnetic LRO? In magnetic insulators, the Hamiltonian is relatively simple and well-defined, which allows the powerful local tools like NMR to access the structure, static and dynamic properties of a spin system. Since $^{31}$P nuclei is strongly coupled to the V$^{4+}$ ions in the crystal structure, one can investigate the properties of the spin chains by probing at the $^{31}$P site.

\subsubsection{NMR Spectra}
\begin{figure}[h]
\includegraphics [width = \linewidth]{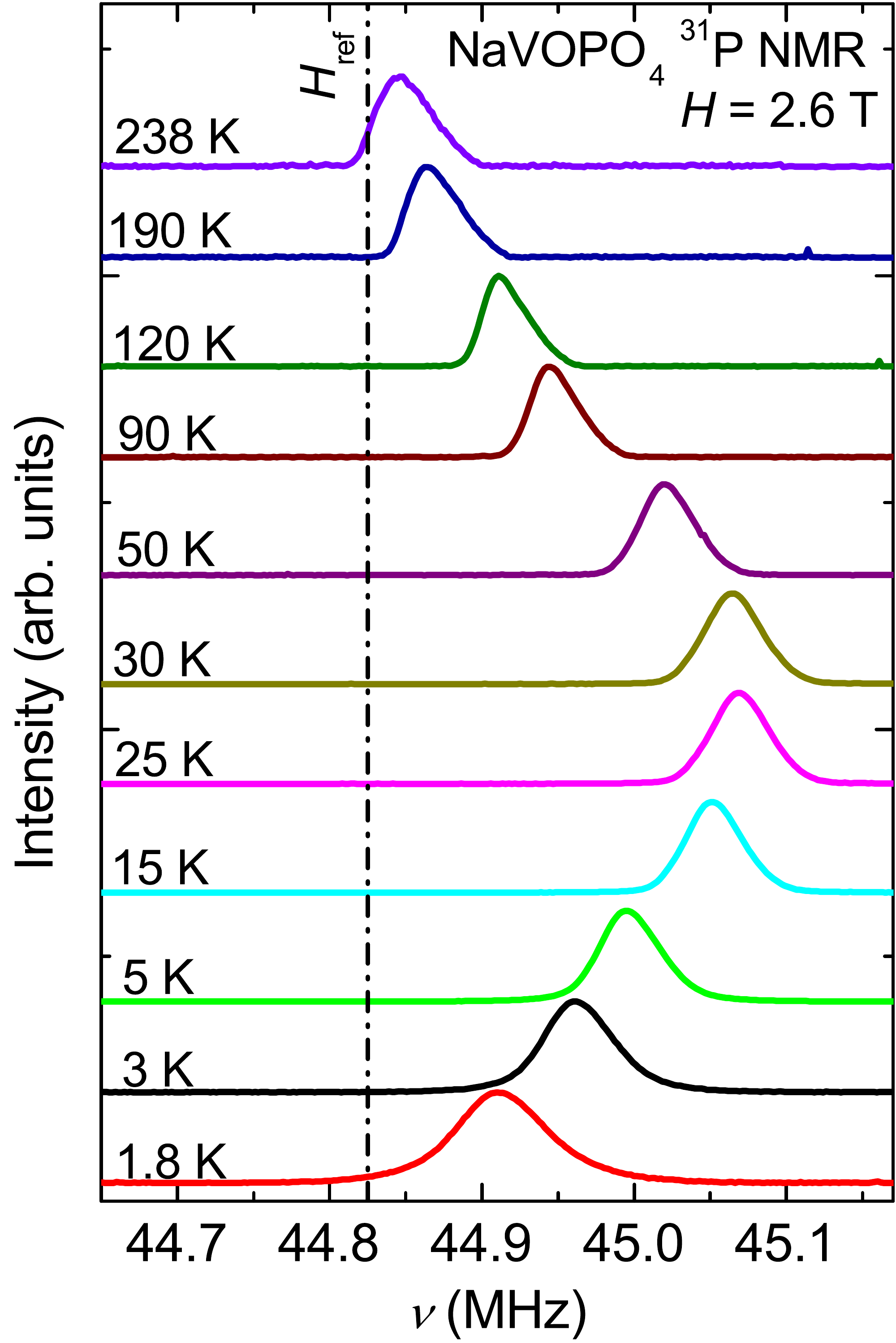}\\
\caption{Temperature-dependent FT $^{31}$P NMR spectra of NaVOPO$_4$ measured at $H = 2.6$~T down to 1.8~K. The dash-dotted line represents the reference field of the non-magnetic H$_{3}$PO$_{4}$}
\label{Fig10}
\end{figure}
 $^{31}$P NMR spectra were measured on the powder sample either by doing Fourier Transform (FT) of the echo signal at a constant magnetic field or by sweeping the magnetic field at a fixed frequency. Since $^{31}$P has the nuclear spin $I=1/2$, a single $^{31}$P NMR line is expected corresponding to one allowed transition. Figure~\ref{Fig10} displays representative NMR spectra taken for 1.8~K~$\leq T \leq$~238~K at $H = 2.6$~T. Indeed, a single spectral line is observed down to 1.8~K and the line width is found to increase systematically with decreasing temperature. This confirms that NaVOPO$_4$ has a single P site, which is commensurate with the crystal structure. Typically, in a magnetically ordered state, the NMR nucleus senses the static internal field and hence the NMR line broadens drastically. As one can see in Fig.~\ref{Fig10}, no significant line broadening was observed around 5~K, ruling out the possibility of any magnetic LRO taking place at this temperature. The line position is found to shift with temperature.
 
 \begin{figure}[h]
 	\includegraphics [height = 8 cm, width = \linewidth]{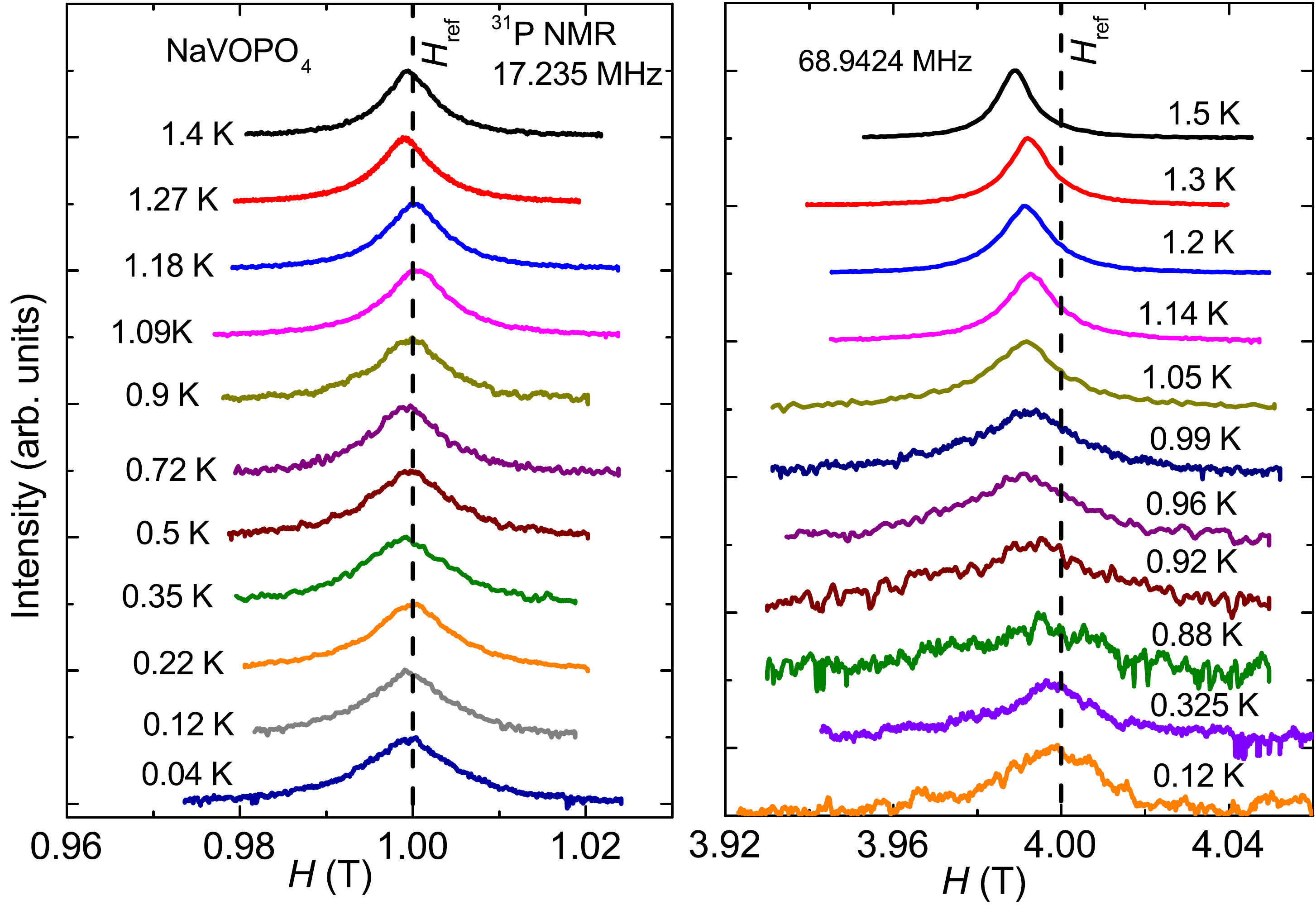}\\
 	\caption{Field-sweep $^{31}$P NMR spectra of NaVOPO$_4$ measured at $\nu = 17.235$~MHz (left panel) and $68.9424$~MHz (right panel) in the low-temperature region.}
 	\label{Fig11}
 \end{figure}
As we have discussed earlier, specific heat in zero field does not show any magnetic LRO down to 0.38~K. In order to ensure that there is no LRO in zero field, NMR spectra were measured down to $40$~mK. As one can see in Fig.~\ref{Fig11}, the NMR line shape remains the same and there is no significant line broadening down to $0.04$~K at 17.235~MHz ($\simeq 1$~T). On the other hand, for $68.9424$~MHz ($\simeq 4$~T), the line broadens abruptly below $\sim 1$~K. This can be taken as an evidence that the system does not undergo magnetic LRO below $1$~T and shows a field-induced magnetic transition at $T_{\rm N} \simeq 1$~K in $4$~T which is consistent with the $\chi(T)$ and $C_{\rm p}(T)$ measurements. The absence of line broadening and any extra feature in the NMR spectra at $1$~T also exclude the possibility of any structural or lattice distortions, consistent with our temperature-dependent powder XRD. This is in contrast with that reported for the spin-Peierls compounds CuGeO$_{3}$ where spin dimerization is accompanied by the lattice distortion and $\alpha ^{'}$- NaV$_{2}$O$_{5}$ where the singlet ground state is driven by the charge ordering at low temperatures.\cite{Hase3651,Fagot4176} $^{63}${Cu}-NMR experiments in the well-known BEC compound TlCuCl$_{3}$ revealed that the field-induced magnetic transition is accompanied by a simultaneous lattice deformation.\cite{Vyaselev207202} This also implies strong spin-phonon coupling at the critical field that drives the BEC phenomenon in TlCuCl$_{3}$. However, in our compound for $H > H_{\rm c1}$ the $^{31}$P NMR line only broadens without any splitting or extra features, ruling out the possibility of significant lattice deformations in the field-induced state.

\subsubsection{\textbf{NMR Shift}}
\begin{figure}[h]
\includegraphics [width = \linewidth]{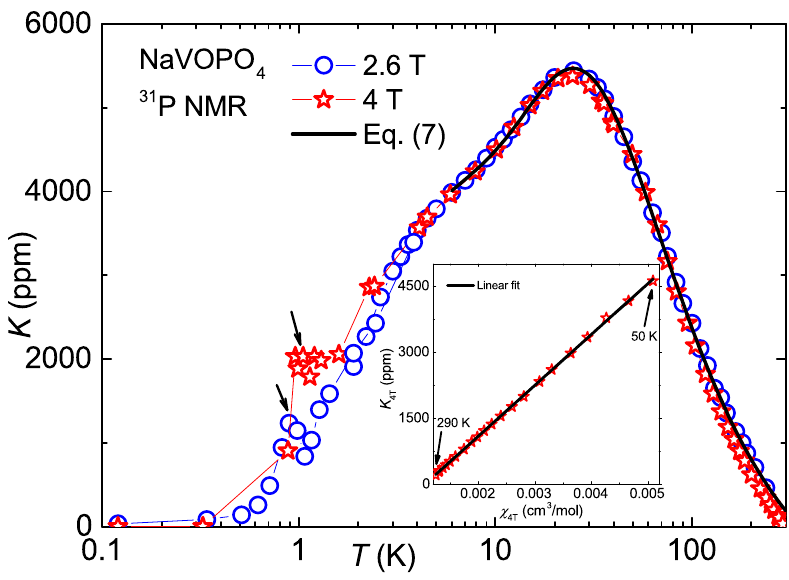}\\
\caption{Temperature-dependent $^{31}$P NMR shift [$K(T)$)] as a function of temperature for two different magnetic fields. The solid line is the fit of $K(T)$ by Eq.~\eqref{K_chi}. The downward arrows point to the field-induced magnetic transitions. Inset: $K$ vs $\chi$ (measured at 4~T) with temperature as an implicit parameter. The solid line is the linear fit.}
\label{Fig12}
\end{figure}

Figure~\ref{Fig12} shows the variation of the NMR shift $K(T)$ with temperature measured at $H = 2.6$~T and $4$~T. Both the data sets resemble each other. With decrease in temperature, $K(T)$ passes through a broad maximum around $25$~K in a similar way as that of the bulk susceptibility. With further decrease in temperature it starts to decrease. A weak but broad hump appears at $T\simeq 4~K$, reflecting that it is an intrinsic feature of the compound. At very low temperatures, the small peaks ($T_{\rm N} \simeq 0.89~$K at $2.6$~T and $1.05$~K at $4$~T) are likely due to field-induced magnetic transitions.
 
Since $K(T)$ is a direct and intrinsic measure of the spin susceptibility $\chi_{\rm spin}(T)$, one can write
\begin{equation}\label{K_chi}
N_{\rm A}K(T) = K_0 + A_{\rm hf}\chi_{\rm spin},
\end{equation}
where $K_0$ is the temperature-independent NMR shift and $A_{\rm hf}$ is the total hyperfine coupling between the $^{31}$P nuclei and V$^{4+}$ spins. $A_{\rm hf}$ is a combination of the transferred hyperfine coupling and the nuclear dipolar coupling contributions, both temperature-independent. The nuclear dipolar coupling is usually very small compared to the transferred hyperfine coupling, and therefore neglected. From Eq.~\eqref{K_chi}, $A_{\rm hf}$ can be calculated from the slope of the linear $K$ vs $\chi$ plot with temperature as an implicit parameter. The inset of Fig.~\ref{Fig12} presents the $K$ vs $\chi$ plot, which obeys a straight line behavior down to 50~K. The data for $T \geq 50$~K were fitted well to a linear function and the slope of the fit yields $K_0 \simeq -1219.65$~ppm and $A_{\rm hf} \simeq 6445$~Oe/$\mu_{\rm B}$. A large value of $A_{\rm hf}$ indicates that the $^{31}$P nucleus is very strongly coupled to the V$^{4+}$ ions. This value is almost twice the hyperfine coupling of As with V$^{4+}$ ions in AgVOAsO$_{4}$ and NaVOAsO$_{4}$.\cite{Arjun014421,Ahmed224433} This also illustrates the fact that the spin chains run along the extended V$^{4+}$--O--P--O--V$^{4+}$ path rather than the short V$^{4+}$--O--V$^{4+}$ pathway.

One advantage of the NMR experiment is that the NMR shift directly probes $\chi_{\rm spin}$ and is inert to impurity contributions. Therefore, in low-dimensional spin systems $K(T)$ data are often used for a reliable estimation of the magnetic parameters instead of the bulk $\chi(T)$. For a tentative estimation of the magnetic parameters, we have fitted the $K(T)$ using Eq.~\eqref{K_chi} down to $6$~K, taking $\chi_{\rm spin}$ for both the alternating-chain and uniform-chain models.\cite{Johnston9558} To minimize the number of fitting parameters, the value of $g$ was fixed to $1.95$ obtained from ESR experiment. The resultant fitting parameters using the alternating-chain model are $K_0 \simeq -1182 $~ppm, $A_{\rm hf} \simeq 6873 $~Oe/$\mu_{\rm B}$, $\alpha \simeq 0.98$, and  $J/k_{B} \simeq 39$~K. Taking these values of $\alpha$ and $J/k_{\rm B}$, we arrive at the spin gap of $\Delta_{\rm 0}/k_{\rm B} \simeq 2.37$~K.\cite{Johnston9558, Barnes11384} On the other hand, the fit using the uniform-chain model yields $K_0 \simeq -1198 $~ppm, $A_{\rm hf} \simeq 6938 $~Oe/$\mu_{\rm B}$, and $J/k_{\rm B} \simeq 39$~K. Thus, approximately the same values of $J/k_{\rm B}$ are obtained from both fits, suggesting that the system approaches the uniform-chain regime but reveals a small spin gap. These findings are quite consistent with our analysis of the $\chi(T)$ data.

\subsubsection{Spin-lattice Relaxation Rate}
\begin{figure}[h]
	\includegraphics[width = \linewidth]{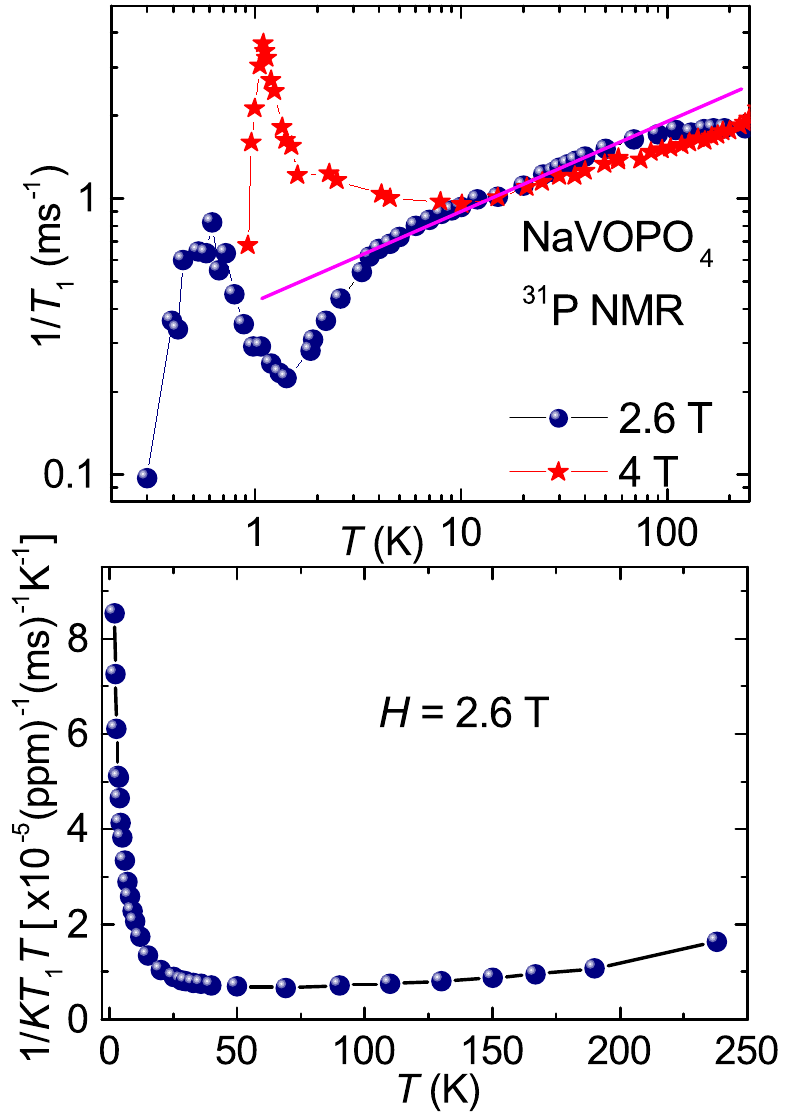}
	\caption{\label{Fig13} Upper panel: $1/T_{1}$ plotted as a function of temperature for two different magnetic fields. The solid line guides through the linear regime in the $H = 2.6$~T data. Lower panel: $\frac{1}{KT_{1}T}$ plotted as a function of $T$ for $H = 2.6$~T.}
\end{figure}
Another crucial quantity determined in the NMR experiment is the spin-lattice relaxation rate ($1/T_{1}$). At the central resonance frequency of each temperature, the sample is irradiated with a single $\pi/2$ saturation pulse and the growth of longitudinal magnetization is monitored for variable delays. These recovery curves are fitted well by a single exponential function
\begin{equation}\label{T_1}
1- \frac{M(t)}{M(0)} = Ae^{t/T_{1}},
\end{equation}
where $M(t)$ stands for the nuclear magnetization at a time $t$ after the saturation pulse, and $M(0)$ is the initial magnetization. Figure~\ref{Fig13} (upper panel) displays the variation of $1/T_{1}$ with temperature for $H = 2.6$~T and $4$~T extracted from the above fit. For both the fields, $1/T_{1}$ is almost temperature-independent down to $T\simeq 50$~K due to random fluctuation of paramagnetic moments at high temperatures.\cite{Moriya23} For $H = 2.6$~T, $1/T_{1}$ decreases linearly down to $\sim 4$ K. Such a behavior is typically observed in 1D spin-$1/2$ systems in the temperature range of $T \sim J/k_{\rm B}$ where $1/T_{1}$ is dominated by the uniform ($q =0$) contribution.\cite{Sachdev13006,Nath174436} At $T \simeq 4~$K, it shows a change in slope similar to that observed in $\chi(T)$, $C_{\rm p}(T)$, $I_{\rm ESR}(T)$, and $K(T)$. At very low temperatures, it exhibits a peak at $T_N \simeq 0.62$~K due to the critical slowing down of the fluctuating moments and reflects the transition to a magnetic LRO. Similarly, the $1/T_{1}$ data for $H =4$~T develops an upward trend below $10$~K due to the growth of strong magnetic correlations and then approaches the magnetic LRO at $T_{\rm N} \simeq 1$~K.

The AFM correlations in a spin system can be assessed by analyzing $1/T_{1}(T)$. The general expression connecting $\frac{1}{T_{1}T}$ with the dynamic susceptibility $\chi_{M}(\vec{q},\omega_{0})$ can be written as \cite{Moriya516}
\begin{equation}
\frac{1}{T_{1}T} = \frac{2\gamma_{N}^{2}k_{B}}{N_{\rm A}^{2}}
\sum\limits_{\vec{q}}\mid A(\vec{q})\mid
^{2}\frac{\chi^{''}_{M}(\vec{q},\omega_{0})}{\omega_{0}},
\label{t1form}
\end{equation}
where $\gamma_{\rm N}$ is the nuclear gyromagnetic ratio, $A(\vec{q})$ is the form factor of the hyperfine interactions between nuclear and electronic spins, $\chi^{''}_{M}(\vec{q},\omega_{0})$ is the imaginary part of the dynamic susceptibility [$\chi_{M}(\vec{q},\omega_{0}) = \chi^{'}_{M}(\vec{q},\omega_{0}) + i \chi^{''}_{M}(\vec{q},\omega_{0})$] at the nuclear Larmor frequency $\omega_{0}$. 
For $\vec{q}$ and $\omega_{0} = 0$, the real component of $\chi^{'}_{M}(\vec{q},\omega_{0})$ corresponds to the uniform static susceptibility $\chi(T)$. As $K(T)$ is a direct measure of the static susceptibility $\chi$, $\chi(T)$ in the above expression can be replaced by $K(T)$. To see the persistent spin correlations, we have plotted $\frac{1}{KT_{1}T}$ against temperature. As one can see in the bottom panel of Fig.~\ref{Fig13}, $\frac{1}{KT_{1}T}$ remains constant down to $T\sim 50$~K as expected in the high-$T$ region. As the temperature decreases and becomes comparable to the energy scale of $J/k_{\rm B}$, $\frac{1}{KT_{1}T}$ increases abruptly due to the growth of AFM correlations.

\subsection{Microscopic magnetic model}
\label{sec:model}
To assess individual magnetic couplings in NaVOPO$_4$, we use two complementary approaches. First, we extract hopping parameters $t_i$ between the V $3d$ states in the uncorrelated band structure, and introduce them into an effective one-orbital Hubbard model that yields antiferromagnetic part of the superexchange as $J_i^{\rm AFM}=4t_i^2/U_{\rm eff}$, where $U_{\rm eff}=4.5$\,eV is the on-site Coulomb repulsion in the V $3d$ bands~\cite{Nath024418,Tsirlin104436}. Second, we use the mapping procedure~\cite{Xiang224429,Tsirlin014405} and obtain exchange couplings $J_i$ from total energies of collinear spin configurations calculated within DFT+$U$.

\begin{table}
\caption{\label{tab:exchange}
The V--V distances (in\,\r A), hopping parameters $t_i$ (in\,meV) of the uncorrelated band structure, and exchange couplings $J_i$ (in\,K) obtained via the DFT+$U$ mapping procedure in NaVOPO$_4$ (this work) and NaVOAsO$_4$ (Ref.~\onlinecite{Arjun014421}). 
}
\begin{ruledtabular}
\begin{tabular}{c@{\hspace{3em}}crr@{\hspace{3em}}crr}
      & \multicolumn{3}{l}{\qquad NaVOPO$_4$} & \multicolumn{3}{c}{NaVOAsO$_4$} \\
			& $d_{\rm V-V}$ & $t_i$ & $J_i$ & $d_{\rm V-V}$ & $t_i$ & $J_i$ \\\hline
$J$   & 5.385 & 89 &  54  & 5.519 & 99 &  57  \\
$J'$  & 5.337 & 86 &  50  & 5.489 & 81 &  54  \\
$J_a$ & 5.952 & 42 &  13  & 6.073 & 25 &   8  \\
$J_c$ & 3.565 & 0 & $-6$  & 3.617 & 9 & $-5$  \\
\end{tabular}
\end{ruledtabular}
\end{table}

The results from both methods are summarized in Table~\ref{tab:exchange}. Similar to AgVOAsO$_4$ and NaVOAsO$_4$, we find leading exchange coupling $J$ and $J'$ along the $110$ and $1\bar 10$ directions oblique to the structural chains of the VO$_6$ octahedra. Only a minor difference between the two couplings is observed, with $(t'/t)^2\simeq 0.93$ and $J'/J\simeq 0.93$, in agreement with the alternation ratio $\alpha$ close to 1.0, as determined experimentally. The interchain couplings are represented by the weakly FM $J_c$ that runs along the structural chains via the V--O--V bridge and by the AFM $J_a$ through the single V--O--P--O--V bridge. These couplings are similar in nature and magnitude to those in the isostructural V$^{4+}$ arsenates~\cite{Tsirlin144412,Arjun014421}.

The hopping parameters in Table~\ref{tab:exchange} confirm that the superexchange pathways for $J$ and $J'$ are quite similar, resulting in the alternation ratio $\alpha$ close 1.0. Microscopically, these couplings are mediated by double bridges of the PO$_4$/AsO$_4$ tetrahedra and may depend on three geometrical parameters~\cite{Roca3167}: i) the in-plane offset of the VO$_6$ octahedra; ii) the out-of-plane offset of the VO$_6$ octahedra, and iii) the tilt of the tetrahedra with respect to the VO$_6$ octahedra. In NaVOPO$_4$-like structures, the in-plane offset plays the crucial role. Indeed, we find similar offsets of $d \simeq 0.534$\,\r A ($J$) and 0.515\,\r A ($J'$) in NaVOPO$_4$ (see Fig.~\ref{Fig1}), where $\alpha$ approaches 1.0, as opposed to the largely different offsets of 0.593\,\r A ($J$) and 0.826\,\r A ($J'$) in NaVOAsO$_4$ ($\alpha\simeq 0.65$), with the stronger coupling $J$ corresponding to the smaller offset. We conclude that the in-plane offsets control the alternation ratio of the spin chains in this family of compounds.

\section{Discussion}
\begin{figure}
	\includegraphics[width = \linewidth]{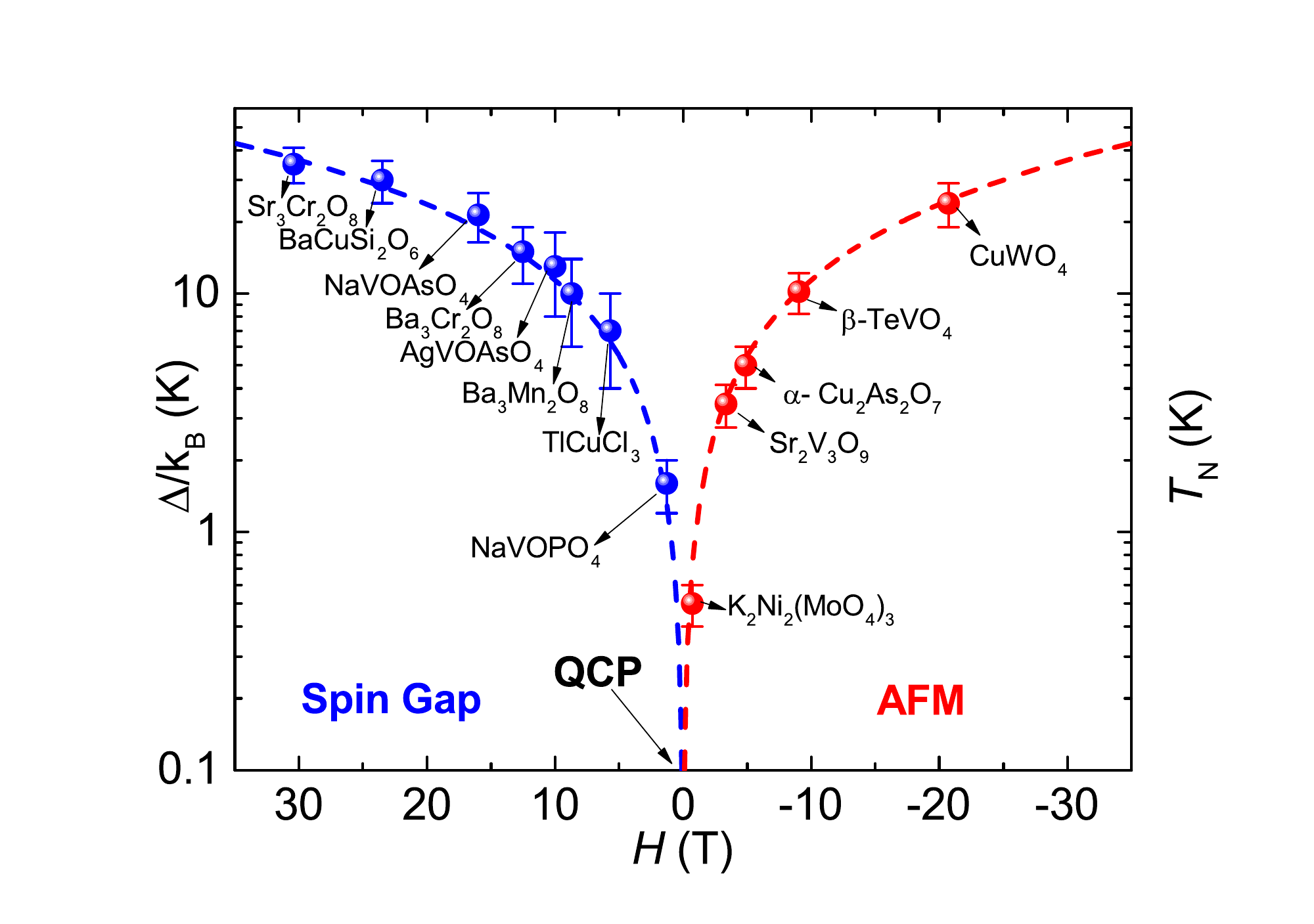}
	\caption{\label{Fig14} Illustration of a conceptual phase diagram showing the spin-gap and AFM regions separated by the QCP. Some reported compounds are placed at their respective positions.}
\end{figure}

The above assessments convincingly demonstrate that NaVOPO$_4$ reveals the magnetism of a spin-$1/2$ alternating chain with a weak alternation of the exchange couplings. This material does not undergo any magnetic LRO down to at least 50~mK, and rather shows a tiny spin gap of $\Delta_{0}/k_{\rm B} \simeq 2~$K in zero field. In Fig.~\ref{Fig14}, we constructed a conceptual phase diagram showing the QCP separating the spin-gap and AFM regions. We have placed some of the rigorously studied gapped and antiferromagneticaly ordered compounds at their respective positions according to the magnitude of their spin gap and $T_{\rm N}$. From the phase diagram we infer that NaVOPO$_{4}$ lies in the proximity of the QCP but in the gapped regime of the phase diagram. This renders NaVOPO$_4$ an ideal candidate for probing field-induced effects in gapped quantum magnets. Further, $\chi(T)$, $C_{\rm p}(T)$, $I_{\rm ESR}(T)$, and $1/T_{1}$ all show a second broad maximum at $\sim 4$~K which is found to be intrinsic to the sample and can be attributed to the effect of disorder and/or magnetic frustration. The same type of feature has been reported in the frustrated spin chain compound LiCuVO$_{4}$, where the broad feature at low temperature is ascribed to the effect of magnetic frustration of the interchain interactions.\cite{Vasil'ev024419,Kegler104418}

For a better understanding of the spin lattice, we compare NaVOPO$_{4}$ with its structural analogs AgVOAsO$_{4}$ and NaVOAsO$_{4}$. Magnetic parameters for all the three compounds are summarized in Table~\ref{Comparison}.
\begin{table}[htbp]
	\caption{Comparison of magnetic parameters for the NaVOPO$_4$-like spin-chain compounds. The $J/k_{\rm B}$ and $\alpha$ values are taken from the $\chi(T)$ analysis, while $\Delta_{\rm 0}/k_{\rm B}$ and $H_{\rm c1}$ are taken from the high-field magnetization data.}
	\label{Comparison}
	\begin{ruledtabular}
	\begin{tabular}{lllll}
		Compounds & $J/k_{\rm B}$~(K) & $\alpha (= J^\prime/J)$ & $\Delta_{\rm 0}/k_{\rm B}$~(K) & $H_{\rm c1}$~(T) \\ \hline
		AgVOAsO$_4$ & $40$  & $0.62$ &  $13$ & $10$  \\ 
	    NaVOAsO$_4$ & $52$  & $0.65$ & $21.4$ & $16$ \\ 
		NaVOPO$_4$  & $39$  & $0.98$  & $2$ & $1.6$ \\ 
	\end{tabular}
	\end{ruledtabular}
\end{table}

\begin{figure}
	\includegraphics[ height = 8 cm, width = \linewidth]{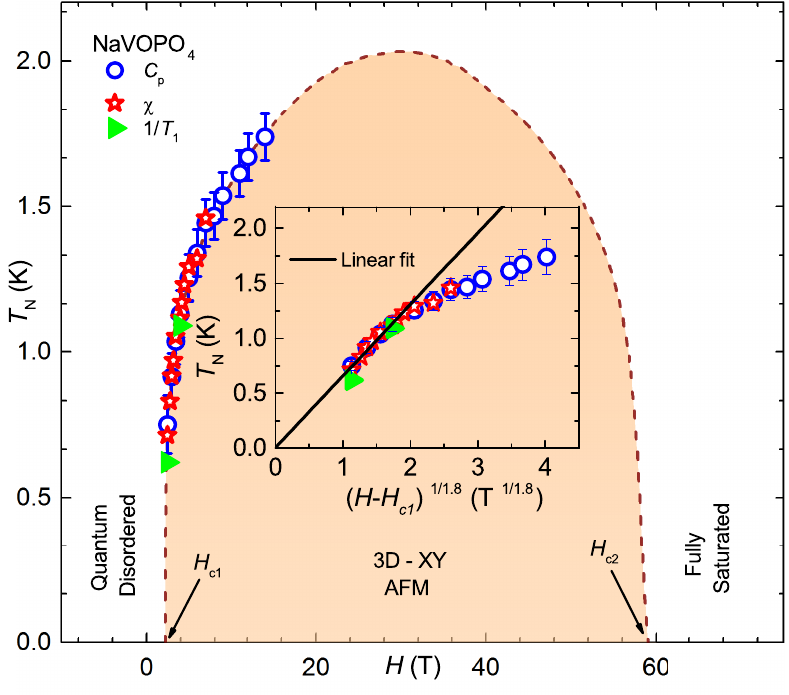}
	\caption{\label{Fig15} $H-T$ phase diagram of NaVOPO$_{4}$ obtained using the data points from the $\chi(T)$, $C_{\rm p}(T)$, and $1/T_1(T)$ measurements. The dashed line traces the anticipated phase boundary between the two critical fields. The lower ($H_{c1}$) and upper ($H_{c2}$) critical fields separate the quantum disordered state from the 3D-XY AFM state and fully saturated state from the the 3D-XY AFM state, respectively. Inset: $T_{\rm N}$ vs ($H-H_{\rm c1})^{1/1.8}$ and the solid line is the linear fit.}
\end{figure}

The most interesting aspect of NaVOPO$_4$ is arguably the field-induced magnetic LRO. The field evolution $T_{\rm N}$ obtained from the $\chi(T)$, $C_{\rm p}(T)$, and $1/T_{1}(T)$ measurements is shown in Fig.~\ref{Fig15}. It traces a single dome-shaped curve in contrast to the double-dome reported for the isostructural compound AgVOAsO$_4$.\cite{Weickert2019} The shape of the $H-T$ phase diagram is quite similar to other spin-gap systems undergoing BEC of triplons under external magnetic field.\cite{Zapf563,Aczel207203,Aczel100409,Jaime087203} The applicability of the BEC scenario is usually verified by the critical exponent obtained from fitting the $H-T$ boundary with the following power-law \cite{Nohadani220402,Giamarchi11398}
\begin{equation}\label{Power-law}
T_{\rm N} \propto {(H - H_{\rm c1})^{\frac{1}{\phi}}},
\end{equation}
where $\phi = d/2$ is the critical exponent, which reflects the universality class of the transition. Here, $H-H_{\rm c1}$ controls the boson density at a given temperature. This power law is typically fitted in the critical regime $T < 0.4 T_{\rm c}^{\rm max}$ in the vicinity of the QCP for a meaningful estimation of $\phi$, where $T_{\rm c}^{\rm max}$ is the maximum temperature of the magnetically ordered regime.\cite{Kawashima3219}  In a 3D system  ($d = 3$), the value of the critical exponent is predicted to be $\phi = 1.5$ \cite{Nikuni5868,Giamarchi198,Zapf563} which has been experimentally verified in quantum magnets TlCuCl$_3$, \cite{Yamada020409,Yamada013701} DTN,      \cite{Zapf077204,Yin1872005} (CH$_{3}$)$_{2}$CHNH$_{3}$CuCl$_{3}$, \cite{Tsujii042217} etc. On the other hand, for a 2D system, the theory predicts $\phi \simeq 1$ over a wide temperature range except at very low temperatures.\cite{Syromyatnikov134421} Indeed, this has been experimentally verified in K$_{2}$CuF$_{4}$.\cite{Hirata174406} At sufficiently low temperatures (i.e. temperatures lower than the interlayer coupling), because of the interlayer interaction, one may expect a crossover from 2D BEC to 3D BEC with $\phi = 3/2$. Of course, exchange anisotropy, which is inherent to real materials, will often alter the universality class of the transition.

To corroborate the above BEC scheme here, the $H-T$ phase boundary of NaVOPO$_4$ was fitted by Eq.~(\ref{Power-law}). Given the limited temperature range available experimentally, we restricted the fitting to $0.75$~K $\leq T \leq 1.13$~K. Our fit returns $H_{\rm c1} \simeq 1.3$~T and $\phi \simeq 1.8$. In the inset of Fig.~\ref{Fig15}, we have plotted $T_{\rm N}$ vs ($H-H_{\rm c1}$)$^{1/1.8}$ in order to highlight the linear regime. The obtained value of $H_{\rm c1} \simeq 1.3$~T leads to  $\Delta_{\rm 0}/k_{\rm B} \simeq 1.6$~K. This value of the spin gap differs slightly from the one obtained from the magnetic isotherm data. Similarly, $\phi \simeq 1.8$ is more close to (though slightly larger than) $1.5$ obtained theoretically.\cite{Nikuni5868,Giamarchi11398,Nohadani220402} Nevertheless, such discrepancies are not uncommon in other quantum magnets, including Ba$_{3}$Cr$_{2}$O$_{8}$, Pb$_{2}$V$_{3}$O$_{9}$, IPA- CuCl$_{3}$ etc undergoing triplon BEC.\cite{Aczel100409,Zheludev054450,Conner132401,Waki3435}

\section{Conclusion}
In conclusion, we have discovered a new spin-$1/2$ quasi-1D compound NaVOPO$_4$, which is well described by an alternating spin chain model with a dominant AFM exchange coupling of $J/k_{\rm B} \simeq 39$~K and a tiny spin gap of $\Delta_{\rm 0}/k_{\rm B} \simeq 2$~K. External magnetic field of $H_{c1} \simeq 1.6$~T closes the spin gap and triggers magnetic LRO. Such a small spin gap and the onset of magnetic LRO already in low magnetic fields place NaVOPO$_{4}$ in the vicinity of the QCP separating the spin-gap and AFM LRO states in the phase diagram. The field-induced magnetic LRO is indeed confirmed from the $\chi(T)$, $C_{\rm p}(T)$, $K(T)$, and $1/T_1(T)$ measurements and found to move toward high temperatures with the field. A power-law fit to the $H-T$ phase boundary yields an exponent $\phi \simeq 1.8$, a possible signature of triplon BEC. This compound appears to be an ideal system for high-field studies. Moreover, it also demands further experimental studies including neutron scattering and NMR on single crystals to elucidate microscopic nature of the field-induced ordered phase. 

\section{Acknowledgement}
PKM and RN would like to acknowledge BRNS, India for financial support bearing sanction No.37(3)/14/26/2017-BRNS. Work at the Ames Laboratory was supported by the U.S. Department of Energy, Office of Science, Basic Energy Sciences, Materials Sciences and Engineering Division. The Ames Laboratory is operated for the U.S. Department of Energy by Iowa State University under Contract No. DEAC02-07CH11358. Y.I. thanks JSPS Program for Fostering Globally Talented Researchers which provided an opportunity to be a visiting scholar at Ames Laboratory. We also thank C. Klausnitzer (MPI-CPfS) for the technical support. AT was funded by the Federal Ministry for Education and Research through the Sofja Kovalevskaya Award of Alexander von Humboldt Foundation.


\begin{thebibliography}{73}%
	\makeatletter
	\providecommand \@ifxundefined [1]{%
		\@ifx{#1\undefined}
	}%
	\providecommand \@ifnum [1]{%
		\ifnum #1\expandafter \@firstoftwo
		\else \expandafter \@secondoftwo
		\fi
	}%
	\providecommand \@ifx [1]{%
		\ifx #1\expandafter \@firstoftwo
		\else \expandafter \@secondoftwo
		\fi
	}%
	\providecommand \natexlab [1]{#1}%
	\providecommand \enquote  [1]{``#1''}%
	\providecommand \bibnamefont  [1]{#1}%
	\providecommand \bibfnamefont [1]{#1}%
	\providecommand \citenamefont [1]{#1}%
	\providecommand \href@noop [0]{\@secondoftwo}%
	\providecommand \href [0]{\begingroup \@sanitize@url \@href}%
	\providecommand \@href[1]{\@@startlink{#1}\@@href}%
	\providecommand \@@href[1]{\endgroup#1\@@endlink}%
	\providecommand \@sanitize@url [0]{\catcode `\\12\catcode `\$12\catcode
		`\&12\catcode `\#12\catcode `\^12\catcode `\_12\catcode `\%12\relax}%
	\providecommand \@@startlink[1]{}%
	\providecommand \@@endlink[0]{}%
	\providecommand \url  [0]{\begingroup\@sanitize@url \@url }%
	\providecommand \@url [1]{\endgroup\@href {#1}{\urlprefix }}%
	\providecommand \urlprefix  [0]{URL }%
	\providecommand \Eprint [0]{\href }%
	\providecommand \doibase [0]{http://dx.doi.org/}%
	\providecommand \selectlanguage [0]{\@gobble}%
	\providecommand \bibinfo  [0]{\@secondoftwo}%
	\providecommand \bibfield  [0]{\@secondoftwo}%
	\providecommand \translation [1]{[#1]}%
	\providecommand \BibitemOpen [0]{}%
	\providecommand \bibitemStop [0]{}%
	\providecommand \bibitemNoStop [0]{.\EOS\space}%
	\providecommand \EOS [0]{\spacefactor3000\relax}%
	\providecommand \BibitemShut  [1]{\csname bibitem#1\endcsname}%
	\let\auto@bib@innerbib\@empty
	\bibitem [{\citenamefont {Sachdev}(2007)}]{Sachdev2007}%
	\BibitemOpen
	\bibfield  {author} {\bibinfo {author} {\bibfnamefont {Subir}\ \bibnamefont
			{Sachdev}},\ }\bibfield  {title} {\enquote {\bibinfo {title} {Quantum phase
				transitions},}\ }\href@noop {} {\bibfield  {journal} {\bibinfo  {journal}
			{Handbook of Magnetism and Advanced Magnetic Materials}\ } (\bibinfo {year}
		{2007})}\BibitemShut {NoStop}%
	\bibitem [{\citenamefont {Sachdev}(2000)}]{Sachdev475}%
	\BibitemOpen
	\bibfield  {author} {\bibinfo {author} {\bibfnamefont {Subir}\ \bibnamefont
			{Sachdev}},\ }\bibfield  {title} {\enquote {\bibinfo {title} {Quantum
				criticality: Competing ground states in low dimensions},}\ }\href {\doibase
		10.1126/science.288.5465.475} {\bibfield  {journal} {\bibinfo  {journal}
			{Science}\ }\textbf {\bibinfo {volume} {288}},\ \bibinfo {pages} {475}
		(\bibinfo {year} {2000})}\BibitemShut {NoStop}%
	\bibitem [{\citenamefont {Sachdev}(2008)}]{Sachdev173}%
	\BibitemOpen
	\bibfield  {author} {\bibinfo {author} {\bibfnamefont {Subir}\ \bibnamefont
			{Sachdev}},\ }\bibfield  {title} {\enquote {\bibinfo {title} {Quantum
				magnetism and criticality},}\ }\href@noop {} {\bibfield  {journal} {\bibinfo
			{journal} {Nat. Phys.}\ }\textbf {\bibinfo {volume} {4}},\ \bibinfo {pages}
		{173} (\bibinfo {year} {2008})}\BibitemShut {NoStop}%
	\bibitem [{\citenamefont {Kojima}\ \emph {et~al.}(1997)\citenamefont {Kojima},
		\citenamefont {Fudamoto}, \citenamefont {Larkin}, \citenamefont {Luke},
		\citenamefont {Merrin}, \citenamefont {Nachumi}, \citenamefont {Uemura},
		\citenamefont {Motoyama}, \citenamefont {Eisaki}, \citenamefont {Uchida},
		\citenamefont {Yamada}, \citenamefont {Endoh}, \citenamefont {Hosoya},
		\citenamefont {Sternlieb},\ and\ \citenamefont {Shirane}}]{Kojima1787}%
	\BibitemOpen
	\bibfield  {author} {\bibinfo {author} {\bibfnamefont {K.~M.}\ \bibnamefont
			{Kojima}}, \bibinfo {author} {\bibfnamefont {Y.}~\bibnamefont {Fudamoto}},
		\bibinfo {author} {\bibfnamefont {M.}~\bibnamefont {Larkin}}, \bibinfo
		{author} {\bibfnamefont {G.~M.}\ \bibnamefont {Luke}}, \bibinfo {author}
		{\bibfnamefont {J.}~\bibnamefont {Merrin}}, \bibinfo {author} {\bibfnamefont
			{B.}~\bibnamefont {Nachumi}}, \bibinfo {author} {\bibfnamefont {Y.~J.}\
			\bibnamefont {Uemura}}, \bibinfo {author} {\bibfnamefont {N.}~\bibnamefont
			{Motoyama}}, \bibinfo {author} {\bibfnamefont {H.}~\bibnamefont {Eisaki}},
		\bibinfo {author} {\bibfnamefont {S.}~\bibnamefont {Uchida}}, \bibinfo
		{author} {\bibfnamefont {K.}~\bibnamefont {Yamada}}, \bibinfo {author}
		{\bibfnamefont {Y.}~\bibnamefont {Endoh}}, \bibinfo {author} {\bibfnamefont
			{S.}~\bibnamefont {Hosoya}}, \bibinfo {author} {\bibfnamefont {B.~J.}\
			\bibnamefont {Sternlieb}}, \ and\ \bibinfo {author} {\bibfnamefont
			{G.}~\bibnamefont {Shirane}},\ }\bibfield  {title} {\enquote {\bibinfo
			{title} {Reduction of ordered moment and n\'eel temperature of
				quasi-one-dimensional antiferromagnets {Sr$_2$CuO$_3$} and
				{Ca$_2$CuO$_3$}},}\ }\href {\doibase 10.1103/PhysRevLett.78.1787} {\bibfield
		{journal} {\bibinfo  {journal} {Phys. Rev. Lett.}\ }\textbf {\bibinfo
			{volume} {78}},\ \bibinfo {pages} {1787} (\bibinfo {year}
		{1997})}\BibitemShut {NoStop}%
	\bibitem [{\citenamefont {Barnes}\ \emph {et~al.}(1999)\citenamefont {Barnes},
		\citenamefont {Riera},\ and\ \citenamefont {Tennant}}]{Barnes11384}%
	\BibitemOpen
	\bibfield  {author} {\bibinfo {author} {\bibfnamefont {T.}~\bibnamefont
			{Barnes}}, \bibinfo {author} {\bibfnamefont {J.}~\bibnamefont {Riera}}, \
		and\ \bibinfo {author} {\bibfnamefont {D.~A.}\ \bibnamefont {Tennant}},\
	}\bibfield  {title} {\enquote {\bibinfo {title} {{$S=\frac{1}{2}$}
			alternating chain using multiprecision methods},}\ }\href {\doibase
	10.1103/PhysRevB.59.11384} {\bibfield  {journal} {\bibinfo  {journal} {Phys.
			Rev. B}\ }\textbf {\bibinfo {volume} {59}},\ \bibinfo {pages} {11384}
	(\bibinfo {year} {1999})}\BibitemShut {NoStop}%
\bibitem [{\citenamefont {Yamauchi}\ \emph {et~al.}(1999)\citenamefont
	{Yamauchi}, \citenamefont {Narumi}, \citenamefont {Kikuchi}, \citenamefont
	{Ueda}, \citenamefont {Tatani}, \citenamefont {Kobayashi}, \citenamefont
	{Kindo},\ and\ \citenamefont {Motoya}}]{Yamauchi3729}%
\BibitemOpen
\bibfield  {author} {\bibinfo {author} {\bibfnamefont {T.}~\bibnamefont
		{Yamauchi}}, \bibinfo {author} {\bibfnamefont {Y.}~\bibnamefont {Narumi}},
	\bibinfo {author} {\bibfnamefont {J.}~\bibnamefont {Kikuchi}}, \bibinfo
	{author} {\bibfnamefont {Y.}~\bibnamefont {Ueda}}, \bibinfo {author}
	{\bibfnamefont {K.}~\bibnamefont {Tatani}}, \bibinfo {author} {\bibfnamefont
		{T.~C.}\ \bibnamefont {Kobayashi}}, \bibinfo {author} {\bibfnamefont
		{K.}~\bibnamefont {Kindo}}, \ and\ \bibinfo {author} {\bibfnamefont
		{K.}~\bibnamefont {Motoya}},\ }\bibfield  {title} {\enquote {\bibinfo {title}
		{Two gaps in {(VO)$_2$P$_2$O$_7$}: Observation using high-field magnetization
			and {NMR}},}\ }\href {\doibase 10.1103/PhysRevLett.83.3729} {\bibfield
	{journal} {\bibinfo  {journal} {Phys. Rev. Lett.}\ }\textbf {\bibinfo
		{volume} {83}},\ \bibinfo {pages} {3729} (\bibinfo {year}
	{1999})}\BibitemShut {NoStop}%
\bibitem [{\citenamefont {Rice}(2002)}]{Rice760}%
\BibitemOpen
\bibfield  {author} {\bibinfo {author} {\bibfnamefont {T.~M.}\ \bibnamefont
		{Rice}},\ }\bibfield  {title} {\enquote {\bibinfo {title} {To condense or not
			to condense},}\ }\href {\doibase 10.1126/science.1078819} {\bibfield
	{journal} {\bibinfo  {journal} {Science}\ }\textbf {\bibinfo {volume}
		{298}},\ \bibinfo {pages} {760} (\bibinfo {year} {2002})}\BibitemShut
{NoStop}%
\bibitem [{\citenamefont {Giamarchi}\ \emph {et~al.}(2008)\citenamefont
	{Giamarchi}, \citenamefont {R{\"u}egg},\ and\ \citenamefont
	{Tchernyshyov}}]{Giamarchi198}%
\BibitemOpen
\bibfield  {author} {\bibinfo {author} {\bibfnamefont {Thierry}\ \bibnamefont
		{Giamarchi}}, \bibinfo {author} {\bibfnamefont {Christian}\ \bibnamefont
		{R{\"u}egg}}, \ and\ \bibinfo {author} {\bibfnamefont {Oleg}\ \bibnamefont
		{Tchernyshyov}},\ }\bibfield  {title} {\enquote {\bibinfo {title}
		{Bose-einstein condensation in magnetic insulators},}\ }\href
{https://doi.org/10.1038/nphys893} {\bibfield  {journal} {\bibinfo  {journal}
		{Nat. Phys.}\ }\textbf {\bibinfo {volume} {4}},\ \bibinfo {pages} {198}
	(\bibinfo {year} {2008})}\BibitemShut {NoStop}%
\bibitem [{\citenamefont {Nikuni}\ \emph {et~al.}(2000)\citenamefont {Nikuni},
	\citenamefont {Oshikawa}, \citenamefont {Oosawa},\ and\ \citenamefont
	{Tanaka}}]{Nikuni5868}%
\BibitemOpen
\bibfield  {author} {\bibinfo {author} {\bibfnamefont {T.}~\bibnamefont
		{Nikuni}}, \bibinfo {author} {\bibfnamefont {M.}~\bibnamefont {Oshikawa}},
	\bibinfo {author} {\bibfnamefont {A.}~\bibnamefont {Oosawa}}, \ and\ \bibinfo
	{author} {\bibfnamefont {H.}~\bibnamefont {Tanaka}},\ }\bibfield  {title}
{\enquote {\bibinfo {title} {{$\rm Bose-Einstein$} condensation of dilute
			magnons in {TlCuCl$_{3}$}},}\ }\href {\doibase 10.1103/PhysRevLett.84.5868}
{\bibfield  {journal} {\bibinfo  {journal} {Phys. Rev. Lett.}\ }\textbf
	{\bibinfo {volume} {84}},\ \bibinfo {pages} {5868} (\bibinfo {year}
	{2000})}\BibitemShut {NoStop}%
\bibitem [{\citenamefont {Aczel}\ \emph
	{et~al.}(2009{\natexlab{a}})\citenamefont {Aczel}, \citenamefont {Kohama},
	\citenamefont {Marcenat}, \citenamefont {Weickert}, \citenamefont {Jaime},
	\citenamefont {Ayala-Valenzuela}, \citenamefont {McDonald}, \citenamefont
	{Selesnic}, \citenamefont {Dabkowska},\ and\ \citenamefont
	{Luke}}]{Aczel207203}%
\BibitemOpen
\bibfield  {author} {\bibinfo {author} {\bibfnamefont {A.~A.}\ \bibnamefont
		{Aczel}}, \bibinfo {author} {\bibfnamefont {Y.}~\bibnamefont {Kohama}},
	\bibinfo {author} {\bibfnamefont {C.}~\bibnamefont {Marcenat}}, \bibinfo
	{author} {\bibfnamefont {F.}~\bibnamefont {Weickert}}, \bibinfo {author}
	{\bibfnamefont {M.}~\bibnamefont {Jaime}}, \bibinfo {author} {\bibfnamefont
		{O.~E.}\ \bibnamefont {Ayala-Valenzuela}}, \bibinfo {author} {\bibfnamefont
		{R.~D.}\ \bibnamefont {McDonald}}, \bibinfo {author} {\bibfnamefont {S.~D.}\
		\bibnamefont {Selesnic}}, \bibinfo {author} {\bibfnamefont {H.~A.}\
		\bibnamefont {Dabkowska}}, \ and\ \bibinfo {author} {\bibfnamefont {G.~M.}\
		\bibnamefont {Luke}},\ }\bibfield  {title} {\enquote {\bibinfo {title}
		{Field-induced {$\rm Bose-Einstein$} condensation of triplons up to {$8$ K}
			in {Sr$_{3}$Cr$_{2}$O$_{8}$}},}\ }\href {\doibase
	10.1103/PhysRevLett.103.207203} {\bibfield  {journal} {\bibinfo  {journal}
		{Phys. Rev. Lett.}\ }\textbf {\bibinfo {volume} {103}},\ \bibinfo {pages}
	{207203} (\bibinfo {year} {2009}{\natexlab{a}})}\BibitemShut {NoStop}%
\bibitem [{\citenamefont {Hirata}\ \emph {et~al.}(2017)\citenamefont {Hirata},
	\citenamefont {Kurita}, \citenamefont {Yamada},\ and\ \citenamefont
	{Tanaka}}]{Hirata174406}%
\BibitemOpen
\bibfield  {author} {\bibinfo {author} {\bibfnamefont {Satoshi}\ \bibnamefont
		{Hirata}}, \bibinfo {author} {\bibfnamefont {Nobuyuki}\ \bibnamefont
		{Kurita}}, \bibinfo {author} {\bibfnamefont {Motoki}\ \bibnamefont {Yamada}},
	\ and\ \bibinfo {author} {\bibfnamefont {Hidekazu}\ \bibnamefont {Tanaka}},\
}\bibfield  {title} {\enquote {\bibinfo {title} {Quasi-two-dimensional
		bose-einstein condensation of lattice bosons in the spin-$\frac{1}{2}$ xxz
		ferromagnet {K$_2$CuF$_4$}},}\ }\href {\doibase 10.1103/PhysRevB.95.174406}
{\bibfield  {journal} {\bibinfo  {journal} {Phys. Rev. B}\ }\textbf {\bibinfo
		{volume} {95}},\ \bibinfo {pages} {174406} (\bibinfo {year}
	{2017})}\BibitemShut {NoStop}%
\bibitem [{\citenamefont {Klanj\ifmmode~\check{s}\else \v{s}\fi{}ek}\ \emph
	{et~al.}(2008)\citenamefont {Klanj\ifmmode~\check{s}\else \v{s}\fi{}ek},
	\citenamefont {Mayaffre}, \citenamefont {Berthier}, \citenamefont
	{Horvati\ifmmode~\acute{c}\else \'{c}\fi{}}, \citenamefont {Chiari},
	\citenamefont {Piovesana}, \citenamefont {Bouillot}, \citenamefont {Kollath},
	\citenamefont {Orignac}, \citenamefont {Citro},\ and\ \citenamefont
	{Giamarchi}}]{Klanjsek137207}%
\BibitemOpen
\bibfield  {author} {\bibinfo {author} {\bibfnamefont {M.}~\bibnamefont
		{Klanj\ifmmode~\check{s}\else \v{s}\fi{}ek}}, \bibinfo {author}
	{\bibfnamefont {H.}~\bibnamefont {Mayaffre}}, \bibinfo {author}
	{\bibfnamefont {C.}~\bibnamefont {Berthier}}, \bibinfo {author}
	{\bibfnamefont {M.}~\bibnamefont {Horvati\ifmmode~\acute{c}\else
			\'{c}\fi{}}}, \bibinfo {author} {\bibfnamefont {B.}~\bibnamefont {Chiari}},
	\bibinfo {author} {\bibfnamefont {O.}~\bibnamefont {Piovesana}}, \bibinfo
	{author} {\bibfnamefont {P.}~\bibnamefont {Bouillot}}, \bibinfo {author}
	{\bibfnamefont {C.}~\bibnamefont {Kollath}}, \bibinfo {author} {\bibfnamefont
		{E.}~\bibnamefont {Orignac}}, \bibinfo {author} {\bibfnamefont
		{R.}~\bibnamefont {Citro}}, \ and\ \bibinfo {author} {\bibfnamefont
		{T.}~\bibnamefont {Giamarchi}},\ }\bibfield  {title} {\enquote {\bibinfo
		{title} {Controlling {Luttinger} liquid physics in spin ladders under a
			magnetic field},}\ }\href {\doibase 10.1103/PhysRevLett.101.137207}
{\bibfield  {journal} {\bibinfo  {journal} {Phys. Rev. Lett.}\ }\textbf
	{\bibinfo {volume} {101}},\ \bibinfo {pages} {137207} (\bibinfo {year}
	{2008})}\BibitemShut {NoStop}%
\bibitem [{\citenamefont {Hong}\ \emph {et~al.}(2010)\citenamefont {Hong},
	\citenamefont {Kim}, \citenamefont {Hotta}, \citenamefont {Takano},
	\citenamefont {Tremelling}, \citenamefont {Turnbull}, \citenamefont {Landee},
	\citenamefont {Kang}, \citenamefont {Christensen}, \citenamefont {Lefmann},
	\citenamefont {Schmidt}, \citenamefont {Uhrig},\ and\ \citenamefont
	{Broholm}}]{Hong137207}%
\BibitemOpen
\bibfield  {author} {\bibinfo {author} {\bibfnamefont {Tao}\ \bibnamefont
		{Hong}}, \bibinfo {author} {\bibfnamefont {Y.~H.}\ \bibnamefont {Kim}},
	\bibinfo {author} {\bibfnamefont {C.}~\bibnamefont {Hotta}}, \bibinfo
	{author} {\bibfnamefont {Y.}~\bibnamefont {Takano}}, \bibinfo {author}
	{\bibfnamefont {G.}~\bibnamefont {Tremelling}}, \bibinfo {author}
	{\bibfnamefont {M.~M.}\ \bibnamefont {Turnbull}}, \bibinfo {author}
	{\bibfnamefont {C.~P.}\ \bibnamefont {Landee}}, \bibinfo {author}
	{\bibfnamefont {H.-J.}\ \bibnamefont {Kang}}, \bibinfo {author}
	{\bibfnamefont {N.~B.}\ \bibnamefont {Christensen}}, \bibinfo {author}
	{\bibfnamefont {K.}~\bibnamefont {Lefmann}}, \bibinfo {author} {\bibfnamefont
		{K.~P.}\ \bibnamefont {Schmidt}}, \bibinfo {author} {\bibfnamefont {G.~S.}\
		\bibnamefont {Uhrig}}, \ and\ \bibinfo {author} {\bibfnamefont
		{C.}~\bibnamefont {Broholm}},\ }\bibfield  {title} {\enquote {\bibinfo
		{title} {Field-induced {Tomonaga-Luttinger} liquid phase of a two-leg
			spin-1/2 ladder with strong leg interactions},}\ }\href {\doibase
	10.1103/PhysRevLett.105.137207} {\bibfield  {journal} {\bibinfo  {journal}
		{Phys. Rev. Lett.}\ }\textbf {\bibinfo {volume} {105}},\ \bibinfo {pages}
	{137207} (\bibinfo {year} {2010})}\BibitemShut {NoStop}%
\bibitem [{\citenamefont {Willenberg}\ \emph {et~al.}(2015)\citenamefont
	{Willenberg}, \citenamefont {Ryll}, \citenamefont {Kiefer}, \citenamefont
	{Tennant}, \citenamefont {Groitl}, \citenamefont {Rolfs}, \citenamefont
	{Manuel}, \citenamefont {Khalyavin}, \citenamefont {Rule}, \citenamefont
	{Wolter},\ and\ \citenamefont {S\"ullow}}]{Willenberg060407}%
\BibitemOpen
\bibfield  {author} {\bibinfo {author} {\bibfnamefont {B.}~\bibnamefont
		{Willenberg}}, \bibinfo {author} {\bibfnamefont {H.}~\bibnamefont {Ryll}},
	\bibinfo {author} {\bibfnamefont {K.}~\bibnamefont {Kiefer}}, \bibinfo
	{author} {\bibfnamefont {D.~A.}\ \bibnamefont {Tennant}}, \bibinfo {author}
	{\bibfnamefont {F.}~\bibnamefont {Groitl}}, \bibinfo {author} {\bibfnamefont
		{K.}~\bibnamefont {Rolfs}}, \bibinfo {author} {\bibfnamefont
		{P.}~\bibnamefont {Manuel}}, \bibinfo {author} {\bibfnamefont
		{D.}~\bibnamefont {Khalyavin}}, \bibinfo {author} {\bibfnamefont {K.~C.}\
		\bibnamefont {Rule}}, \bibinfo {author} {\bibfnamefont {A.~U.~B.}\
		\bibnamefont {Wolter}}, \ and\ \bibinfo {author} {\bibfnamefont
		{S.}~\bibnamefont {S\"ullow}},\ }\bibfield  {title} {\enquote {\bibinfo
		{title} {Luttinger liquid behavior in the alternating spin-chain system
			copper nitrate},}\ }\href {\doibase 10.1103/PhysRevB.91.060407} {\bibfield
	{journal} {\bibinfo  {journal} {Phys. Rev. B}\ }\textbf {\bibinfo {volume}
		{91}},\ \bibinfo {pages} {060407} (\bibinfo {year} {2015})}\BibitemShut
{NoStop}%
\bibitem [{\citenamefont {Horsch}\ \emph {et~al.}(2005)\citenamefont {Horsch},
	\citenamefont {Sofin}, \citenamefont {Mayr},\ and\ \citenamefont
	{Jansen}}]{Horsch076403}%
\BibitemOpen
\bibfield  {author} {\bibinfo {author} {\bibfnamefont {P.}~\bibnamefont
		{Horsch}}, \bibinfo {author} {\bibfnamefont {M.}~\bibnamefont {Sofin}},
	\bibinfo {author} {\bibfnamefont {M.}~\bibnamefont {Mayr}}, \ and\ \bibinfo
	{author} {\bibfnamefont {M.}~\bibnamefont {Jansen}},\ }\bibfield  {title}
{\enquote {\bibinfo {title} {Wigner crystallization in {Na$_3$Cu$_2$O$_4$}
			and {Na$_8$Cu$_5$O$_{10}$} chain compounds},}\ }\href {\doibase
	10.1103/PhysRevLett.94.076403} {\bibfield  {journal} {\bibinfo  {journal}
		{Phys. Rev. Lett.}\ }\textbf {\bibinfo {volume} {94}},\ \bibinfo {pages}
	{076403} (\bibinfo {year} {2005})}\BibitemShut {NoStop}%
\bibitem [{\citenamefont {Kageyama}\ \emph {et~al.}(1999)\citenamefont
	{Kageyama}, \citenamefont {Yoshimura}, \citenamefont {Stern}, \citenamefont
	{Mushnikov}, \citenamefont {Onizuka}, \citenamefont {Kato}, \citenamefont
	{Kosuge}, \citenamefont {Slichter}, \citenamefont {Goto},\ and\ \citenamefont
	{Ueda}}]{Kageyama3168}%
\BibitemOpen
\bibfield  {author} {\bibinfo {author} {\bibfnamefont {H.}~\bibnamefont
		{Kageyama}}, \bibinfo {author} {\bibfnamefont {K.}~\bibnamefont {Yoshimura}},
	\bibinfo {author} {\bibfnamefont {R.}~\bibnamefont {Stern}}, \bibinfo
	{author} {\bibfnamefont {N.~V.}\ \bibnamefont {Mushnikov}}, \bibinfo {author}
	{\bibfnamefont {K.}~\bibnamefont {Onizuka}}, \bibinfo {author} {\bibfnamefont
		{M.}~\bibnamefont {Kato}}, \bibinfo {author} {\bibfnamefont {K.}~\bibnamefont
		{Kosuge}}, \bibinfo {author} {\bibfnamefont {C.~P.}\ \bibnamefont
		{Slichter}}, \bibinfo {author} {\bibfnamefont {T.}~\bibnamefont {Goto}}, \
	and\ \bibinfo {author} {\bibfnamefont {Y.}~\bibnamefont {Ueda}},\ }\bibfield
{title} {\enquote {\bibinfo {title} {Exact dimer ground state and quantized
			magnetization plateaus in the two-dimensional spin system
			{SrCu$_2$({BO$_3$})$_2$}},}\ }\href {\doibase 10.1103/PhysRevLett.82.3168}
{\bibfield  {journal} {\bibinfo  {journal} {Phys. Rev. Lett.}\ }\textbf
	{\bibinfo {volume} {82}},\ \bibinfo {pages} {3168} (\bibinfo {year}
	{1999})}\BibitemShut {NoStop}%
\bibitem [{\citenamefont {Mukhopadhyay}\ \emph {et~al.}(2012)\citenamefont
	{Mukhopadhyay}, \citenamefont {Klanj\ifmmode~\check{s}\else \v{s}\fi{}ek},
	\citenamefont {Grbi\ifmmode~\acute{c}\else \'{c}\fi{}}, \citenamefont
	{Blinder}, \citenamefont {Mayaffre}, \citenamefont {Berthier}, \citenamefont
	{Horvati\ifmmode~\acute{c}\else \'{c}\fi{}}, \citenamefont {Continentino},
	\citenamefont {Paduan-Filho}, \citenamefont {Chiari},\ and\ \citenamefont
	{Piovesana}}]{Mukhopadhyay177206}%
\BibitemOpen
\bibfield  {author} {\bibinfo {author} {\bibfnamefont {S.}~\bibnamefont
		{Mukhopadhyay}}, \bibinfo {author} {\bibfnamefont {M.}~\bibnamefont
		{Klanj\ifmmode~\check{s}\else \v{s}\fi{}ek}}, \bibinfo {author}
	{\bibfnamefont {M.~S.}\ \bibnamefont {Grbi\ifmmode~\acute{c}\else
			\'{c}\fi{}}}, \bibinfo {author} {\bibfnamefont {R.}~\bibnamefont {Blinder}},
	\bibinfo {author} {\bibfnamefont {H.}~\bibnamefont {Mayaffre}}, \bibinfo
	{author} {\bibfnamefont {C.}~\bibnamefont {Berthier}}, \bibinfo {author}
	{\bibfnamefont {M.}~\bibnamefont {Horvati\ifmmode~\acute{c}\else
			\'{c}\fi{}}}, \bibinfo {author} {\bibfnamefont {M.~A.}\ \bibnamefont
		{Continentino}}, \bibinfo {author} {\bibfnamefont {A.}~\bibnamefont
		{Paduan-Filho}}, \bibinfo {author} {\bibfnamefont {B.}~\bibnamefont
		{Chiari}}, \ and\ \bibinfo {author} {\bibfnamefont {O.}~\bibnamefont
		{Piovesana}},\ }\bibfield  {title} {\enquote {\bibinfo {title}
		{Quantum-critical spin dynamics in quasi-one-dimensional antiferromagnets},}\
}\href {\doibase 10.1103/PhysRevLett.109.177206} {\bibfield  {journal}
{\bibinfo  {journal} {Phys. Rev. Lett.}\ }\textbf {\bibinfo {volume} {109}},\
\bibinfo {pages} {177206} (\bibinfo {year} {2012})}\BibitemShut {NoStop}%
\bibitem [{\citenamefont {Thielemann}\ \emph {et~al.}(2009)\citenamefont
	{Thielemann}, \citenamefont {R\"uegg}, \citenamefont {Kiefer}, \citenamefont
	{R\o{}nnow}, \citenamefont {Normand}, \citenamefont {Bouillot}, \citenamefont
	{Kollath}, \citenamefont {Orignac}, \citenamefont {Citro}, \citenamefont
	{Giamarchi}, \citenamefont {L\"auchli}, \citenamefont {Biner}, \citenamefont
	{Kr\"amer}, \citenamefont {Wolff-Fabris}, \citenamefont {Zapf}, \citenamefont
	{Jaime}, \citenamefont {Stahn}, \citenamefont {Christensen}, \citenamefont
	{Grenier}, \citenamefont {McMorrow},\ and\ \citenamefont
	{Mesot}}]{Thielemann020408}%
\BibitemOpen
\bibfield  {author} {\bibinfo {author} {\bibfnamefont {B.}~\bibnamefont
		{Thielemann}}, \bibinfo {author} {\bibfnamefont {Ch.}\ \bibnamefont
		{R\"uegg}}, \bibinfo {author} {\bibfnamefont {K.}~\bibnamefont {Kiefer}},
	\bibinfo {author} {\bibfnamefont {H.~M.}\ \bibnamefont {R\o{}nnow}}, \bibinfo
	{author} {\bibfnamefont {B.}~\bibnamefont {Normand}}, \bibinfo {author}
	{\bibfnamefont {P.}~\bibnamefont {Bouillot}}, \bibinfo {author}
	{\bibfnamefont {C.}~\bibnamefont {Kollath}}, \bibinfo {author} {\bibfnamefont
		{E.}~\bibnamefont {Orignac}}, \bibinfo {author} {\bibfnamefont
		{R.}~\bibnamefont {Citro}}, \bibinfo {author} {\bibfnamefont
		{T.}~\bibnamefont {Giamarchi}}, \bibinfo {author} {\bibfnamefont {A.~M.}\
		\bibnamefont {L\"auchli}}, \bibinfo {author} {\bibfnamefont {D.}~\bibnamefont
		{Biner}}, \bibinfo {author} {\bibfnamefont {K.~W.}\ \bibnamefont {Kr\"amer}},
	\bibinfo {author} {\bibfnamefont {F.}~\bibnamefont {Wolff-Fabris}}, \bibinfo
	{author} {\bibfnamefont {V.~S.}\ \bibnamefont {Zapf}}, \bibinfo {author}
	{\bibfnamefont {M.}~\bibnamefont {Jaime}}, \bibinfo {author} {\bibfnamefont
		{J.}~\bibnamefont {Stahn}}, \bibinfo {author} {\bibfnamefont {N.~B.}\
		\bibnamefont {Christensen}}, \bibinfo {author} {\bibfnamefont
		{B.}~\bibnamefont {Grenier}}, \bibinfo {author} {\bibfnamefont {D.~F.}\
		\bibnamefont {McMorrow}}, \ and\ \bibinfo {author} {\bibfnamefont
		{J.}~\bibnamefont {Mesot}},\ }\bibfield  {title} {\enquote {\bibinfo {title}
		{Field-controlled magnetic order in the quantum spin-ladder system
			{(Hpip)$_2$CuBr$_4$}},}\ }\href {\doibase 10.1103/PhysRevB.79.020408}
{\bibfield  {journal} {\bibinfo  {journal} {Phys. Rev. B}\ }\textbf {\bibinfo
		{volume} {79}},\ \bibinfo {pages} {020408} (\bibinfo {year}
	{2009})}\BibitemShut {NoStop}%
\bibitem [{\citenamefont {Jaime}\ \emph {et~al.}(2004)\citenamefont {Jaime},
	\citenamefont {Correa}, \citenamefont {Harrison}, \citenamefont {Batista},
	\citenamefont {Kawashima}, \citenamefont {Kazuma}, \citenamefont {Jorge},
	\citenamefont {Stern}, \citenamefont {Heinmaa}, \citenamefont {Zvyagin},
	\citenamefont {Sasago},\ and\ \citenamefont {Uchinokura}}]{Jaime087203}%
\BibitemOpen
\bibfield  {author} {\bibinfo {author} {\bibfnamefont {M.}~\bibnamefont
		{Jaime}}, \bibinfo {author} {\bibfnamefont {V.~F.}\ \bibnamefont {Correa}},
	\bibinfo {author} {\bibfnamefont {N.}~\bibnamefont {Harrison}}, \bibinfo
	{author} {\bibfnamefont {C.~D.}\ \bibnamefont {Batista}}, \bibinfo {author}
	{\bibfnamefont {N.}~\bibnamefont {Kawashima}}, \bibinfo {author}
	{\bibfnamefont {Y.}~\bibnamefont {Kazuma}}, \bibinfo {author} {\bibfnamefont
		{G.~A.}\ \bibnamefont {Jorge}}, \bibinfo {author} {\bibfnamefont
		{R.}~\bibnamefont {Stern}}, \bibinfo {author} {\bibfnamefont
		{I.}~\bibnamefont {Heinmaa}}, \bibinfo {author} {\bibfnamefont {S.~A.}\
		\bibnamefont {Zvyagin}}, \bibinfo {author} {\bibfnamefont {Y.}~\bibnamefont
		{Sasago}}, \ and\ \bibinfo {author} {\bibfnamefont {K.}~\bibnamefont
		{Uchinokura}},\ }\bibfield  {title} {\enquote {\bibinfo {title}
		{Magnetic-field-induced condensation of triplons in han purple pigment
			{BaCuSi$_2$O$_6$}},}\ }\href {\doibase 10.1103/PhysRevLett.93.087203}
{\bibfield  {journal} {\bibinfo  {journal} {Phys. Rev. Lett.}\ }\textbf
	{\bibinfo {volume} {93}},\ \bibinfo {pages} {087203} (\bibinfo {year}
	{2004})}\BibitemShut {NoStop}%
\bibitem [{\citenamefont {Aczel}\ \emph
	{et~al.}(2009{\natexlab{b}})\citenamefont {Aczel}, \citenamefont {Kohama},
	\citenamefont {Jaime}, \citenamefont {Ninios}, \citenamefont {Chan},
	\citenamefont {Balicas}, \citenamefont {Dabkowska},\ and\ \citenamefont
	{Luke}}]{Aczel100409}%
\BibitemOpen
\bibfield  {author} {\bibinfo {author} {\bibfnamefont {A.~A.}\ \bibnamefont
		{Aczel}}, \bibinfo {author} {\bibfnamefont {Y.}~\bibnamefont {Kohama}},
	\bibinfo {author} {\bibfnamefont {M.}~\bibnamefont {Jaime}}, \bibinfo
	{author} {\bibfnamefont {K.}~\bibnamefont {Ninios}}, \bibinfo {author}
	{\bibfnamefont {H.~B.}\ \bibnamefont {Chan}}, \bibinfo {author}
	{\bibfnamefont {L.}~\bibnamefont {Balicas}}, \bibinfo {author} {\bibfnamefont
		{H.~A.}\ \bibnamefont {Dabkowska}}, \ and\ \bibinfo {author} {\bibfnamefont
		{G.~M.}\ \bibnamefont {Luke}},\ }\bibfield  {title} {\enquote {\bibinfo
		{title} {{$\rm Bose-Einstein$} condensation of triplons in
			{Sr$_{3}$Cr$_{2}$O$_{8}$}},}\ }\href {\doibase 10.1103/PhysRevB.79.100409}
{\bibfield  {journal} {\bibinfo  {journal} {Phys. Rev. B}\ }\textbf {\bibinfo
		{volume} {79}},\ \bibinfo {pages} {100409} (\bibinfo {year}
	{2009}{\natexlab{b}})}\BibitemShut {NoStop}%
\bibitem [{\citenamefont {Samulon}\ \emph {et~al.}(2009)\citenamefont
	{Samulon}, \citenamefont {Kohama}, \citenamefont {McDonald}, \citenamefont
	{Shapiro}, \citenamefont {Al-Hassanieh}, \citenamefont {Batista},
	\citenamefont {Jaime},\ and\ \citenamefont {Fisher}}]{Samulon047202}%
\BibitemOpen
\bibfield  {author} {\bibinfo {author} {\bibfnamefont {E.~C.}\ \bibnamefont
		{Samulon}}, \bibinfo {author} {\bibfnamefont {Y.}~\bibnamefont {Kohama}},
	\bibinfo {author} {\bibfnamefont {R.~D.}\ \bibnamefont {McDonald}}, \bibinfo
	{author} {\bibfnamefont {M.~C.}\ \bibnamefont {Shapiro}}, \bibinfo {author}
	{\bibfnamefont {K.~A.}\ \bibnamefont {Al-Hassanieh}}, \bibinfo {author}
	{\bibfnamefont {C.~D.}\ \bibnamefont {Batista}}, \bibinfo {author}
	{\bibfnamefont {M.}~\bibnamefont {Jaime}}, \ and\ \bibinfo {author}
	{\bibfnamefont {I.~R.}\ \bibnamefont {Fisher}},\ }\bibfield  {title}
{\enquote {\bibinfo {title} {Asymmetric quintuplet condensation in the
			frustrated {$S=1$} spin dimer compound {Ba$_{3}$Mn$_{2}$O$_{8}$}},}\ }\href
{\doibase 10.1103/PhysRevLett.103.047202} {\bibfield  {journal} {\bibinfo
		{journal} {Phys. Rev. Lett.}\ }\textbf {\bibinfo {volume} {103}},\ \bibinfo
	{pages} {047202} (\bibinfo {year} {2009})}\BibitemShut {NoStop}%
\bibitem [{\citenamefont {Arjun}\ \emph {et~al.}(2019)\citenamefont {Arjun},
	\citenamefont {Ranjith}, \citenamefont {Koo}, \citenamefont {Sichelschmidt},
	\citenamefont {Skourski}, \citenamefont {Baenitz}, \citenamefont {Tsirlin},\
	and\ \citenamefont {Nath}}]{Arjun014421}%
\BibitemOpen
\bibfield  {author} {\bibinfo {author} {\bibfnamefont {U.}~\bibnamefont
		{Arjun}}, \bibinfo {author} {\bibfnamefont {K.~M.}\ \bibnamefont {Ranjith}},
	\bibinfo {author} {\bibfnamefont {B.}~\bibnamefont {Koo}}, \bibinfo {author}
	{\bibfnamefont {J.}~\bibnamefont {Sichelschmidt}}, \bibinfo {author}
	{\bibfnamefont {Y.}~\bibnamefont {Skourski}}, \bibinfo {author}
	{\bibfnamefont {M.}~\bibnamefont {Baenitz}}, \bibinfo {author} {\bibfnamefont
		{A.~A.}\ \bibnamefont {Tsirlin}}, \ and\ \bibinfo {author} {\bibfnamefont
		{R.}~\bibnamefont {Nath}},\ }\bibfield  {title} {\enquote {\bibinfo {title}
		{Singlet ground state in the alternating spin-$\frac{1}{2}$ chain compound
			{NaVOAsO$_{4}$}},}\ }\href {\doibase 10.1103/PhysRevB.99.014421} {\bibfield
	{journal} {\bibinfo  {journal} {Phys. Rev. B}\ }\textbf {\bibinfo {volume}
		{99}},\ \bibinfo {pages} {014421} (\bibinfo {year} {2019})}\BibitemShut
{NoStop}%
\bibitem [{\citenamefont {Ahmed}\ \emph {et~al.}(2017)\citenamefont {Ahmed},
	\citenamefont {Khuntia}, \citenamefont {Ranjith}, \citenamefont {Rosner},
	\citenamefont {Baenitz}, \citenamefont {Tsirlin},\ and\ \citenamefont
	{Nath}}]{Ahmed224433}%
\BibitemOpen
\bibfield  {author} {\bibinfo {author} {\bibfnamefont {N.}~\bibnamefont
		{Ahmed}}, \bibinfo {author} {\bibfnamefont {P.}~\bibnamefont {Khuntia}},
	\bibinfo {author} {\bibfnamefont {K.~M.}\ \bibnamefont {Ranjith}}, \bibinfo
	{author} {\bibfnamefont {H.}~\bibnamefont {Rosner}}, \bibinfo {author}
	{\bibfnamefont {M.}~\bibnamefont {Baenitz}}, \bibinfo {author} {\bibfnamefont
		{A.~A.}\ \bibnamefont {Tsirlin}}, \ and\ \bibinfo {author} {\bibfnamefont
		{R.}~\bibnamefont {Nath}},\ }\bibfield  {title} {\enquote {\bibinfo {title}
		{Alternating spin chain compound {AgVOAsO$_4$} probed by {$^{75}${As}}
			{NMR}},}\ }\href {\doibase 10.1103/PhysRevB.96.224423} {\bibfield  {journal}
	{\bibinfo  {journal} {Phys. Rev. B}\ }\textbf {\bibinfo {volume} {96}},\
	\bibinfo {pages} {224423} (\bibinfo {year} {2017})}\BibitemShut {NoStop}%
\bibitem [{\citenamefont {Tsirlin}\ \emph {et~al.}(2011)\citenamefont
	{Tsirlin}, \citenamefont {Nath}, \citenamefont {Sichelschmidt}, \citenamefont
	{Skourski}, \citenamefont {Geibel},\ and\ \citenamefont
	{Rosner}}]{Tsirlin144412}%
\BibitemOpen
\bibfield  {author} {\bibinfo {author} {\bibfnamefont {Alexander~A.}\
		\bibnamefont {Tsirlin}}, \bibinfo {author} {\bibfnamefont {Ramesh}\
		\bibnamefont {Nath}}, \bibinfo {author} {\bibfnamefont {J\"org}\ \bibnamefont
		{Sichelschmidt}}, \bibinfo {author} {\bibfnamefont {Yurii}\ \bibnamefont
		{Skourski}}, \bibinfo {author} {\bibfnamefont {Christoph}\ \bibnamefont
		{Geibel}}, \ and\ \bibinfo {author} {\bibfnamefont {Helge}\ \bibnamefont
		{Rosner}},\ }\bibfield  {title} {\enquote {\bibinfo {title} {Frustrated
			couplings between alternating spin-$\frac{1}{2}$ chains in
			{AgVOAsO$_{4}$}},}\ }\href {\doibase 10.1103/PhysRevB.83.144412} {\bibfield
	{journal} {\bibinfo  {journal} {Phys. Rev. B}\ }\textbf {\bibinfo {volume}
		{83}},\ \bibinfo {pages} {144412} (\bibinfo {year} {2011})}\BibitemShut
{NoStop}%
\bibitem [{\citenamefont {Weickert}\ \emph {et~al.}(2019)\citenamefont
	{Weickert}, \citenamefont {Aczel}, \citenamefont {Stone}, \citenamefont
	{Garlea}, \citenamefont {Dong}, \citenamefont {Kohama}, \citenamefont
	{Movshovich}, \citenamefont {Demuer}, \citenamefont {Harrison}, \citenamefont
	{Gamza} \emph {et~al.}}]{Weickert2019}%
\BibitemOpen
\bibfield  {author} {\bibinfo {author} {\bibfnamefont {Franziska}\
		\bibnamefont {Weickert}}, \bibinfo {author} {\bibfnamefont {Adam~A}\
		\bibnamefont {Aczel}}, \bibinfo {author} {\bibfnamefont {Matthew~B}\
		\bibnamefont {Stone}}, \bibinfo {author} {\bibfnamefont {V~Ovidiu}\
		\bibnamefont {Garlea}}, \bibinfo {author} {\bibfnamefont {Chao}\ \bibnamefont
		{Dong}}, \bibinfo {author} {\bibfnamefont {Yoshimitsu}\ \bibnamefont
		{Kohama}}, \bibinfo {author} {\bibfnamefont {Roman}\ \bibnamefont
		{Movshovich}}, \bibinfo {author} {\bibfnamefont {Albin}\ \bibnamefont
		{Demuer}}, \bibinfo {author} {\bibfnamefont {Neil}\ \bibnamefont {Harrison}},
	\bibinfo {author} {\bibfnamefont {Monika~B}\ \bibnamefont {Gamza}},  \emph
	{et~al.},\ }\bibfield  {title} {\enquote {\bibinfo {title} {Field-induced
			double dome and {Bose-Einstein} condensation in the crossing quantum spin
			chain system {AgVOAsO$_4$}},}\ }\href@noop {} {\bibfield  {journal} {\bibinfo
		{journal} {arXiv preprint arXiv:1902.04633}\ } (\bibinfo {year}
	{2019})}\BibitemShut {NoStop}%
\bibitem [{\citenamefont {Rodríguez-Carvajal}(1993)}]{Carvajal55}%
\BibitemOpen
\bibfield  {author} {\bibinfo {author} {\bibfnamefont {Juan}\ \bibnamefont
		{Rodríguez-Carvajal}},\ }\bibfield  {title} {\enquote {\bibinfo {title}
		{Recent advances in magnetic structure determination by neutron powder
			diffraction},}\ }\href {\doibase
	https://doi.org/10.1016/0921-4526(93)90108-I} {\bibfield  {journal} {\bibinfo
		{journal} {Physica B: Condens. Matter}\ }\textbf {\bibinfo {volume} {192}},\
	\bibinfo {pages} {55} (\bibinfo {year} {1993})}\BibitemShut {NoStop}%
\bibitem [{\citenamefont {Tsirlin}\ \emph {et~al.}(2009)\citenamefont
	{Tsirlin}, \citenamefont {Schmidt}, \citenamefont {Skourski}, \citenamefont
	{Nath}, \citenamefont {Geibel},\ and\ \citenamefont
	{Rosner}}]{Tsirlin132407}%
\BibitemOpen
\bibfield  {author} {\bibinfo {author} {\bibfnamefont {A.A.}\ \bibnamefont
		{Tsirlin}}, \bibinfo {author} {\bibfnamefont {B.}~\bibnamefont {Schmidt}},
	\bibinfo {author} {\bibfnamefont {Y.}~\bibnamefont {Skourski}}, \bibinfo
	{author} {\bibfnamefont {R.}~\bibnamefont {Nath}}, \bibinfo {author}
	{\bibfnamefont {C.}~\bibnamefont {Geibel}}, \ and\ \bibinfo {author}
	{\bibfnamefont {H.}~\bibnamefont {Rosner}},\ }\bibfield  {title} {\enquote
	{\bibinfo {title} {Exploring the spin-$\frac{1}{2}$ frustrated square lattice
			model with high-field magnetization studies},}\ }\href {\doibase
	10.1103/PhysRevB.80.132407} {\bibfield  {journal} {\bibinfo  {journal} {Phys.
			Rev. B}\ }\textbf {\bibinfo {volume} {80}},\ \bibinfo {pages} {132407}
	(\bibinfo {year} {2009})}\BibitemShut {NoStop}%
\bibitem [{\citenamefont {Skourski}\ \emph {et~al.}(2011)\citenamefont
	{Skourski}, \citenamefont {Kuz'min}, \citenamefont {Skokov}, \citenamefont
	{Andreev},\ and\ \citenamefont {Wosnitza}}]{Skourski214420}%
\BibitemOpen
\bibfield  {author} {\bibinfo {author} {\bibfnamefont {Y.}~\bibnamefont
		{Skourski}}, \bibinfo {author} {\bibfnamefont {M.~D.}\ \bibnamefont
		{Kuz'min}}, \bibinfo {author} {\bibfnamefont {K.~P.}\ \bibnamefont {Skokov}},
	\bibinfo {author} {\bibfnamefont {A.~V.}\ \bibnamefont {Andreev}}, \ and\
	\bibinfo {author} {\bibfnamefont {J.}~\bibnamefont {Wosnitza}},\ }\bibfield
{title} {\enquote {\bibinfo {title} {High-field magnetization of
			{Ho$_2$Fe$_{17}$}},}\ }\href {\doibase 10.1103/PhysRevB.83.214420} {\bibfield
	{journal} {\bibinfo  {journal} {Phys. Rev. B}\ }\textbf {\bibinfo {volume}
		{83}},\ \bibinfo {pages} {214420} (\bibinfo {year} {2011})}\BibitemShut
{NoStop}%
\bibitem [{\citenamefont {Koepernik}\ and\ \citenamefont
	{Eschrig}(1999)}]{Koepernik1743}%
\BibitemOpen
\bibfield  {author} {\bibinfo {author} {\bibfnamefont {K.}~\bibnamefont
		{Koepernik}}\ and\ \bibinfo {author} {\bibfnamefont {H.}~\bibnamefont
		{Eschrig}},\ }\bibfield  {title} {\enquote {\bibinfo {title} {Full-potential
			nonorthogonal local-orbital minimum-basis band-structure scheme},}\ }\href
{\doibase 10.1103/PhysRevB.59.1743} {\bibfield  {journal} {\bibinfo
		{journal} {Phys. Rev. B}\ }\textbf {\bibinfo {volume} {59}},\ \bibinfo
	{pages} {1743} (\bibinfo {year} {1999})}\BibitemShut {NoStop}%
\bibitem [{\citenamefont {Perdew}\ \emph {et~al.}(1996)\citenamefont {Perdew},
	\citenamefont {Burke},\ and\ \citenamefont {Ernzerhof}}]{Perdew3865}%
\BibitemOpen
\bibfield  {author} {\bibinfo {author} {\bibfnamefont {J.~P.}\ \bibnamefont
		{Perdew}}, \bibinfo {author} {\bibfnamefont {K.}~\bibnamefont {Burke}}, \
	and\ \bibinfo {author} {\bibfnamefont {M.}~\bibnamefont {Ernzerhof}},\
}\bibfield  {title} {\enquote {\bibinfo {title} {Generalized gradient
		approximation made simple},}\ }\href {\doibase 10.1103/PhysRevLett.77.3865}
{\bibfield  {journal} {\bibinfo  {journal} {Phys. Rev. Lett.}\ }\textbf
	{\bibinfo {volume} {77}},\ \bibinfo {pages} {3865} (\bibinfo {year}
	{1996})}\BibitemShut {NoStop}%
\bibitem [{\citenamefont {Lii}\ \emph {et~al.}(1991)\citenamefont {Lii},
	\citenamefont {Li}, \citenamefont {Chen},\ and\ \citenamefont
	{Wang}}]{Lii67}%
\BibitemOpen
\bibfield  {author} {\bibinfo {author} {\bibfnamefont {KH}~\bibnamefont
		{Lii}}, \bibinfo {author} {\bibfnamefont {CH}~\bibnamefont {Li}}, \bibinfo
	{author} {\bibfnamefont {TM}~\bibnamefont {Chen}}, \ and\ \bibinfo {author}
	{\bibfnamefont {SL}~\bibnamefont {Wang}},\ }\bibfield  {title} {\enquote
	{\bibinfo {title} {Synthesis and structural characterization of sodium
			vanadyl (iv) orthophosphate {NaVOPO$_4$}},}\ }\href@noop {} {\bibfield
	{journal} {\bibinfo  {journal} {Z. Kristallogr. Cryst. Mater.}\ }\textbf
	{\bibinfo {volume} {197}},\ \bibinfo {pages} {67} (\bibinfo {year}
	{1991})}\BibitemShut {NoStop}%
\bibitem [{\citenamefont {Isobe}\ \emph {et~al.}(2002)\citenamefont {Isobe},
	\citenamefont {Ninomiya}, \citenamefont {Vasil'ev},\ and\ \citenamefont
	{Ueda}}]{Isobe1423}%
\BibitemOpen
\bibfield  {author} {\bibinfo {author} {\bibfnamefont {Masahiko}\
		\bibnamefont {Isobe}}, \bibinfo {author} {\bibfnamefont {Emi}\ \bibnamefont
		{Ninomiya}}, \bibinfo {author} {\bibfnamefont {Alexander~N}\ \bibnamefont
		{Vasil'ev}}, \ and\ \bibinfo {author} {\bibfnamefont {Yutaka}\ \bibnamefont
		{Ueda}},\ }\bibfield  {title} {\enquote {\bibinfo {title} {Novel phase
			transition in spin-1/2 linear chain systems: {NaTiSi$_2$O$_6$} and
			{LiTiSi$_2$O$_6$}},}\ }\href@noop {} {\bibfield  {journal} {\bibinfo
		{journal} {J. Phys. Soc. Jpn.}\ }\textbf {\bibinfo {volume} {71}},\ \bibinfo
	{pages} {1423} (\bibinfo {year} {2002})}\BibitemShut {NoStop}%
\bibitem [{\citenamefont {Hirota}\ \emph {et~al.}(1994)\citenamefont {Hirota},
	\citenamefont {Cox}, \citenamefont {Lorenzo}, \citenamefont {Shirane},
	\citenamefont {Tranquada}, \citenamefont {Hase}, \citenamefont {Uchinokura},
	\citenamefont {Kojima}, \citenamefont {Shibuya},\ and\ \citenamefont
	{Tanaka}}]{Hirota736}%
\BibitemOpen
\bibfield  {author} {\bibinfo {author} {\bibfnamefont {K.}~\bibnamefont
		{Hirota}}, \bibinfo {author} {\bibfnamefont {D.~E.}\ \bibnamefont {Cox}},
	\bibinfo {author} {\bibfnamefont {J.~E.}\ \bibnamefont {Lorenzo}}, \bibinfo
	{author} {\bibfnamefont {G.}~\bibnamefont {Shirane}}, \bibinfo {author}
	{\bibfnamefont {J.~M.}\ \bibnamefont {Tranquada}}, \bibinfo {author}
	{\bibfnamefont {M.}~\bibnamefont {Hase}}, \bibinfo {author} {\bibfnamefont
		{K.}~\bibnamefont {Uchinokura}}, \bibinfo {author} {\bibfnamefont
		{H.}~\bibnamefont {Kojima}}, \bibinfo {author} {\bibfnamefont
		{Y.}~\bibnamefont {Shibuya}}, \ and\ \bibinfo {author} {\bibfnamefont
		{I.}~\bibnamefont {Tanaka}},\ }\bibfield  {title} {\enquote {\bibinfo {title}
		{Dimerization of {CuGeO$_3$} in the spin-{P}eierls state},}\ }\href {\doibase
	10.1103/PhysRevLett.73.736} {\bibfield  {journal} {\bibinfo  {journal} {Phys.
			Rev. Lett.}\ }\textbf {\bibinfo {volume} {73}},\ \bibinfo {pages} {736}
	(\bibinfo {year} {1994})}\BibitemShut {NoStop}%
\bibitem [{\citenamefont {Isobe}\ and\ \citenamefont {Ueda}(1996)}]{Isobe1178}%
\BibitemOpen
\bibfield  {author} {\bibinfo {author} {\bibfnamefont {Masahiko}\
		\bibnamefont {Isobe}}\ and\ \bibinfo {author} {\bibfnamefont {Yutaka}\
		\bibnamefont {Ueda}},\ }\bibfield  {title} {\enquote {\bibinfo {title}
		{Magnetic susceptibility of quasi-one-dimensional compound
			{$\alpha$}-{NaV$_2$O$_5$} –possible spin-peierls compound with high
			critical temperature of 34 {K}–},}\ }\href {\doibase 10.1143/JPSJ.65.1178}
{\bibfield  {journal} {\bibinfo  {journal} {J. Phys. Soc. Jpn.}\ }\textbf
	{\bibinfo {volume} {65}},\ \bibinfo {pages} {1178} (\bibinfo {year}
	{1996})}\BibitemShut {NoStop}%
\bibitem [{\citenamefont {L\'epine}\ \emph {et~al.}(1978)\citenamefont
	{L\'epine}, \citenamefont {Caill\'e},\ and\ \citenamefont
	{Larochelle}}]{Lepine3585}%
\BibitemOpen
\bibfield  {author} {\bibinfo {author} {\bibfnamefont {Y.}~\bibnamefont
		{L\'epine}}, \bibinfo {author} {\bibfnamefont {A.}~\bibnamefont {Caill\'e}},
	\ and\ \bibinfo {author} {\bibfnamefont {V.}~\bibnamefont {Larochelle}},\
}\bibfield  {title} {\enquote {\bibinfo {title}
	{Potassium-tetracyanoquinodimethane {(K-TCNQ)}: A spin-peierls system},}\
}\href {\doibase 10.1103/PhysRevB.18.3585} {\bibfield  {journal} {\bibinfo
	{journal} {Phys. Rev. B}\ }\textbf {\bibinfo {volume} {18}},\ \bibinfo
{pages} {3585} (\bibinfo {year} {1978})}\BibitemShut {NoStop}%
\bibitem [{\citenamefont {Islam}\ \emph {et~al.}(2018)\citenamefont {Islam},
	\citenamefont {Ranjith}, \citenamefont {Baenitz}, \citenamefont {Skourski},
	\citenamefont {Tsirlin},\ and\ \citenamefont {Nath}}]{Islam174432}%
\BibitemOpen
\bibfield  {author} {\bibinfo {author} {\bibfnamefont {S.~S.}\ \bibnamefont
		{Islam}}, \bibinfo {author} {\bibfnamefont {K.~M.}\ \bibnamefont {Ranjith}},
	\bibinfo {author} {\bibfnamefont {M.}~\bibnamefont {Baenitz}}, \bibinfo
	{author} {\bibfnamefont {Y.}~\bibnamefont {Skourski}}, \bibinfo {author}
	{\bibfnamefont {A.~A.}\ \bibnamefont {Tsirlin}}, \ and\ \bibinfo {author}
	{\bibfnamefont {R.}~\bibnamefont {Nath}},\ }\bibfield  {title} {\enquote
	{\bibinfo {title} {Frustration of square cupola in
			{Sr(TiO)Cu$_4$(PO$_4$)$_4$}},}\ }\href {\doibase 10.1103/PhysRevB.97.174432}
{\bibfield  {journal} {\bibinfo  {journal} {Phys. Rev. B}\ }\textbf {\bibinfo
		{volume} {97}},\ \bibinfo {pages} {174432} (\bibinfo {year}
	{2018})}\BibitemShut {NoStop}%
\bibitem [{\citenamefont {Bag}\ \emph {et~al.}(2018)\citenamefont {Bag},
	\citenamefont {Baral},\ and\ \citenamefont {Nath}}]{Bag144436}%
\BibitemOpen
\bibfield  {author} {\bibinfo {author} {\bibfnamefont {Pallab}\ \bibnamefont
		{Bag}}, \bibinfo {author} {\bibfnamefont {P.~R.}\ \bibnamefont {Baral}}, \
	and\ \bibinfo {author} {\bibfnamefont {R.}~\bibnamefont {Nath}},\ }\bibfield
{title} {\enquote {\bibinfo {title} {Cluster spin-glass behavior and memory
			effect in {Cr$_{0.5}$Fe$_{0.5}$Ga}},}\ }\href {\doibase
	10.1103/PhysRevB.98.144436} {\bibfield  {journal} {\bibinfo  {journal} {Phys.
			Rev. B}\ }\textbf {\bibinfo {volume} {98}},\ \bibinfo {pages} {144436}
	(\bibinfo {year} {2018})}\BibitemShut {NoStop}%
\bibitem [{\citenamefont {Wollny}\ \emph {et~al.}(2011)\citenamefont {Wollny},
	\citenamefont {Fritz},\ and\ \citenamefont {Vojta}}]{Wolly137204}%
\BibitemOpen
\bibfield  {author} {\bibinfo {author} {\bibfnamefont {Alexander}\
		\bibnamefont {Wollny}}, \bibinfo {author} {\bibfnamefont {Lars}\ \bibnamefont
		{Fritz}}, \ and\ \bibinfo {author} {\bibfnamefont {Matthias}\ \bibnamefont
		{Vojta}},\ }\bibfield  {title} {\enquote {\bibinfo {title} {Fractional
			impurity moments in two-dimensional noncollinear magnets},}\ }\href {\doibase
	10.1103/PhysRevLett.107.137204} {\bibfield  {journal} {\bibinfo  {journal}
		{Phys. Rev. Lett.}\ }\textbf {\bibinfo {volume} {107}},\ \bibinfo {pages}
	{137204} (\bibinfo {year} {2011})}\BibitemShut {NoStop}%
\bibitem [{\citenamefont {Selwood}(2013)}]{Selwood2013}%
\BibitemOpen
\bibfield  {author} {\bibinfo {author} {\bibfnamefont {Pierce~W}\
		\bibnamefont {Selwood}},\ }\href@noop {} {\emph {\bibinfo {title}
		{Magnetochemistry}}}\ (\bibinfo  {publisher} {Read Books Ltd},\ \bibinfo
{year} {2013})\BibitemShut {NoStop}%
\bibitem [{\citenamefont {Johnston}\ \emph {et~al.}(2000)\citenamefont
	{Johnston}, \citenamefont {Kremer}, \citenamefont {Troyer}, \citenamefont
	{Wang}, \citenamefont {Kl{\"u}mper}, \citenamefont {Budko}, \citenamefont
	{Panchula},\ and\ \citenamefont {Canfield}}]{Johnston9558}%
\BibitemOpen
\bibfield  {author} {\bibinfo {author} {\bibfnamefont {D.~C.}\ \bibnamefont
		{Johnston}}, \bibinfo {author} {\bibfnamefont {R.~K.}\ \bibnamefont
		{Kremer}}, \bibinfo {author} {\bibfnamefont {M.}~\bibnamefont {Troyer}},
	\bibinfo {author} {\bibfnamefont {X.}~\bibnamefont {Wang}}, \bibinfo {author}
	{\bibfnamefont {A.}~\bibnamefont {Kl{\"u}mper}}, \bibinfo {author}
	{\bibfnamefont {S.~L.}\ \bibnamefont {Budko}}, \bibinfo {author}
	{\bibfnamefont {A.~F.}\ \bibnamefont {Panchula}}, \ and\ \bibinfo {author}
	{\bibfnamefont {P.~C.}\ \bibnamefont {Canfield}},\ }\bibfield  {title}
{\enquote {\bibinfo {title} {Thermodynamics of spin {$S= 1/2$}
			antiferromagnetic uniform and alternating-exchange heisenberg chains},}\
}\href {\doibase 10.1103/PhysRevB.61.9558} {\bibfield  {journal} {\bibinfo
	{journal} {Phys. Rev. B}\ }\textbf {\bibinfo {volume} {61}},\ \bibinfo
{pages} {9558} (\bibinfo {year} {2000})}\BibitemShut {NoStop}%
\bibitem [{\citenamefont {Lebernegg}\ \emph {et~al.}(2011)\citenamefont
	{Lebernegg}, \citenamefont {Tsirlin}, \citenamefont {Janson}, \citenamefont
	{Nath}, \citenamefont {Sichelschmidt}, \citenamefont {Skourski},
	\citenamefont {Amthauer},\ and\ \citenamefont {Rosner}}]{Lebernegg174436}%
\BibitemOpen
\bibfield  {author} {\bibinfo {author} {\bibfnamefont {S.}~\bibnamefont
		{Lebernegg}}, \bibinfo {author} {\bibfnamefont {A.~A.}\ \bibnamefont
		{Tsirlin}}, \bibinfo {author} {\bibfnamefont {O.}~\bibnamefont {Janson}},
	\bibinfo {author} {\bibfnamefont {R.}~\bibnamefont {Nath}}, \bibinfo {author}
	{\bibfnamefont {J.}~\bibnamefont {Sichelschmidt}}, \bibinfo {author}
	{\bibfnamefont {Yu.}\ \bibnamefont {Skourski}}, \bibinfo {author}
	{\bibfnamefont {G.}~\bibnamefont {Amthauer}}, \ and\ \bibinfo {author}
	{\bibfnamefont {H.}~\bibnamefont {Rosner}},\ }\bibfield  {title} {\enquote
	{\bibinfo {title} {Magnetic model for {A$_{2}$CuP$_{2}$O$_{7}$}
			({$A$=Na,Li}): One-dimensional versus two-dimensional behavior},}\ }\href
{\doibase 10.1103/PhysRevB.84.174436} {\bibfield  {journal} {\bibinfo
		{journal} {Phys. Rev. B}\ }\textbf {\bibinfo {volume} {84}},\ \bibinfo
	{pages} {174436} (\bibinfo {year} {2011})}\BibitemShut {NoStop}%
\bibitem [{\citenamefont {Nath}\ \emph {et~al.}(2009)\citenamefont {Nath},
	\citenamefont {Singh},\ and\ \citenamefont {Johnston}}]{Nath174513}%
\BibitemOpen
\bibfield  {author} {\bibinfo {author} {\bibfnamefont {R.}~\bibnamefont
		{Nath}}, \bibinfo {author} {\bibfnamefont {Yogesh}\ \bibnamefont {Singh}}, \
	and\ \bibinfo {author} {\bibfnamefont {D.~C.}\ \bibnamefont {Johnston}},\
}\bibfield  {title} {\enquote {\bibinfo {title} {Magnetic, thermal, and
		transport properties of layered arsenides {BaRu$_2$As$_2$} and
		{SrRu$_2$As$_2$}},}\ }\href {\doibase 10.1103/PhysRevB.79.174513} {\bibfield
{journal} {\bibinfo  {journal} {Phys. Rev. B}\ }\textbf {\bibinfo {volume}
	{79}},\ \bibinfo {pages} {174513} (\bibinfo {year} {2009})}\BibitemShut
{NoStop}%
\bibitem [{\citenamefont {Kittel}(1986)}]{Kittel}%
\BibitemOpen
\bibfield  {author} {\bibinfo {author} {\bibfnamefont {Charles}\ \bibnamefont
		{Kittel}},\ }\href@noop {} {\emph {\bibinfo {title} {{Introduction to Solid
				State Physics}}}},\ \bibinfo {edition} {6th}\ ed.\ (\bibinfo  {publisher}
{John Wiley \& Sons, Inc.},\ \bibinfo {address} {New York},\ \bibinfo {year}
{1986})\BibitemShut {NoStop}%
\bibitem [{\citenamefont {Arjun}\ \emph {et~al.}(2017)\citenamefont {Arjun},
	\citenamefont {Kumar}, \citenamefont {Anjana}, \citenamefont {Thirumurugan},
	\citenamefont {Sichelschmidt}, \citenamefont {Mahajan},\ and\ \citenamefont
	{Nath}}]{Arjun174421}%
\BibitemOpen
\bibfield  {author} {\bibinfo {author} {\bibfnamefont {U.}~\bibnamefont
		{Arjun}}, \bibinfo {author} {\bibfnamefont {Vinod}\ \bibnamefont {Kumar}},
	\bibinfo {author} {\bibfnamefont {P.~K.}\ \bibnamefont {Anjana}}, \bibinfo
	{author} {\bibfnamefont {A.}~\bibnamefont {Thirumurugan}}, \bibinfo {author}
	{\bibfnamefont {J.}~\bibnamefont {Sichelschmidt}}, \bibinfo {author}
	{\bibfnamefont {A.~V.}\ \bibnamefont {Mahajan}}, \ and\ \bibinfo {author}
	{\bibfnamefont {R.}~\bibnamefont {Nath}},\ }\bibfield  {title} {\enquote
	{\bibinfo {title} {Singlet ground state in the spin-$\frac12$ weakly coupled
			dimer compound
			{NH$_4$[(V$_2$O$_3$)$_2$(4,4$^\prime$-bpy)$_2$(H$_2$PO$_4$)(PO$_4$)$_2$]$\cdot$
				0.5H$_2$O}},}\ }\href {\doibase 10.1103/PhysRevB.95.174421} {\bibfield
	{journal} {\bibinfo  {journal} {Phys. Rev. B}\ }\textbf {\bibinfo {volume}
		{95}},\ \bibinfo {pages} {174421} (\bibinfo {year} {2017})}\BibitemShut
{NoStop}%
\bibitem [{\citenamefont {Ahmed}\ \emph {et~al.}(2015)\citenamefont {Ahmed},
	\citenamefont {Tsirlin},\ and\ \citenamefont {Nath}}]{Niyaz214413}%
\BibitemOpen
\bibfield  {author} {\bibinfo {author} {\bibfnamefont {N.}~\bibnamefont
		{Ahmed}}, \bibinfo {author} {\bibfnamefont {A.~A.}\ \bibnamefont {Tsirlin}},
	\ and\ \bibinfo {author} {\bibfnamefont {R.}~\bibnamefont {Nath}},\
}\bibfield  {title} {\enquote {\bibinfo {title} {Multiple magnetic
		transitions in the spin-$\frac{1}{2}$ chain antiferromagnet
		{SrCuTe$_2$O$_6$}},}\ }\href {\doibase 10.1103/PhysRevB.91.214413} {\bibfield
{journal} {\bibinfo  {journal} {Phys. Rev. B}\ }\textbf {\bibinfo {volume}
	{91}},\ \bibinfo {pages} {214413} (\bibinfo {year} {2015})}\BibitemShut
{NoStop}%
\bibitem [{\citenamefont {Bernu}\ and\ \citenamefont
	{Misguich}(2001)}]{Bernu134409}%
\BibitemOpen
\bibfield  {author} {\bibinfo {author} {\bibfnamefont {B.}~\bibnamefont
		{Bernu}}\ and\ \bibinfo {author} {\bibfnamefont {G.}~\bibnamefont
		{Misguich}},\ }\bibfield  {title} {\enquote {\bibinfo {title} {Specific heat
			and high-temperature series of lattice models: Interpolation scheme and
			examples on quantum spin systems in one and two dimensions},}\ }\href
{\doibase 10.1103/PhysRevB.63.134409} {\bibfield  {journal} {\bibinfo
		{journal} {Phys. Rev. B}\ }\textbf {\bibinfo {volume} {63}},\ \bibinfo
	{pages} {134409} (\bibinfo {year} {2001})}\BibitemShut {NoStop}%
\bibitem [{\citenamefont {Hase}\ \emph {et~al.}(1993)\citenamefont {Hase},
	\citenamefont {Terasaki},\ and\ \citenamefont {Uchinokura}}]{Hase3651}%
\BibitemOpen
\bibfield  {author} {\bibinfo {author} {\bibfnamefont {Masashi}\ \bibnamefont
		{Hase}}, \bibinfo {author} {\bibfnamefont {Ichiro}\ \bibnamefont {Terasaki}},
	\ and\ \bibinfo {author} {\bibfnamefont {Kunimitsu}\ \bibnamefont
		{Uchinokura}},\ }\bibfield  {title} {\enquote {\bibinfo {title} {Observation
			of the {spin-Peierls} transition in linear {Cu$^{2+}$} (spin-1/2) chains in
			an inorganic compound {CuGeO$_3$}},}\ }\href {\doibase
	10.1103/PhysRevLett.70.3651} {\bibfield  {journal} {\bibinfo  {journal}
		{Phys. Rev. Lett.}\ }\textbf {\bibinfo {volume} {70}},\ \bibinfo {pages}
	{3651} (\bibinfo {year} {1993})}\BibitemShut {NoStop}%
\bibitem [{\citenamefont {Fagot-Revurat}\ \emph {et~al.}(2000)\citenamefont
	{Fagot-Revurat}, \citenamefont {Mehring},\ and\ \citenamefont
	{Kremer}}]{Fagot4176}%
\BibitemOpen
\bibfield  {author} {\bibinfo {author} {\bibfnamefont {Y.}~\bibnamefont
		{Fagot-Revurat}}, \bibinfo {author} {\bibfnamefont {M.}~\bibnamefont
		{Mehring}}, \ and\ \bibinfo {author} {\bibfnamefont {R.~K.}\ \bibnamefont
		{Kremer}},\ }\bibfield  {title} {\enquote {\bibinfo {title}
		{Charge-order-driven spin-peierls transition in
			{$\alpha$}-{Na$_x$V$_2$O$_5$}},}\ }\href {\doibase
	10.1103/PhysRevLett.84.4176} {\bibfield  {journal} {\bibinfo  {journal}
		{Phys. Rev. Lett.}\ }\textbf {\bibinfo {volume} {84}},\ \bibinfo {pages}
	{4176} (\bibinfo {year} {2000})}\BibitemShut {NoStop}%
\bibitem [{\citenamefont {Vyaselev}\ \emph {et~al.}(2004)\citenamefont
	{Vyaselev}, \citenamefont {Takigawa}, \citenamefont {Vasiliev}, \citenamefont
	{Oosawa},\ and\ \citenamefont {Tanaka}}]{Vyaselev207202}%
\BibitemOpen
\bibfield  {author} {\bibinfo {author} {\bibfnamefont {O.}~\bibnamefont
		{Vyaselev}}, \bibinfo {author} {\bibfnamefont {M.}~\bibnamefont {Takigawa}},
	\bibinfo {author} {\bibfnamefont {A.}~\bibnamefont {Vasiliev}}, \bibinfo
	{author} {\bibfnamefont {A.}~\bibnamefont {Oosawa}}, \ and\ \bibinfo {author}
	{\bibfnamefont {H.}~\bibnamefont {Tanaka}},\ }\bibfield  {title} {\enquote
	{\bibinfo {title} {Field-induced magnetic order and simultaneous lattice
			deformation in {TlCuCl$_3$}},}\ }\href {\doibase
	10.1103/PhysRevLett.92.207202} {\bibfield  {journal} {\bibinfo  {journal}
		{Phys. Rev. Lett.}\ }\textbf {\bibinfo {volume} {92}},\ \bibinfo {pages}
	{207202} (\bibinfo {year} {2004})}\BibitemShut {NoStop}%
\bibitem [{\citenamefont {Moriya}(1956)}]{Moriya23}%
\BibitemOpen
\bibfield  {author} {\bibinfo {author} {\bibfnamefont {T.}~\bibnamefont
		{Moriya}},\ }\bibfield  {title} {\enquote {\bibinfo {title} {{Nuclear
				Magnetic Relaxation in Antiferromagnetics}},}\ }\href {\doibase
	10.1143/PTP.16.23} {\bibfield  {journal} {\bibinfo  {journal} {Prog. Theor.
			Phys.}\ }\textbf {\bibinfo {volume} {16}},\ \bibinfo {pages} {23} (\bibinfo
	{year} {1956})}\BibitemShut {NoStop}%
\bibitem [{\citenamefont {Sachdev}(1994)}]{Sachdev13006}%
\BibitemOpen
\bibfield  {author} {\bibinfo {author} {\bibfnamefont {Subir}\ \bibnamefont
		{Sachdev}},\ }\bibfield  {title} {\enquote {\bibinfo {title} {{NMR}
			relaxation in half-integer antiferromagnetic spin chains},}\ }\href {\doibase
	10.1103/PhysRevB.50.13006} {\bibfield  {journal} {\bibinfo  {journal} {Phys.
			Rev. B}\ }\textbf {\bibinfo {volume} {50}},\ \bibinfo {pages} {13006}
	(\bibinfo {year} {1994})}\BibitemShut {NoStop}%
\bibitem [{\citenamefont {Nath}\ \emph {et~al.}(2005)\citenamefont {Nath},
	\citenamefont {Mahajan}, \citenamefont {B\"uttgen}, \citenamefont {Kegler},
	\citenamefont {Loidl},\ and\ \citenamefont {Bobroff}}]{Nath174436}%
\BibitemOpen
\bibfield  {author} {\bibinfo {author} {\bibfnamefont {R.}~\bibnamefont
		{Nath}}, \bibinfo {author} {\bibfnamefont {A.~V.}\ \bibnamefont {Mahajan}},
	\bibinfo {author} {\bibfnamefont {N.}~\bibnamefont {B\"uttgen}}, \bibinfo
	{author} {\bibfnamefont {C.}~\bibnamefont {Kegler}}, \bibinfo {author}
	{\bibfnamefont {A.}~\bibnamefont {Loidl}}, \ and\ \bibinfo {author}
	{\bibfnamefont {J.}~\bibnamefont {Bobroff}},\ }\bibfield  {title} {\enquote
	{\bibinfo {title} {Study of one-dimensional nature of {$S=\frac{1}{2}$}
			{(Sr,Ba)$_2$Cu(PO$_4$)$_2$} and {BaCuP$_2$O$_7$} via {$^{31}${P}} {NMR}},}\
}\href {\doibase 10.1103/PhysRevB.71.174436} {\bibfield  {journal} {\bibinfo
	{journal} {Phys. Rev. B}\ }\textbf {\bibinfo {volume} {71}},\ \bibinfo
{pages} {174436} (\bibinfo {year} {2005})}\BibitemShut {NoStop}%
\bibitem [{\citenamefont {Moriya}(1963)}]{Moriya516}%
\BibitemOpen
\bibfield  {author} {\bibinfo {author} {\bibfnamefont {T.}~\bibnamefont
		{Moriya}},\ }\bibfield  {title} {\enquote {\bibinfo {title} {The effect of
			electron-electron interaction on the nuclear spin relaxation in metals},}\
}\href {\doibase 10.1143/JPSJ.18.516} {\bibfield  {journal} {\bibinfo
	{journal} {J. Phys. Soc. Jpn.}\ }\textbf {\bibinfo {volume} {18}},\ \bibinfo
{pages} {516} (\bibinfo {year} {1963})}\BibitemShut {NoStop}%
\bibitem [{\citenamefont {Nath}\ \emph {et~al.}(2008)\citenamefont {Nath},
	\citenamefont {Tsirlin}, \citenamefont {Kaul}, \citenamefont {Baenitz},
	\citenamefont {B\"uttgen}, \citenamefont {Geibel},\ and\ \citenamefont
	{Rosner}}]{Nath024418}%
\BibitemOpen
\bibfield  {author} {\bibinfo {author} {\bibfnamefont {R.}~\bibnamefont
		{Nath}}, \bibinfo {author} {\bibfnamefont {A.A.}\ \bibnamefont {Tsirlin}},
	\bibinfo {author} {\bibfnamefont {E.E.}\ \bibnamefont {Kaul}}, \bibinfo
	{author} {\bibfnamefont {M.}~\bibnamefont {Baenitz}}, \bibinfo {author}
	{\bibfnamefont {N.}~\bibnamefont {B\"uttgen}}, \bibinfo {author}
	{\bibfnamefont {C.}~\bibnamefont {Geibel}}, \ and\ \bibinfo {author}
	{\bibfnamefont {H.}~\bibnamefont {Rosner}},\ }\bibfield  {title} {\enquote
	{\bibinfo {title} {Strong frustration due to competing ferromagnetic and
			antiferromagnetic interactions: Magnetic properties of {M(VO)$_2$(PO$_4)_2$
				(M = Ca and Sr)}},}\ }\href {\doibase 10.1103/PhysRevB.78.024418} {\bibfield
	{journal} {\bibinfo  {journal} {Phys. Rev. B}\ }\textbf {\bibinfo {volume}
		{78}},\ \bibinfo {pages} {024418} (\bibinfo {year} {2008})}\BibitemShut
{NoStop}%
\bibitem [{\citenamefont {Tsirlin}\ \emph {et~al.}(2008)\citenamefont
	{Tsirlin}, \citenamefont {Nath}, \citenamefont {Geibel},\ and\ \citenamefont
	{Rosner}}]{Tsirlin104436}%
\BibitemOpen
\bibfield  {author} {\bibinfo {author} {\bibfnamefont {A.A.}\ \bibnamefont
		{Tsirlin}}, \bibinfo {author} {\bibfnamefont {R.}~\bibnamefont {Nath}},
	\bibinfo {author} {\bibfnamefont {C.}~\bibnamefont {Geibel}}, \ and\ \bibinfo
	{author} {\bibfnamefont {H.}~\bibnamefont {Rosner}},\ }\bibfield  {title}
{\enquote {\bibinfo {title} {Magnetic properties of {Ag$_2$VOP$_2$O$_7$}: An
			unexpected spin dimer system},}\ }\href {\doibase 10.1103/PhysRevB.77.104436}
{\bibfield  {journal} {\bibinfo  {journal} {Phys. Rev. B}\ }\textbf {\bibinfo
		{volume} {77}},\ \bibinfo {pages} {104436} (\bibinfo {year}
	{2008})}\BibitemShut {NoStop}%
\bibitem [{\citenamefont {Xiang}\ \emph {et~al.}(2011)\citenamefont {Xiang},
	\citenamefont {Kan}, \citenamefont {Wei}, \citenamefont {Whangbo},\ and\
	\citenamefont {Gong}}]{Xiang224429}%
\BibitemOpen
\bibfield  {author} {\bibinfo {author} {\bibfnamefont {H.~J.}\ \bibnamefont
		{Xiang}}, \bibinfo {author} {\bibfnamefont {E.~J.}\ \bibnamefont {Kan}},
	\bibinfo {author} {\bibfnamefont {S.-H.}\ \bibnamefont {Wei}}, \bibinfo
	{author} {\bibfnamefont {M.-H.}\ \bibnamefont {Whangbo}}, \ and\ \bibinfo
	{author} {\bibfnamefont {X.~G.}\ \bibnamefont {Gong}},\ }\bibfield  {title}
{\enquote {\bibinfo {title} {Predicting the spin-lattice order of frustrated
			systems from first principles},}\ }\href {\doibase
	10.1103/PhysRevB.84.224429} {\bibfield  {journal} {\bibinfo  {journal} {Phys.
			Rev. B}\ }\textbf {\bibinfo {volume} {84}},\ \bibinfo {pages} {224429}
	(\bibinfo {year} {2011})}\BibitemShut {NoStop}%
\bibitem [{\citenamefont {Tsirlin}(2014)}]{Tsirlin014405}%
\BibitemOpen
\bibfield  {author} {\bibinfo {author} {\bibfnamefont {A.A.}\ \bibnamefont
		{Tsirlin}},\ }\bibfield  {title} {\enquote {\bibinfo {title} {Spin-chain
			magnetism and uniform {Dzyaloshinsky-Moriya} anisotropy in {BaV$_3$O$_8$}},}\
}\href {\doibase 10.1103/PhysRevB.89.014405} {\bibfield  {journal} {\bibinfo
	{journal} {Phys. Rev. B}\ }\textbf {\bibinfo {volume} {89}},\ \bibinfo
{pages} {014405} (\bibinfo {year} {2014})}\BibitemShut {NoStop}%
\bibitem [{\citenamefont {Roca}\ \emph {et~al.}(1998)\citenamefont {Roca},
	\citenamefont {Amor\'os}, \citenamefont {Cano}, \citenamefont {{Dolores
			Marcos}}, \citenamefont {Alamo}, \citenamefont {Beltr\'an-Porter},\ and\
	\citenamefont {Beltr\'an-Porter}}]{Roca3167}%
\BibitemOpen
\bibfield  {author} {\bibinfo {author} {\bibfnamefont {M.}~\bibnamefont
		{Roca}}, \bibinfo {author} {\bibfnamefont {P.}~\bibnamefont {Amor\'os}},
	\bibinfo {author} {\bibfnamefont {J.}~\bibnamefont {Cano}}, \bibinfo {author}
	{\bibfnamefont {M.}~\bibnamefont {{Dolores Marcos}}}, \bibinfo {author}
	{\bibfnamefont {J.}~\bibnamefont {Alamo}}, \bibinfo {author} {\bibfnamefont
		{A.}~\bibnamefont {Beltr\'an-Porter}}, \ and\ \bibinfo {author}
	{\bibfnamefont {D.}~\bibnamefont {Beltr\'an-Porter}},\ }\bibfield  {title}
{\enquote {\bibinfo {title} {Prediction of magnetic properties in
			{oxovanadium(IV)} phosphates: The role of the bridging {PO$_4$} anions},}\
}\href {\doibase 10.1021/ic971210o} {\bibfield  {journal} {\bibinfo
	{journal} {Inorg. Chem.}\ }\textbf {\bibinfo {volume} {37}},\ \bibinfo
{pages} {3167} (\bibinfo {year} {1998})}\BibitemShut {NoStop}%
\bibitem [{\citenamefont {Vasil'ev}\ \emph {et~al.}(2001)\citenamefont
	{Vasil'ev}, \citenamefont {Ponomarenko}, \citenamefont {Manaka},
	\citenamefont {Yamada}, \citenamefont {Isobe},\ and\ \citenamefont
	{Ueda}}]{Vasil'ev024419}%
\BibitemOpen
\bibfield  {author} {\bibinfo {author} {\bibfnamefont {A.~N.}\ \bibnamefont
		{Vasil'ev}}, \bibinfo {author} {\bibfnamefont {L.~A.}\ \bibnamefont
		{Ponomarenko}}, \bibinfo {author} {\bibfnamefont {H.}~\bibnamefont {Manaka}},
	\bibinfo {author} {\bibfnamefont {I.}~\bibnamefont {Yamada}}, \bibinfo
	{author} {\bibfnamefont {M.}~\bibnamefont {Isobe}}, \ and\ \bibinfo {author}
	{\bibfnamefont {Y.}~\bibnamefont {Ueda}},\ }\bibfield  {title} {\enquote
	{\bibinfo {title} {Magnetic and resonant properties of quasi-one-dimensional
			antiferromagnet {LiCuVO$_4$}},}\ }\href {\doibase 10.1103/PhysRevB.64.024419}
{\bibfield  {journal} {\bibinfo  {journal} {Phys. Rev. B}\ }\textbf {\bibinfo
		{volume} {64}},\ \bibinfo {pages} {024419} (\bibinfo {year}
	{2001})}\BibitemShut {NoStop}%
\bibitem [{\citenamefont {Kegler}\ \emph {et~al.}(2006)\citenamefont {Kegler},
	\citenamefont {B\"uttgen}, \citenamefont {Krug~von Nidda}, \citenamefont
	{Loidl}, \citenamefont {Nath}, \citenamefont {Mahajan}, \citenamefont
	{Prokofiev},\ and\ \citenamefont {A\ss{}mus}}]{Kegler104418}%
\BibitemOpen
\bibfield  {author} {\bibinfo {author} {\bibfnamefont {C.}~\bibnamefont
		{Kegler}}, \bibinfo {author} {\bibfnamefont {N.}~\bibnamefont {B\"uttgen}},
	\bibinfo {author} {\bibfnamefont {H.-A.}\ \bibnamefont {Krug~von Nidda}},
	\bibinfo {author} {\bibfnamefont {A.}~\bibnamefont {Loidl}}, \bibinfo
	{author} {\bibfnamefont {R.}~\bibnamefont {Nath}}, \bibinfo {author}
	{\bibfnamefont {A.~V.}\ \bibnamefont {Mahajan}}, \bibinfo {author}
	{\bibfnamefont {A.~V.}\ \bibnamefont {Prokofiev}}, \ and\ \bibinfo {author}
	{\bibfnamefont {W.}~\bibnamefont {A\ss{}mus}},\ }\bibfield  {title} {\enquote
	{\bibinfo {title} {{NMR} study of lineshifts and relaxation rates of the
			one-dimensional antiferromagnet {LiCuVO$_4$}},}\ }\href {\doibase
	10.1103/PhysRevB.73.104418} {\bibfield  {journal} {\bibinfo  {journal} {Phys.
			Rev. B}\ }\textbf {\bibinfo {volume} {73}},\ \bibinfo {pages} {104418}
	(\bibinfo {year} {2006})}\BibitemShut {NoStop}%
\bibitem [{\citenamefont {Zapf}\ \emph {et~al.}(2014)\citenamefont {Zapf},
	\citenamefont {Jaime},\ and\ \citenamefont {Batista}}]{Zapf563}%
\BibitemOpen
\bibfield  {author} {\bibinfo {author} {\bibfnamefont {Vivien}\ \bibnamefont
		{Zapf}}, \bibinfo {author} {\bibfnamefont {Marcelo}\ \bibnamefont {Jaime}}, \
	and\ \bibinfo {author} {\bibfnamefont {C.~D.}\ \bibnamefont {Batista}},\
}\bibfield  {title} {\enquote {\bibinfo {title} {{$\rm Bose-Einstein$}
		condensation in quantum magnets},}\ }\href {\doibase
10.1103/RevModPhys.86.563} {\bibfield  {journal} {\bibinfo  {journal} {Rev.
		Mod. Phys.}\ }\textbf {\bibinfo {volume} {86}},\ \bibinfo {pages} {563}
(\bibinfo {year} {2014})}\BibitemShut {NoStop}%
\bibitem [{\citenamefont {Nohadani}\ \emph {et~al.}(2004)\citenamefont
	{Nohadani}, \citenamefont {Wessel}, \citenamefont {Normand},\ and\
	\citenamefont {Haas}}]{Nohadani220402}%
\BibitemOpen
\bibfield  {author} {\bibinfo {author} {\bibfnamefont {O.}~\bibnamefont
		{Nohadani}}, \bibinfo {author} {\bibfnamefont {S.}~\bibnamefont {Wessel}},
	\bibinfo {author} {\bibfnamefont {B.}~\bibnamefont {Normand}}, \ and\
	\bibinfo {author} {\bibfnamefont {S.}~\bibnamefont {Haas}},\ }\bibfield
{title} {\enquote {\bibinfo {title} {Universal scaling at field-induced
			magnetic phase transitions},}\ }\href {\doibase 10.1103/PhysRevB.69.220402}
{\bibfield  {journal} {\bibinfo  {journal} {Phys. Rev. B}\ }\textbf {\bibinfo
		{volume} {69}},\ \bibinfo {pages} {220402} (\bibinfo {year}
	{2004})}\BibitemShut {NoStop}%
\bibitem [{\citenamefont {Giamarchi}\ and\ \citenamefont
	{Tsvelik}(1999)}]{Giamarchi11398}%
\BibitemOpen
\bibfield  {author} {\bibinfo {author} {\bibfnamefont {T.}~\bibnamefont
		{Giamarchi}}\ and\ \bibinfo {author} {\bibfnamefont {A.~M.}\ \bibnamefont
		{Tsvelik}},\ }\bibfield  {title} {\enquote {\bibinfo {title} {Coupled ladders
			in a magnetic field},}\ }\href {\doibase 10.1103/PhysRevB.59.11398}
{\bibfield  {journal} {\bibinfo  {journal} {Phys. Rev. B}\ }\textbf {\bibinfo
		{volume} {59}},\ \bibinfo {pages} {11398} (\bibinfo {year}
	{1999})}\BibitemShut {NoStop}%
\bibitem [{\citenamefont {Kawashima}(2004)}]{Kawashima3219}%
\BibitemOpen
\bibfield  {author} {\bibinfo {author} {\bibfnamefont {N.}~\bibnamefont
		{Kawashima}},\ }\bibfield  {title} {\enquote {\bibinfo {title} {Quantum
			critical point of the {$\rm XY$} model and condensation of field-induced
			quasiparticles in dimer compounds},}\ }\href {\doibase 10.1143/JPSJ.73.3219}
{\bibfield  {journal} {\bibinfo  {journal} {J. Phys. Soc. Jpn.}\ }\textbf
	{\bibinfo {volume} {73}},\ \bibinfo {pages} {3219} (\bibinfo {year}
	{2004})}\BibitemShut {NoStop}%
\bibitem [{\citenamefont {Yamada}\ \emph {et~al.}(2011)\citenamefont {Yamada},
	\citenamefont {Tanaka}, \citenamefont {Ono},\ and\ \citenamefont
	{Nojiri}}]{Yamada020409}%
\BibitemOpen
\bibfield  {author} {\bibinfo {author} {\bibfnamefont {F.}~\bibnamefont
		{Yamada}}, \bibinfo {author} {\bibfnamefont {H.}~\bibnamefont {Tanaka}},
	\bibinfo {author} {\bibfnamefont {T.}~\bibnamefont {Ono}}, \ and\ \bibinfo
	{author} {\bibfnamefont {H.}~\bibnamefont {Nojiri}},\ }\bibfield  {title}
{\enquote {\bibinfo {title} {Transition from {$\rm Bose$} glass to a
			condensate of triplons in {Tl$_{1-x}$K$_x$CuCl$_3$}},}\ }\href {\doibase
	10.1103/PhysRevB.83.020409} {\bibfield  {journal} {\bibinfo  {journal} {Phys.
			Rev. B}\ }\textbf {\bibinfo {volume} {83}},\ \bibinfo {pages} {020409}
	(\bibinfo {year} {2011})}\BibitemShut {NoStop}%
\bibitem [{\citenamefont {Yamada}\ \emph {et~al.}(2008)\citenamefont {Yamada},
	\citenamefont {Ono}, \citenamefont {Tanaka}, \citenamefont {Misguich},
	\citenamefont {Oshikawa},\ and\ \citenamefont {Sakakibara}}]{Yamada013701}%
\BibitemOpen
\bibfield  {author} {\bibinfo {author} {\bibfnamefont {F.}~\bibnamefont
		{Yamada}}, \bibinfo {author} {\bibfnamefont {T.}~\bibnamefont {Ono}},
	\bibinfo {author} {\bibfnamefont {H.}~\bibnamefont {Tanaka}}, \bibinfo
	{author} {\bibfnamefont {G.}~\bibnamefont {Misguich}}, \bibinfo {author}
	{\bibfnamefont {M.}~\bibnamefont {Oshikawa}}, \ and\ \bibinfo {author}
	{\bibfnamefont {T.}~\bibnamefont {Sakakibara}},\ }\bibfield  {title}
{\enquote {\bibinfo {title} {Magnetic-field induced {$\rm Bose-Einstein$}
			condensation of magnons and critical behavior in interacting spin dimer
			system {TlCuCl$_3$}},}\ }\href {\doibase 10.1143/JPSJ.77.013701} {\bibfield
	{journal} {\bibinfo  {journal} {J. Phys. Soc. Jpn.}\ }\textbf {\bibinfo
		{volume} {77}},\ \bibinfo {pages} {013701} (\bibinfo {year}
	{2008})}\BibitemShut {NoStop}%
\bibitem [{\citenamefont {Zapf}\ \emph {et~al.}(2006)\citenamefont {Zapf},
	\citenamefont {Zocco}, \citenamefont {Hansen}, \citenamefont {Jaime},
	\citenamefont {Harrison}, \citenamefont {Batista}, \citenamefont
	{Kenzelmann}, \citenamefont {Niedermayer}, \citenamefont {Lacerda},\ and\
	\citenamefont {Paduan-Filho}}]{Zapf077204}%
\BibitemOpen
\bibfield  {author} {\bibinfo {author} {\bibfnamefont {V.~S.}\ \bibnamefont
		{Zapf}}, \bibinfo {author} {\bibfnamefont {D.}~\bibnamefont {Zocco}},
	\bibinfo {author} {\bibfnamefont {B.~R.}\ \bibnamefont {Hansen}}, \bibinfo
	{author} {\bibfnamefont {M.}~\bibnamefont {Jaime}}, \bibinfo {author}
	{\bibfnamefont {N.}~\bibnamefont {Harrison}}, \bibinfo {author}
	{\bibfnamefont {C.~D.}\ \bibnamefont {Batista}}, \bibinfo {author}
	{\bibfnamefont {M.}~\bibnamefont {Kenzelmann}}, \bibinfo {author}
	{\bibfnamefont {C.}~\bibnamefont {Niedermayer}}, \bibinfo {author}
	{\bibfnamefont {A.}~\bibnamefont {Lacerda}}, \ and\ \bibinfo {author}
	{\bibfnamefont {A.}~\bibnamefont {Paduan-Filho}},\ }\bibfield  {title}
{\enquote {\bibinfo {title} {{$\rm Bose-Einstein$} condensation of {$S=1$}
			nickel spin degrees of freedom in {NiCl$_2$-4SC(NH$_2$)$_2$}},}\ }\href
{\doibase 10.1103/PhysRevLett.96.077204} {\bibfield  {journal} {\bibinfo
		{journal} {Phys. Rev. Lett.}\ }\textbf {\bibinfo {volume} {96}},\ \bibinfo
	{pages} {077204} (\bibinfo {year} {2006})}\BibitemShut {NoStop}%
\bibitem [{\citenamefont {Yin}\ \emph {et~al.}(2008)\citenamefont {Yin},
	\citenamefont {Xia}, \citenamefont {Zapf}, \citenamefont {Sullivan},\ and\
	\citenamefont {Paduan-Filho}}]{Yin1872005}%
\BibitemOpen
\bibfield  {author} {\bibinfo {author} {\bibfnamefont {L.}~\bibnamefont
		{Yin}}, \bibinfo {author} {\bibfnamefont {J.~S.}\ \bibnamefont {Xia}},
	\bibinfo {author} {\bibfnamefont {V.~S.}\ \bibnamefont {Zapf}}, \bibinfo
	{author} {\bibfnamefont {N.~S.}\ \bibnamefont {Sullivan}}, \ and\ \bibinfo
	{author} {\bibfnamefont {A.}~\bibnamefont {Paduan-Filho}},\ }\bibfield
{title} {\enquote {\bibinfo {title} {Direct measurement of the {$\rm
				Bose-Einstein$} condensation universality class in {NiCl$_2$-4SC(NH$_2$)$_2$}
			at ultralow temperatures},}\ }\href {\doibase 10.1103/PhysRevLett.101.187205}
{\bibfield  {journal} {\bibinfo  {journal} {Phys. Rev. Lett.}\ }\textbf
	{\bibinfo {volume} {101}},\ \bibinfo {pages} {187205} (\bibinfo {year}
	{2008})}\BibitemShut {NoStop}%
\bibitem [{\citenamefont {Tsujii}\ \emph {et~al.}(2009)\citenamefont {Tsujii},
	\citenamefont {Kim}, \citenamefont {Yoshida}, \citenamefont {Takano},
	\citenamefont {Murphy}, \citenamefont {Kanada}, \citenamefont {Saito},
	\citenamefont {Oosawa},\ and\ \citenamefont {Goto}}]{Tsujii042217}%
\BibitemOpen
\bibfield  {author} {\bibinfo {author} {\bibfnamefont {H.}~\bibnamefont
		{Tsujii}}, \bibinfo {author} {\bibfnamefont {Y.~H.}\ \bibnamefont {Kim}},
	\bibinfo {author} {\bibfnamefont {Y.}~\bibnamefont {Yoshida}}, \bibinfo
	{author} {\bibfnamefont {Y.}~\bibnamefont {Takano}}, \bibinfo {author}
	{\bibfnamefont {T.~P.}\ \bibnamefont {Murphy}}, \bibinfo {author}
	{\bibfnamefont {K.}~\bibnamefont {Kanada}}, \bibinfo {author} {\bibfnamefont
		{T.}~\bibnamefont {Saito}}, \bibinfo {author} {\bibfnamefont
		{A.}~\bibnamefont {Oosawa}}, \ and\ \bibinfo {author} {\bibfnamefont
		{T.}~\bibnamefont {Goto}},\ }\bibfield  {title} {\enquote {\bibinfo {title}
		{Magnetic phase diagram of the {$S = 1/2$} antiferromagnetic ladder
			{(CH$_3$)$_2$CHNH$_3$CuCl$_3$}},}\ }\href {\doibase
	10.1088/1742-6596/150/4/042217} {\bibfield  {journal} {\bibinfo  {journal}
		{J. Phys. Conf. Ser.}\ }\textbf {\bibinfo {volume} {150}},\ \bibinfo {pages}
	{042217} (\bibinfo {year} {2009})}\BibitemShut {NoStop}%
\bibitem [{\citenamefont {Syromyatnikov}(2007)}]{Syromyatnikov134421}%
\BibitemOpen
\bibfield  {author} {\bibinfo {author} {\bibfnamefont {A.~V.}\ \bibnamefont
		{Syromyatnikov}},\ }\bibfield  {title} {\enquote {\bibinfo {title} {{$\rm
				Bose-Einstein$} condensation of magnons in magnets with predominant
			ferromagnetic interactions},}\ }\href {\doibase 10.1103/PhysRevB.75.134421}
{\bibfield  {journal} {\bibinfo  {journal} {Phys. Rev. B}\ }\textbf {\bibinfo
		{volume} {75}},\ \bibinfo {pages} {134421} (\bibinfo {year}
	{2007})}\BibitemShut {NoStop}%
\bibitem [{\citenamefont {Zheludev}\ \emph {et~al.}(2007)\citenamefont
	{Zheludev}, \citenamefont {Garlea}, \citenamefont {Masuda}, \citenamefont
	{Manaka}, \citenamefont {Regnault}, \citenamefont {Ressouche}, \citenamefont
	{Grenier}, \citenamefont {Chung}, \citenamefont {Qiu}, \citenamefont
	{Habicht}, \citenamefont {Kiefer},\ and\ \citenamefont
	{Boehm}}]{Zheludev054450}%
\BibitemOpen
\bibfield  {author} {\bibinfo {author} {\bibfnamefont {A.}~\bibnamefont
		{Zheludev}}, \bibinfo {author} {\bibfnamefont {V.~O.}\ \bibnamefont
		{Garlea}}, \bibinfo {author} {\bibfnamefont {T.}~\bibnamefont {Masuda}},
	\bibinfo {author} {\bibfnamefont {H.}~\bibnamefont {Manaka}}, \bibinfo
	{author} {\bibfnamefont {L.-P.}\ \bibnamefont {Regnault}}, \bibinfo {author}
	{\bibfnamefont {E.}~\bibnamefont {Ressouche}}, \bibinfo {author}
	{\bibfnamefont {B.}~\bibnamefont {Grenier}}, \bibinfo {author} {\bibfnamefont
		{J.-H.}\ \bibnamefont {Chung}}, \bibinfo {author} {\bibfnamefont
		{Y.}~\bibnamefont {Qiu}}, \bibinfo {author} {\bibfnamefont {K.}~\bibnamefont
		{Habicht}}, \bibinfo {author} {\bibfnamefont {K.}~\bibnamefont {Kiefer}}, \
	and\ \bibinfo {author} {\bibfnamefont {M.}~\bibnamefont {Boehm}},\ }\bibfield
{title} {\enquote {\bibinfo {title} {Dynamics of quantum spin liquid and
			spin solid phases in {IPA-CuCl$_3$} under an applied magnetic field studied
			with neutron scattering},}\ }\href {\doibase 10.1103/PhysRevB.76.054450}
{\bibfield  {journal} {\bibinfo  {journal} {Phys. Rev. B}\ }\textbf {\bibinfo
		{volume} {76}},\ \bibinfo {pages} {054450} (\bibinfo {year}
	{2007})}\BibitemShut {NoStop}%
\bibitem [{\citenamefont {Conner}\ \emph {et~al.}(2010)\citenamefont {Conner},
	\citenamefont {Zhou}, \citenamefont {Jo}, \citenamefont {Balicas},
	\citenamefont {Wiebe}, \citenamefont {Carlo}, \citenamefont {Uemura},
	\citenamefont {Aczel}, \citenamefont {Williams},\ and\ \citenamefont
	{Luke}}]{Conner132401}%
\BibitemOpen
\bibfield  {author} {\bibinfo {author} {\bibfnamefont {B.~S.}\ \bibnamefont
		{Conner}}, \bibinfo {author} {\bibfnamefont {H.~D.}\ \bibnamefont {Zhou}},
	\bibinfo {author} {\bibfnamefont {Y.~J.}\ \bibnamefont {Jo}}, \bibinfo
	{author} {\bibfnamefont {L.}~\bibnamefont {Balicas}}, \bibinfo {author}
	{\bibfnamefont {C.~R.}\ \bibnamefont {Wiebe}}, \bibinfo {author}
	{\bibfnamefont {J.~P.}\ \bibnamefont {Carlo}}, \bibinfo {author}
	{\bibfnamefont {Y.~J.}\ \bibnamefont {Uemura}}, \bibinfo {author}
	{\bibfnamefont {A.~A.}\ \bibnamefont {Aczel}}, \bibinfo {author}
	{\bibfnamefont {T.~J.}\ \bibnamefont {Williams}}, \ and\ \bibinfo {author}
	{\bibfnamefont {G.~M.}\ \bibnamefont {Luke}},\ }\bibfield  {title} {\enquote
	{\bibinfo {title} {Possible {Bose-Einstein} condensate of magnons in
			single-crystalline {Pb$_2$V$_3$O$_9$}},}\ }\href {\doibase
	10.1103/PhysRevB.81.132401} {\bibfield  {journal} {\bibinfo  {journal} {Phys.
			Rev. B}\ }\textbf {\bibinfo {volume} {81}},\ \bibinfo {pages} {132401}
	(\bibinfo {year} {2010})}\BibitemShut {NoStop}%
\bibitem [{\citenamefont {Waki}\ \emph {et~al.}(2004)\citenamefont {Waki},
	\citenamefont {Morimoto}, \citenamefont {Michioka}, \citenamefont {Kato},
	\citenamefont {Kageyama}, \citenamefont {Yoshimura}, \citenamefont
	{Nakatsuji}, \citenamefont {Sakai}, \citenamefont {Maeno}, \citenamefont
	{Mitamura},\ and\ \citenamefont {Goto}}]{Waki3435}%
\BibitemOpen
\bibfield  {author} {\bibinfo {author} {\bibfnamefont {T.}~\bibnamefont
		{Waki}}, \bibinfo {author} {\bibfnamefont {Y.}~\bibnamefont {Morimoto}},
	\bibinfo {author} {\bibfnamefont {C.}~\bibnamefont {Michioka}}, \bibinfo
	{author} {\bibfnamefont {M.}~\bibnamefont {Kato}}, \bibinfo {author}
	{\bibfnamefont {H.}~\bibnamefont {Kageyama}}, \bibinfo {author}
	{\bibfnamefont {K.}~\bibnamefont {Yoshimura}}, \bibinfo {author}
	{\bibfnamefont {S.}~\bibnamefont {Nakatsuji}}, \bibinfo {author}
	{\bibfnamefont {O.}~\bibnamefont {Sakai}}, \bibinfo {author} {\bibfnamefont
		{Y.}~\bibnamefont {Maeno}}, \bibinfo {author} {\bibfnamefont
		{H.}~\bibnamefont {Mitamura}}, \ and\ \bibinfo {author} {\bibfnamefont
		{T.}~\bibnamefont {Goto}},\ }\bibfield  {title} {\enquote {\bibinfo {title}
		{Observation of {$\rm Bose-Einstein$} condensation of triplons in quasi {1D}
			spin-gap system {Pb$_2$V$_3$O$_9$}},}\ }\href {\doibase 10.1143/JPSJ.73.3435}
{\bibfield  {journal} {\bibinfo  {journal} {J. Phys. Soc. Jpn.}\ }\textbf
	{\bibinfo {volume} {73}},\ \bibinfo {pages} {3435--3438} (\bibinfo {year}
	{2004})}\BibitemShut {NoStop}%
\end{thebibliography}
%
\end{document}